\documentclass[12pt,twoside, onecolumn,journal,draftclsnofoot]{IEEEtran}

\usepackage{cuted}
\usepackage{mathrsfs,amssymb,amsmath,latexsym,array, arydshln, version,bm}
\usepackage{cite}
\usepackage{graphicx, epsfig, epic, eepic, color, subfigure}

\newtheorem{definition}{Definition}
\newtheorem{lemma}{Lemma}
\newtheorem{theorem}{Theorem}
\newtheorem{proposition}{Proposition}
\newtheorem{corollary}{Corollary}

\DeclareMathOperator{\tr}{tr}

\begin{document}

\title{Asymptotic FRESH Properizer for Block Processing of Improper-Complex Second-Order Cyclostationary Random Processes}
\author{Jeongho~Yeo and~Joon~Ho~Cho \emph{Member, IEEE}
\thanks{

%

The material in this paper was presented in part at the IEEE Military Communications Conference (MILCOM), Orlando, FL, 29 Oct.-1 Nov. 2012 and in part at the IEEE Wireless Communications and Networking Conference (WCNC), Shanghai, China, 7-10 Apr. 2013.}
\thanks{The authors are with the Department of Electrical Engineering, Pohang University of Science and Technology (POSTECH), Pohang, Gyeongbuk 790-784, Korea (e-mail: \{yjh2304, jcho\}@postech.ac.kr).}}

\date{ }
\markboth{Submitted to IEEE Transactions on Information Theory (Ver 1.0)}{Submitted to IEEE Transactions on Information Theory (Ver 1.0)}

\maketitle
\vspace{-0.3in}
\begin{abstract}
In this paper, the block processing of a discrete-time (DT) improper-complex second-order cyclostationary (SOCS) random process is considered.
In particular, it is of interest to find a pre-processing operation that enables computationally efficient near-optimal post-processing.
An invertible linear-conjugate linear (LCL) operator named the DT FREquency Shift (FRESH) properizer is first proposed.
It is shown that the DT FRESH properizer converts a DT improper-complex SOCS random process input to an equivalent DT proper-complex SOCS random process output by utilizing the information only about the cycle period of the input.
An invertible LCL block processing operator named the asymptotic FRESH properizer is then proposed that mimics the operation of the DT FRESH properizer but processes a finite number of consecutive samples of a DT improper-complex SOCS random process.
It is shown that the output of the asymptotic FRESH properizer is not proper but asymptotically proper and that its frequency-domain covariance matrix converges to a highly-structured block matrix with diagonal blocks as the block size tends to infinity.
Two representative estimation and detection problems are presented to demonstrate that asymptotically optimal low-complexity post-processors can be easily designed by exploiting these asymptotic second-order properties of the output of the asymptotic FRESH properizer.
\end{abstract}
\begin{IEEEkeywords}
Asymptotic analysis, complexity reduction, improper-complex random process, properization, second-order cyclostationarity
\end{IEEEkeywords}
\section{Introduction}

\IEEEPARstart{I}{t} is well known that the complex envelope of a real-valued bandpass wide-sense stationary (WSS) random process is proper, i.e., its complementary auto-covariance (a.k.a.~the pseudo-covariance) function vanishes \cite{Massey_93}.
However, there are a lot of important improper-complex random processes that do not have vanishing complementary auto-covariance functions \cite{Scharf_03, Schreier_08, Taubock_12}.
For example, the complex envelope of a real-valued bandpass nonstationary signal is not necessarily proper and even the complex envelope of a real-valued bandpass WSS signal becomes improper in the presence of the imbalance between its in-phase and quadrature components \cite{Schreier_10}.
In order to fully capture the statistical properties of a complex-valued signal, including all the second-order statistics, the filtering of augmented signals has been proposed.
This so-called widely linear (WL) filtering processes either the signal augmented by its complex conjugate or the real part of the signal augmented by the imaginary part \cite{Picinbono_95, Schreier_10}, where the former is referred to as the linear-conjugate linear (LCL) filtering \cite{Brown_69}.

On the other hand, many digitally modulated signals are well modeled by wide-sense cyclostationary (WSCS) random processes \cite{Giannakis_05, Gardner_06}, whose complex envelopes possess periodicity in all variables of the mean and the auto-covariance functions.
To efficiently extract the correlation structure of a WSCS random process in the time and the frequency domains, the translation series representation (TSR) and the harmonic series representation (HSR) are proposed \cite{Gardner_75}, which are linear periodically time-varying (PTV) processings.
Using these representations, it is shown \cite{Gardner_75} that a proper-complex WSCS scalar random process can be converted to an equivalent proper-complex WSS vector random process.
Such second-order structure of a proper-complex WSCS random process has long been exploited in the design of many communications and signal processing systems including presence detectors \cite{Gardner_88, Gardner_92}, estimators \cite{Gardner_87_2, Gardner_89, Giannakis_97}, and optimal transceivers \cite{Mitra_01, Cho_04, Yun_10} under various criteria.

The complex envelopes of the majority of digitally modulated signals are proper and WSCS.
However, as well documented in \cite{Schreier_10}, there still remain many other digitally modulated signals such as pulse amplitude modulation (PAM), offset quaternary phase-shift keying (OQPSK), and Gaussian minimum shift keying (GMSK), that are not only improper-complex WSCS but also second-order cyclostationary (SOCS) \cite{Han_12}, i.e., the complementary auto-covariance function is also periodic in all its variables with the same period as the mean and the auto-covariance functions.
To exploit both cyclostationarity and impropriety, the LCL FREquency Shift (FRESH) filter has been proposed in \cite{Gardner_93} that combines signal augmentation and linear PTV processing.

Recently, another LCL PTV operator called the properizing FRESH (p-FRESH) vectorizer is proposed in \cite{Han_12, Han_12_2} by non-trivially extending the HSR.
The p-FRESH vectorizer converts an \emph{improper-complex} SOCS scalar random process to an equivalent \emph{proper-complex} WSS vector random process by exploiting the frequency-domain correlation and complementary correlation structures that are rigorously examined in \cite{Schreier_08}.
By successfully deriving the capacity of an SOCS Gaussian noise channel, it is shown that the optimal channel input is improper-complex SOCS in general when the interfering signal is improper-complex SOCS.
It is also well demonstrated that such properization provides the advantage of enabling the adoption of the conventional signal processing techniques and algorithms that utilize only the correlation but not the complementary correlation structure.
These results warrant further research in communications and signal processing on the efficient construction and processing of improper-complex SOCS random processes.

In this paper, we consider the block processing of a complex-valued random vector that is obtained by taking a finite number of consecutive samples from a discrete-time (DT) improper-complex SOCS random process.
To proceed, the p-FRESH vectorization developed for continuous-time (CT) SOCS random processes is first extended to DT improper-complex SOCS random processes.
Instead of straightforwardly modifying the CT p-FRESH vectorizer that properizes as well as vectorizes the input CT improper-complex SOCS random process, an invertible LCL operator is proposed in this paper that does not vectorize but only properizes the input DT improper-complex SOCS random process.
Thus, the LCL operator is named the DT FRESH properizer and its output does not need vector processing.
Specifically, the DT FRESH properizer transforms a DT improper-complex SOCS scalar random process into an equivalent proper-complex SOCS scalar random process with the cycle period that is twice the cycle period of the input process.

To make this idea of pre-processing by properization better suited for digital signal processing, an invertible LCL block processing operator is then proposed that mimics the operation of the DT FRESH properizer.
Although the augmentation of the improper-complex random vector by its complex conjugate can generate a sufficient statistic \cite{Schreier_10}, it is not only a redundant information of twice the length of the original observation vector but also improper.
Although the strong uncorrelating transform (SUT) of the observation vector can generate a highly-structured sufficient statistic of the same length as the original observation vector \cite{Schreier_10, Eriksson_06}, the output is still improper and, moreover, it requires the information about both the correlation and the complementary correlation matrices of the improper-complex random vector input for the transformation.
Thus, neither the augmenting pre-processor nor the SUT allows the direct application of the conventional techniques and algorithms dedicated to the block processing of proper-complex random vectors.
Motivated by how the DT FRESH properizer works in the frequency domain, the pre-processor proposed in this paper utilizes the centered discrete Fourier transform (DFT) and the information only about the cycle period of the DT improper-complex SOCS random process in order to convert a finite number of consecutive samples of the random process to an equivalent random vector.
Unlike the output of the DT FRESH properizer, the output of this LCL block processing operator is not proper but approximately proper for sufficiently large block size.
Thus, the pre-processor is named the asymptotic FRESH properizer.
Specifically, the asymptotic FRESH properizer makes the sequence of the complementary covariance matrices of the output asymptotically equivalent \cite{Gray_book} to the sequence of all-zero matrices.

In \cite{Yoo_10}, it is shown that the complex-valued random vector consisting of a finite number of consecutive samples of a DT \emph{proper-complex} SOCS random process has its frequency-domain covariance matrix that approaches a block matrix with diagonal blocks as the number of samples increases.
This is because the periodicity in the second-order statistics of the DT SOCS random process naturally leads to a sequence of block Toeplitz covariance matrices as the number of samples increases, the sequence of block Toeplitz matrices is asymptotically equivalent to a sequence of block circulant matrices, and a block circulant matrix becomes a block matrix with diagonal blocks when pre- and post-multiplied by DFT and inverse DFT matrices, respectively.
Since the asymptotic FRESH properizer mimics the DT FRESH properizer and the DT FRESH properizer outputs a DT \emph{proper-complex} SOCS random process, it naturally becomes of interest to examine the asymptotic property of the frequency-domain covariance matrix of the asymptotic FRESH properizer output.
It turns out that the output of the asymptotic FRESH properizer has the same property discovered in \cite{Yoo_10}.

Such properties of the covariance and the complementary covariance matrices may be used in designing, under various optimality criteria, low-complexity post-processors that follow the asymptotic FRESH properizer.
In particular, a post-processor can be developed that approximates the output of the asymptotic FRESH properizer by a proper-complex random vector having the block matrix with diagonal blocks that is asymptotically equivalent to the exact frequency-domain covariance matrix as its frequency-domain covariance matrix.
Of course, this technique makes the post-processor suboptimal that performs most of the main operations in the frequency domain.
As the two asymptotic properties strongly suggest, however, if the block size is large enough, then the performance degradation may be negligible.
It is already shown in \cite{Yoo_10} that this is the case for the asymptotic property only of the covariance matrix, where a suboptimal frequency-domain equalizer is proposed that approximates the frequency-domain covariance matrix of a proper-complex SOCS interference by an asymptotically equivalent block matrix with diagonal blocks.
It turns out that this equalizer not only achieves significantly lower computational complexity than the exact linear minimum mean-square error (LMMSE) equalizer by exploiting the block structure of the frequency-domain covariance matrix, but also is asymptotically optimal in the sense that its average mean-squared error (MSE) approaches that of the LMMSE equalizer as the number of samples tends to infinity.

To demonstrate the simultaneous achievability of asymptotic optimality and low complexity by employing the post-processor that processes the output of the asymptotic FRESH properizer and exploits the two asymptotic properties, we consider two representative estimation and detection problems.
First, for a DT improper-complex SOCS random signal in additive proper-complex white noise, a low-complexity signal estimator is proposed that is a linear function of the output of the asymptotic FRESH properizer.
It is shown that the average MSE performance of the estimator approaches that of the widely linear minimum mean-squared error (WLMMSE) estimator as the number of samples tends to infinity.
Second, for a DT improper-complex SOCS Gaussian random signal in additive proper-complex white Gaussian noise, a low-complexity signal presence detector is proposed whose test statistic is a quadratic function of the output of the asymptotic FRESH properizer
It is shown that the test statistic converges to the exact likelihood ratio test (LRT) statistic that is a quadratic function of the augmented observation vector with probability one (w.p.~$1$) as the number of samples tends to infinity.
Note that in both cases the asymptotic FRESH properizer as the pre-processor utilizes only the information about the cycle period.
Thus, the adoption of adaptive estimation and detection algorithms may be possible that are developed for the processing of proper-complex random vectors.
This advantage of employing the asymptotic FRESH properizer may be taken in other communications and signal processing problems involving the block processing of DT improper-complex SOCS random processes.

The rest of this paper is organized as follows.
In Section~II, definitions and lemmas related to SOCS random processes are provided and the DT FRESH properizer is proposed.
In Section~III, the asymptotic FRESH properizer is proposed and the second-order properties of its output are analyzed.
In Sections~IV and V, the application of the asymptotic FRESH properizer is considered to exemplary estimation and detection problems, respectively.
Finally, concluding remarks are offered in Section~VI.

Throughout this paper, the operator $\mathbf{E}\{\cdot\}$ denotes the expectation, the operator $\mathscr{F}\{x[n]\}\triangleq \sum_{n=-\infty}^{\infty}x[n] e^{-{\rm j 2} \pi fn }$ denotes the discrete-time Fourier transform (DTFT) of an absolutely summa-ble sequence $(x[n])_n$, and the function $\delta(\cdot)$ denotes the Dirac delta function.
The sets ${\mathbb{Z}}$ and ${\mathbb{N}}$ are the sets of all integers and of all positive integers, respectively.
The operator $\times$ denotes the Cartesian product between two sets.
The superscripts $^*$, $^\mathcal{T}$, and $^\mathcal{H}$ denote the complex conjugation, the transpose, and the Hermitian transpose, respectively.
The operator $*$ denotes the convolution.\footnote{There should be no confusion from the superscript $^*$ that denotes the complex conjugation.}
The operators $\odot$ and $\otimes$ denote the Hadamard product and the Kronecker product, respectively.
The matrices $\bm{1}_{N}$, $\bm{I}_{N}$, $\bm{O}_{N}$, and $\bm{O}_{M,N}$ denote the $N$-by-$N$ all-one matrix, the $N$-by-$N$ identity matrix, the $N$-by-$N$ all-zero matrix, and the $M$-by-$N$ all-zero matrix, respectively.
The matrix $\bm{P}_{N}$ denotes the $N$-by-$N$ backward identity matrix whose $(m,n)$th entry is given by $1$ for $m+n=N+1$, and $0$ otherwise.
The operator $[\bm{A}]_{m,n}$ and $\rm{tr}\{ \bm{A} \}$ denote the $(m,n)$th entry and the trace of a matrix $\bm{A}$, respectively.
To describe the computational complexity, we will use the big-O notation $\mathcal{O}(g(N))$ defined as $f(N) = \mathcal{O}(g(N))$ if and only if there exist a positive constant $M$ and a real number $N_0$ such that $|f(N)| \leq M|g(N)|, \;\forall N>N_0$.

\section{DT FRESH Properizer}

In this section, the notion of DT second-order cyclostationarity is introduced and an LCL PTV operator is proposed that converts a DT improper-complex SOCS random process into an equivalent DT proper-complex SOCS random process.
Similar to the p-FRESH vectorizer proposed in \cite{Han_12}, where an input CT SOCS random process is converted to an equivalent CT proper-complex WSS vector random process, this operator as a pre-processor enables the adoption of the conventional signal processing techniques and algorithms that utilize only the correlation but not the complementary correlation structure of the signal.
Note that, unlike the p-FRESH vectorizer, this DT operator does not vectorize but only properizes the input improper-complex SOCS random process.
A CT version of this properizer can be found in \cite{Yeo_12}.

\subsection{DT SOCS Random Processes}

In this subsection, definitions and lemmas related to the DT SOCS random processes are provided.

\begin{definition}\label{definition: random process}
Given a DT complex-valued random process $X[n]$ with a finite power, i.e., $\mathbf{E}\{|X[n]|^2\}<\infty, \forall n$, the mean, the auto-correlation, the auto-covariance, the complementary auto-correlation, and the complementary auto-covariance functions of $X[n]$ are defined, respectively, as
\begin{IEEEeqnarray}{rCl}
\mu_X[n]&\triangleq& \mathbf{E}\{X[n]\},\IEEEeqnarraynumspace\IEEEyessubnumber\\
r_X[n,m]&\triangleq& \mathbf{E}\{X[n]X[m]^*\}, \IEEEeqnarraynumspace\IEEEyessubnumber\\
c_X[n,m]&\triangleq& \mathbf{E}\{(X[n]-\mu_X[n])(X[m]-\mu_X[m])^*\}, \IEEEeqnarraynumspace\IEEEyessubnumber\\
\tilde{r}_X[n,m]&\triangleq& \mathbf{E}\{X[n]X[m]\}, \rm{ and} \IEEEeqnarraynumspace\IEEEyessubnumber\\
\tilde{c}_X[n,m]&\triangleq& \mathbf{E}\{(X[n]-\mu_X[n])(X[m]-\mu_X[m])\}.\IEEEeqnarraynumspace\IEEEyessubnumber
\end{IEEEeqnarray}
\end{definition}

Throughout this paper, all DT complex-valued random processes are assumed to be of finite power, i.e., $|r_X[n,m]|\leq{\mathbf{E}}\{|X[n]|^2\} <\infty, \forall n, \forall m$.

\begin{definition}\label{definition: 2D Fourier}
The two-dimensional (2-D) power spectral density (PSD) $R_X(f,f')$ and the 2-D complementary PSD $\tilde{R}_X(f,f')$ of a DT complex-valued random process $X[n]$ are defined as
\begin{IEEEeqnarray}{rCl}
R_X(f,f')&\triangleq & \sum_{m=-\infty}^{\infty}\sum_{n=-\infty}^{\infty} r_X[n,m] e^{-{\rm j 2} \pi (fn-f'm) } \IEEEeqnarraynumspace\IEEEyessubnumber\label{eq: def_R_X_a}\\
\noalign{\noindent{\text{and}}\vspace{\jot}}
\tilde{R}_X(f,f')&\triangleq &\sum_{m=-\infty}^{\infty}\sum_{n=-\infty}^{\infty}\tilde{r}_X[n,m]e^{-{\rm j 2} \pi (fn-f'm) },\IEEEeqnarraynumspace\IEEEyessubnumber\label{eq: def_R_X_b}
\end{IEEEeqnarray}
respectively, if they exist.
\end{definition}

Since the 2-D PSD and the 2-D complementary PSD are the DT double Fourier transforms of $r_X[n,m]$ and $\tilde{r}_X[n,m]$, respectively, they are always periodic in both variables $f$ and $f'$ with the common period $1$.
The set of all DT complex-valued random processes can be partitioned into two subsets by using the following definition.

\begin{definition}{\cite[Definition~2]{Massey_93}} %
A DT complex-valued random process $X[n]$ is proper if its complementary auto-covariance function vanishes, i.e., $\tilde{c}_X[n,m]= 0, \forall n, \forall m,$ and is improper otherwise.
\end{definition}

Two types of stationarity can be defined as follows by using the second-order moments of a DT complex-valued random process.

\begin{definition}\label{definition: SOS}{\cite[Section~II-B]{Picinbono_97}} %
A DT complex-valued random process $X[n]$ is second-order stationary (SOS) if, $\forall m, \forall n,$
\begin{IEEEeqnarray}{rCl}
\mu_X[n]&=&\mu_X[0], \IEEEeqnarraynumspace\IEEEyessubnumber\label{eq: SOS_1}\\
c_X[n,m] &=&c_X[n-m,0], \rm{ and} \IEEEeqnarraynumspace\IEEEyessubnumber\label{eq: SOS_2}\\
\tilde{c}_X[n,m]&=&\tilde{c}_X[n-m,0].\IEEEeqnarraynumspace\IEEEyessubnumber\label{eq: SOS_3}
\end{IEEEeqnarray}
\end{definition}

\begin{definition}\label{definition: SOCS}
A DT complex-valued random process $X[n]$ is SOCS with cycle period $M\in \mathbb{N}$ if, $\forall n, \forall m,$
\begin{IEEEeqnarray}{rCl}
\mu_X[n]&=&\mu_X[n+M], \IEEEeqnarraynumspace\IEEEyessubnumber\\
c_X[n,m]&=&c_X[n+M, m+M], \rm{ and} \IEEEeqnarraynumspace\IEEEyessubnumber\\
\tilde{c}_X[n,m]&=&\tilde{c}_X[n+M, m+M].\IEEEeqnarraynumspace\IEEEyessubnumber\label{eq: SOCS}
\end{IEEEeqnarray}
\end{definition}

We are mainly interested in the DT SOCS random processes in this paper.
Note that a DT SOS random process can be viewed as a special DT SOCS random process with cycle period $1$.
Note also that the above definition of a DT SOCS random process is a straightforward extension of the definition of a CT SOCS random process in \cite{Han_12}.
For the ease of comparison with the results in \cite{Han_12}, we use the time indexes $m$ and $n$ in the orders appearing in Definitions~\ref{definition: random process}-\ref{definition: SOCS}.

In the following lemmas, the implications of the second-order cyclostationarity are provided in the time and the frequency domains, respectively.

\begin{lemma}\label{lemma: series time}
For a DT SOCS random process $X[n]$ with cycle period $M\in{\mathbb{N}}$, there exist $(r_X^{(k)}[n])_{k=0}^{M-1}$ and $(\tilde{r}_X^{(k)}[n])_{k=0}^{M-1}$ such that
\begin{IEEEeqnarray}{rCl}
r_X[n,m]&=&\sum_{k=0}^{M-1}r_X^{(k)}[n-m]e^{{\rm j} 2\pi k n / M},\IEEEeqnarraynumspace\IEEEyessubnumber\label{eq: correl_expand}\\
\noalign{\noindent{\text{and}}\vspace{\jot}}
\tilde{r}_X[n,m]&=&\sum_{k=0}^{M-1}\tilde{r}_X^{(k)}[n-m]e^{{\rm j} 2\pi k n / M}.\IEEEeqnarraynumspace\IEEEyessubnumber\label{eq: com_correl_expand}
\end{IEEEeqnarray}
\end{lemma}

\begin{IEEEproof}
Since $r'_X[n,l]\triangleq r_X[n,n-l]$ is periodic in $n$ with period $M$ and is finite, there exist the DT Fourier series coefficients $(r_X^{(k)}[l])_{k=0}^{M-1}$ for each $l$ such that $r'_X[n,l]=\sum_{k=0}^{M-1}r_X^{(k)}[l]e^{{\rm j} 2\pi k n / M}$.
By replacing $l$ with $n-m$, we obtain (\ref{eq: correl_expand}).
Similarly, we obtain (\ref{eq: com_correl_expand}).
Therefore, the conclusion follows.
\end{IEEEproof}

\begin{lemma}\label{lemma: PSD_impulse}
For a DT SOCS random process $X[n]$ with cycle period $M\in{\mathbb{N}}$, the 2-D PSD and the 2-D complementary PSD are given, respectively, by
\begin{IEEEeqnarray}{rCl}\label{eq: R_X_all}
R_X(f,f')&=&\!\sum_{l=-\infty}^{\infty}\!\sum_{k=0}^{M-1}R_X^{(k)}\!\!\left( f-\frac{k}{M}\right) \! \delta \! \left( f- f'-\frac{k}{M}-l\right) \nonumber\\
&& \IEEEeqnarraynumspace\IEEEyessubnumber\label{eq: R_X_a}\\
\noalign{\noindent{\text{and}}\vspace{\jot}}
\tilde{R}_X(f,f')&=&\!\sum_{l=-\infty}^{\infty}\!\sum_{k=0}^{M-1}\tilde{R}_X^{(k)}\!\!\left( f-\frac{k}{M}\right) \!\delta \! \left(f-f'-\frac{k}{M}-l\right) \!,\nonumber\\
&& \IEEEeqnarraynumspace\IEEEyessubnumber\label{eq: R_X_b}
\end{IEEEeqnarray}
where $R_X^{(k)}(f) \triangleq  \mathscr{F}\{r_X^{(k)}[n]\}$ and $\tilde{R}_X^{(k)}(f) \triangleq  \mathscr{F}\{\tilde{r}_X^{(k)}[n]\}$.
\end{lemma}

\begin{IEEEproof}
It is straightforward by applying the definitions (\ref{eq: def_R_X_a}) and (\ref{eq: def_R_X_b}) in Definition~\ref{definition: 2D Fourier}, respectively, to (\ref{eq: correl_expand}) and (\ref{eq: com_correl_expand}) in Lemma~\ref{lemma: series time}.
\end{IEEEproof}

Note from (\ref{eq: R_X_all}) that both $R_X(f,f')$ and $\tilde{R}_X(f,f')$ consist of $1/M$-spaced impulse fences along the lines $f=f' \pm k/M, \forall k\in\mathbb{Z}$.
The above lemmas are the extensions of the results in \cite{Han_12} for a CT SOCS random process.

\subsection{DT FRESH Properizer}

In this subsection, an LCL PTV operator is proposed that always outputs an equivalent DT proper-complex random process, regardless of the propriety of the input DT SOCS random process.
In what follows, all SOCS random processes are DT processes unless otherwise specified.

To proceed, the following frequency-selective filter is defined that has half of the entire frequency band as its stopband.

\begin{definition}\label{def: FD-RSW}
Given a reference frequency $1/M$, a linear time-invariant system with impulse response $g_M[n]$ is called the frequency-domain raised square wave (FD-RSW) filter if its frequency response $G_M(f)=\mathscr{F}\{g_M[n]\}$ is given by
\begin{equation}\label{eq: FD-RSW}
G_M(f)\triangleq\left\{
\begin{array}{ll}
0, & \text{for }\; -\frac{1}{2M} \leq f < 0,\\
1, & \text{for }\; 0 \leq f < \frac{1}{2M}, \rm{ and}\\
G_{M}\left(f + \frac{1}{M} \right), & \rm{elsewhere}.
\end{array}
\right.
\end{equation}
\end{definition}

\begin{figure}[tbp]
\setlength{\unitlength}{0.7pt}
{
\begin{center}
\begin{picture}(620, 190)(0,0)

        \put(30,45){\vector(1,0){550}}
        \put(310,35){\vector(0,1){115}}

        \put(590,45){\makebox(0,0){$f$}}%
        \put(395,165){\makebox(0,0){$G_M(f)\triangleq \mathscr{F}\{g_M[n]\}$}}%

        \put(310,45){\path(-245,50)(-225,50)(-225,0)(-150,0)(-150,50)(-75,50)(-75,0)(0,0)(0,50)(75,50)(75,0)(150,0)(150,50)(225,50)(225,0)}
        \put(30,70){\makebox(0,0){$\cdots$}}%
        \put(570,70){\makebox(0,0){$\cdots$}}%
        \put(320,110){\makebox(0,0){$1$}}%

        \put(85,20){\makebox(0,0){$-\frac{3}{2M}$}}%
        \put(160,20){\makebox(0,0){$-\frac{2}{2M}$}}%
        \put(235,20){\makebox(0,0){$-\frac{1}{2M}$}}%
        \put(325,20){\makebox(0,0){$0$}}%
        \put(395,20){\makebox(0,0){$\frac{1}{2M}$}}%
        \put(470,20){\makebox(0,0){$\frac{2}{2M}$}}%
        \put(545,20){\makebox(0,0){$\frac{3}{2M}$}}%
\end{picture}
\end{center}}
\caption{Frequency response of the FD-RSW filter with reference frequency $1/M$.}
\label{Fig: FDSW pulse}
\end{figure}
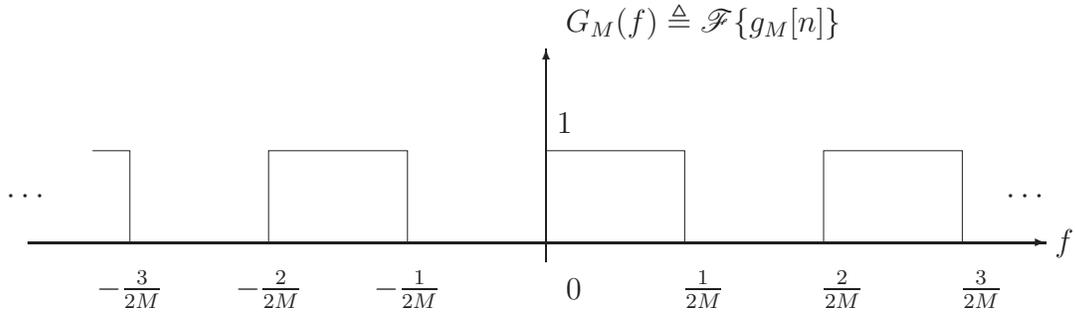

Fig.~\ref{Fig: FDSW pulse} shows the frequency response of the FD-RSW filter with reference frequency $1/M$, which alternately passes the frequency components of the input signal in every other interval of bandwidth $1/(2M)$.
Hereafter, $\mathcal{G}_{M}$ denotes the support of this FD-RSW filter.
By using the impulse response $g_M[n]$ of the FD-RSW filter with reference frequency $1/M$, we can define an LCL PTV filter as follows.

\begin{definition}\label{def: FRESH properizer}
Given a reference frequency $1/M$ and an input $X[n]$, the DT FRESH properizer is defined as a single-input single-output LCL PTV system, whose output is given by
\begin{equation}\label{eq: def_output}
Y[n] \triangleq X[n]*g_M[n] + \left(X[n]^**g_M[n]\right)e^{-{\rm j}2\pi n/(2M)}.
\end{equation}
\end{definition}

\begin{figure}[tbp]
\setlength{\unitlength}{0.7pt}
{
\begin{center}
\begin{picture}(620,290)(0,0)

        \put(220,250){DT FRESH properizer} 
        \put(80,10){\dashbox{10}(425,225)}

        \put(5,178){$X[n]$}
        \put(50,185){\line(1,0){165}}
        \put(220,185){\vector(1,0){0}}
        \put(220,165){\framebox(80,40){$g_M[n]$}}%
        \put(300,185){\line(1,0){157}}
        \put(463,185){\vector(1,0){0}}
        \put(358,160){$X_1[n]$}
        \put(461.5,181){$\bigoplus$}
        \put(478.5,185){\line(1,0){59}}
        \put(543,185){\vector(1,0){0}}
        \put(550,178){$Y[n]$}

        \put(135,185){\line(0,-1){25}}
        \put(135,155){\vector(0,-1){0}}
        \put(110,105){\framebox(50,50){$\;(\cdot)^*$}}%
        \put(135,105){\line(0,-1){20.4}}	
        \put(135,85){\line(1,0){80}}
        \put(220,85){\vector(1,0){0}}
        \put(220,65){\framebox(80,40){$g_M[n]$}}%
        \put(300,85){\line(1,0){67}}
        \put(373,85){\vector(1,0){0}}
        \put(371.5,81.5){$\bigotimes$}%
        \put(381,46){\line(0,1){26}}
        \put(381,78){\vector(0,1){0}}
        \put(342,25){$e^{-{\rm j}2\pi n/(2M)}$}%
        \put(405,63){$X_2[n]$}
        \put(389,85){\line(1,0){82.3}}
        \put(470.8,85){\line(0,1){86}}
        \put(470.8,177){\vector(0,1){0}}

\end{picture}
\end{center}}
\caption{DT FRESH properizer with reference frequency $1/M$ viewed in the time domain. }
\label{Fig: FRESH properizer}
\end{figure}
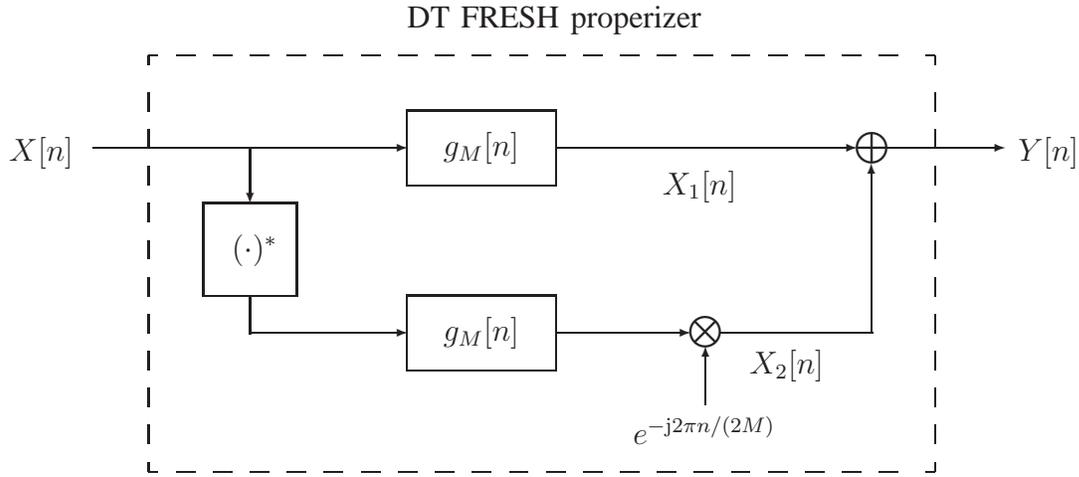

Fig.~\ref{Fig: FRESH properizer} shows how the DT FRESH properizer works in the time domain, where $X[n]$ is the input, $Y[n]$ is the output, and $X_1[n]$ and $X_2[n]$ are, respectively, the first and the second terms on the right side of (\ref{eq: def_output}).
Note that, unlike the upper branch where $X[n]$ is processed by the FD-RSW filter to generate $X_1[n]$, $X[n]^*$ is processed in the lower branch by the FD-RSW filter and multiplied by a complex-exponential function $e^{-{\rm j}2\pi n/(2M)}$ to generate $X_2[n]$.
Even though $-1/(2M)$ is chosen as the frequency of the complex-exponential function, any $(2k+1)/(2M)$ for $k\in \mathbb{Z}$ can be chosen.
The reason why this is so becomes clear once the DT FRESH properization is viewed in the frequency domain.

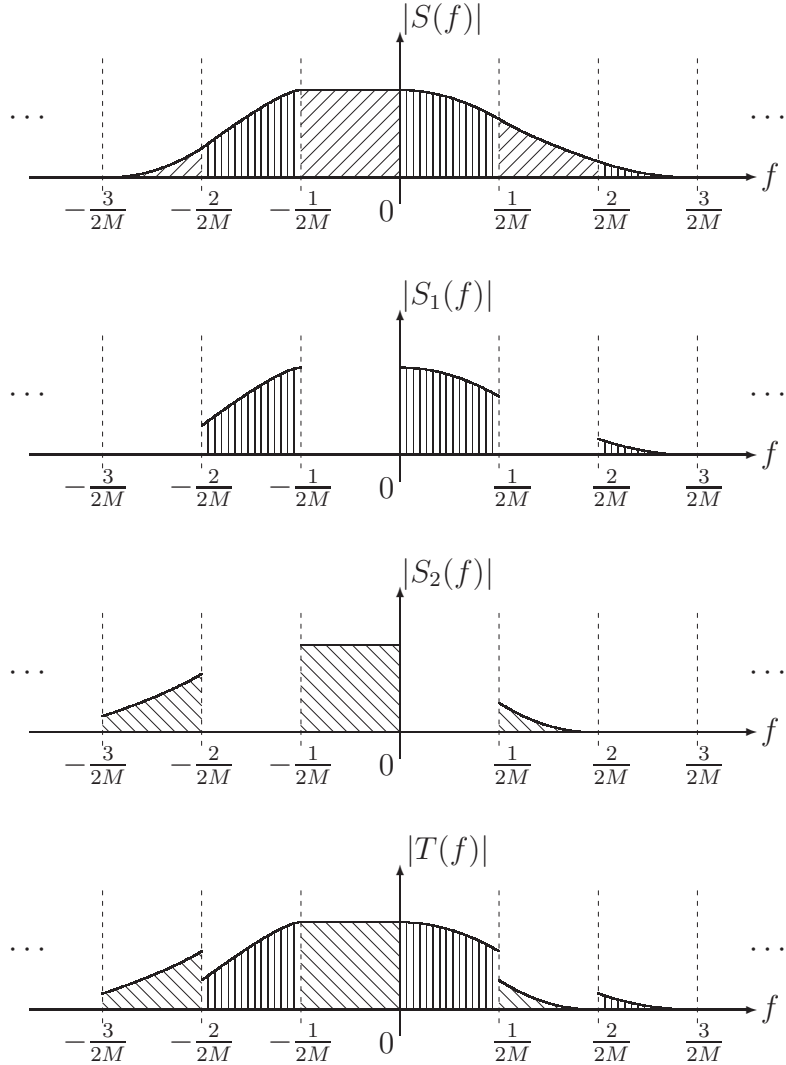
\begin{figure}[tbp]
\setlength{\unitlength}{0.5pt}
{
\begin{center}
\begin{picture}(620, 840)(0,0)
		
        \put(0,630){
            \put(30,65){\line(1,0){543}}
            \put(580,65){\vector(1,0){0}}
            \put(310,45){\line(0,1){123}}
            \put(310,175){\vector(0,1){0}}
            \put(590,65){\makebox(0,0){$f$}}
            \put(342,185){\makebox(0,0){$|S(f)|$}}

	    \put(80,40){\makebox(0,0){$-\frac{3}{2M}$}}	
            \put(160,40){\makebox(0,0){$-\frac{2}{2M}$}}
            \put(235,40){\makebox(0,0){$-\frac{1}{2M}$}}
            \put(300,40){\makebox(0,0){$0$}}
            \put(395,40){\makebox(0,0){$\frac{1}{2M}$}}
            \put(470,40){\makebox(0,0){$\frac{2}{2M}$}}
            \put(540,40){\makebox(0,0){$\frac{3}{2M}$}}	
            \put(590,110){\makebox(0,0){$\cdots$}}
	    \put(30,110){\makebox(0,0){$\cdots$}}
		
            \put(310,65){
                \dashline[30]{4}(-225,-10)(-225, 90)
                \dashline[30]{4}(-150,-10)(-150, 90)
                \dashline[30]{4}(-75,-10)(-75, 90)
                \dashline[30]{4}(75,-10)(75, 90)
                \dashline[30]{4}(150,-10)(150, 90)
                \dashline[30]{4}(225,-10)(225, 90)
            }

            \put(310,65){

                \bezier{1000}(-220,0)(-185,0)(-150,22)
                \bezier{1000}(-75,66)(-90,66)(-150,22)%
                \bezier{1000}(0,66)(-37,66)(-75,66)%
                \bezier{1000}(0,66)(37,66)(75,44)%
                \bezier{1000}(75,44)(102,28)(150,12)%
                \bezier{1000}(220,0)(185,0)(150,12)%

                \linethickness{0.2pt}
                \put(-145,0){\line(0,1){26}}\put(-140,0){\line(0,1){29}}\put(-135,0){\line(0,1){33}}\put(-130,0){\line(0,1){36}}
                \put(-125,0){\line(0,1){40}}\put(-120,0){\line(0,1){43}}\put(-115,0){\line(0,1){46}}\put(-110,0){\line(0,1){50}}
                \put(-105,0){\line(0,1){53}}\put(-100,0){\line(0,1){56}}\put(-95,0){\line(0,1){59}}\put(-90,0){\line(0,1){61}}
                \put(-85,0){\line(0,1){63}}\put(-80,0){\line(0,1){65}}

                \put(5,0){\line(0,1){66}}\put(10,0){\line(0,1){65}}\put(15,0){\line(0,1){65}}\put(20,0){\line(0,1){64}}
                \put(25,0){\line(0,1){63}}\put(30,0){\line(0,1){62}}\put(35,0){\line(0,1){61}}\put(40,0){\line(0,1){60}}
                \put(45,0){\line(0,1){58}}\put(50,0){\line(0,1){56}}\put(55,0){\line(0,1){54}}\put(60,0){\line(0,1){52}}
                \put(65,0){\line(0,1){49}}\put(70,0){\line(0,1){47}}

                \put(155,0){\line(0,1){10}}\put(160,0){\line(0,1){9}}\put(165,0){\line(0,1){8}}\put(170,0){\line(0,1){6}}
                \put(175,0){\line(0,1){5}}\put(180,0){\line(0,1){4}}\put(185,0){\line(0,1){3}}\put(190,0){\line(0,1){2}}
                \put(195,0){\line(0,1){1}}\put(200,0){\line(0,1){1}}

                \put(-155,0){\line(1,1){5}}\put(-165,0){\line(1,1){15}}
                \put(-175,0){\line(1,1){17}}\put(-185,0){\line(1,1){9}}
                \put(-195,0){\line(1,1){4}}\put(-205,0){\line(1,1){1}}

                \put(-5,0){\line(1,1){5}}\put(-15,0){\line(1,1){15}}
                \put(-25,0){\line(1,1){25}}\put(-35,0){\line(1,1){35}}
                \put(-45,0){\line(1,1){45}}\put(-55,0){\line(1,1){55}}
                \put(-65,0){\line(1,1){65}}\put(-75,0){\line(1,1){66}}
                \put(-75,10){\line(1,1){56}}\put(-75,20){\line(1,1){46}}
                \put(-75,30){\line(1,1){36}}\put(-75,40){\line(1,1){26}}
                \put(-75,50){\line(1,1){16}}\put(-75,60){\line(1,1){6}}

                \put(75,0){\line(1,1){29}}\put(75,10){\line(1,1){22.5}}
                \put(75,20){\line(1,1){16}}\put(75,30){\line(1,1){9}}
                \put(75,40){\line(1,1){3}}
                \put(85,0){\line(1,1){26.7}}\put(95,0){\line(1,1){24}}
                \put(105,0){\line(1,1){21}}\put(115,0){\line(1,1){18}}
                \put(125,0){\line(1,1){15}}\put(135,0){\line(1,1){13}}
                \put(145,0){\line(1,1){5}}
            }
        }

        \put(0,420){
            \put(30,65){\line(1,0){543}}
            \put(580,65){\vector(1,0){0}}
            \put(310,45){\line(0,1){123}}
            \put(310,175){\vector(0,1){0}}
            \put(590,65){\makebox(0,0){$f$}}
            \put(346,185){\makebox(0,0){$|S_1(f)|$}}

            \put(80,40){\makebox(0,0){$-\frac{3}{2M}$}}	
            \put(160,40){\makebox(0,0){$-\frac{2}{2M}$}}
            \put(235,40){\makebox(0,0){$-\frac{1}{2M}$}}
            \put(300,40){\makebox(0,0){$0$}}
            \put(395,40){\makebox(0,0){$\frac{1}{2M}$}}
            \put(470,40){\makebox(0,0){$\frac{2}{2M}$}}
            \put(540,40){\makebox(0,0){$\frac{3}{2M}$}}	
            \put(590,110){\makebox(0,0){$\cdots$}}
	    \put(30,110){\makebox(0,0){$\cdots$}}
		
            \put(310,65){
                \dashline[30]{4}(-225,-10)(-225, 90)
                \dashline[30]{4}(-150,-10)(-150, 90)
                \dashline[30]{4}(-75,-10)(-75, 90)
                \dashline[30]{4}(75,-10)(75, 90)
                \dashline[30]{4}(150,-10)(150, 90)
                \dashline[30]{4}(225,-10)(225, 90)
            }

            \put(310,65){ %
                \bezier{1000}(-75,66)(-90,66)(-150,22)%
                \bezier{1000}(0,66)(37,66)(75,44)%
                \bezier{1000}(220,0)(185,0)(150,12)%

                \linethickness{0.2pt}
                \put(-145,0){\line(0,1){26}}\put(-140,0){\line(0,1){29}}\put(-135,0){\line(0,1){33}}\put(-130,0){\line(0,1){36}}
                \put(-125,0){\line(0,1){40}}\put(-120,0){\line(0,1){43}}\put(-115,0){\line(0,1){46}}\put(-110,0){\line(0,1){50}}
                \put(-105,0){\line(0,1){53}}\put(-100,0){\line(0,1){56}}\put(-95,0){\line(0,1){59}}\put(-90,0){\line(0,1){61}}
                \put(-85,0){\line(0,1){63}}\put(-80,0){\line(0,1){65}}

                \put(5,0){\line(0,1){66}}\put(10,0){\line(0,1){65}}\put(15,0){\line(0,1){65}}\put(20,0){\line(0,1){64}}
                \put(25,0){\line(0,1){63}}\put(30,0){\line(0,1){62}}\put(35,0){\line(0,1){61}}\put(40,0){\line(0,1){60}}
                \put(45,0){\line(0,1){58}}\put(50,0){\line(0,1){56}}\put(55,0){\line(0,1){54}}\put(60,0){\line(0,1){52}}
                \put(65,0){\line(0,1){49}}\put(70,0){\line(0,1){47}}

                \put(155,0){\line(0,1){10}}\put(160,0){\line(0,1){9}}\put(165,0){\line(0,1){8}}\put(170,0){\line(0,1){6}}
                \put(175,0){\line(0,1){5}}\put(180,0){\line(0,1){4}}\put(185,0){\line(0,1){3}}\put(190,0){\line(0,1){2}}
                \put(195,0){\line(0,1){1}}\put(200,0){\line(0,1){1}}

            }
        }

        \put(0,210){
            \put(30,65){\line(1,0){543}}
            \put(580,65){\vector(1,0){0}}
            \put(310,45){\line(0,1){123}}
            \put(310,175){\vector(0,1){0}}
            \put(590,65){\makebox(0,0){$f$}}
            \put(346,185){\makebox(0,0){$|S_2(f)|$}}

            \put(80,40){\makebox(0,0){$-\frac{3}{2M}$}}	
            \put(160,40){\makebox(0,0){$-\frac{2}{2M}$}}
            \put(235,40){\makebox(0,0){$-\frac{1}{2M}$}}
            \put(300,40){\makebox(0,0){$0$}}
            \put(395,40){\makebox(0,0){$\frac{1}{2M}$}}
            \put(470,40){\makebox(0,0){$\frac{2}{2M}$}}
            \put(540,40){\makebox(0,0){$\frac{3}{2M}$}}	
            \put(590,110){\makebox(0,0){$\cdots$}}
	    \put(30,110){\makebox(0,0){$\cdots$}}
		
            \put(310,65){
                \dashline[30]{4}(-225,-10)(-225, 90)
                \dashline[30]{4}(-150,-10)(-150, 90)
                \dashline[30]{4}(-75,-10)(-75, 90)
                \dashline[30]{4}(75,-10)(75, 90)
                \dashline[30]{4}(150,-10)(150, 90)
                \dashline[30]{4}(225,-10)(225, 90)
            }

            \put(235,65){ %
                \bezier{1000}(220,0)(185,0)(150,22)%
                \bezier{1000}(0,66)(37,66)(75,66)%
                \bezier{1000}(-75,44)(-102,28)(-150,12)%

                %
                \put(155,0){\line(-1,1){5}}\put(165,0){\line(-1,1){15}}
                \put(175,0){\line(-1,1){17}}\put(185,0){\line(-1,1){9}}
                \put(195,0){\line(-1,1){4}}\put(205,0){\line(-1,1){1}}

                \put(5,0){\line(-1,1){5}}\put(15,0){\line(-1,1){15}}
                \put(25,0){\line(-1,1){25}}\put(35,0){\line(-1,1){35}}
                \put(45,0){\line(-1,1){45}}\put(55,0){\line(-1,1){55}}
                \put(65,0){\line(-1,1){65}}\put(75,0){\line(-1,1){66}}
                \put(75,10){\line(-1,1){56}}\put(75,20){\line(-1,1){46}}
                \put(75,30){\line(-1,1){36}}\put(75,40){\line(-1,1){26}}
                \put(75,50){\line(-1,1){16}}\put(75,60){\line(-1,1){6}}

                \put(-75,0){\line(-1,1){29}}\put(-75,10){\line(-1,1){22.5}}
                \put(-75,20){\line(-1,1){16}}\put(-75,30){\line(-1,1){9}}
                \put(-75,40){\line(-1,1){3}}
                \put(-85,0){\line(-1,1){26.7}}\put(-95,0){\line(-1,1){24}}
                \put(-105,0){\line(-1,1){21}}\put(-115,0){\line(-1,1){18}}
                \put(-125,0){\line(-1,1){15}}\put(-135,0){\line(-1,1){13}}
                \put(-145,0){\line(-1,1){5}}
            }
        }

        \put(0,0){
            \put(30,65){\line(1,0){543}}
            \put(580,65){\vector(1,0){0}}
            \put(310,45){\line(0,1){123}}
            \put(310,175){\vector(0,1){0}}
            \put(590,65){\makebox(0,0){$f$}}
            \put(346,185){\makebox(0,0){$|T(f)|$}}

            \put(80,40){\makebox(0,0){$-\frac{3}{2M}$}}	
            \put(160,40){\makebox(0,0){$-\frac{2}{2M}$}}
            \put(235,40){\makebox(0,0){$-\frac{1}{2M}$}}
            \put(300,40){\makebox(0,0){$0$}}
            \put(395,40){\makebox(0,0){$\frac{1}{2M}$}}
            \put(470,40){\makebox(0,0){$\frac{2}{2M}$}}
            \put(540,40){\makebox(0,0){$\frac{3}{2M}$}}	
            \put(590,110){\makebox(0,0){$\cdots$}}
	    \put(30,110){\makebox(0,0){$\cdots$}}
		
            \put(310,65){
                \dashline[30]{4}(-225,-10)(-225, 90)
                \dashline[30]{4}(-150,-10)(-150, 90)
                \dashline[30]{4}(-75,-10)(-75, 90)
                \dashline[30]{4}(75,-10)(75, 90)
                \dashline[30]{4}(150,-10)(150, 90)
                \dashline[30]{4}(225,-10)(225, 90)
            }

            \put(235,65){ %
                \bezier{1000}(220,0)(185,0)(150,22)%
                \bezier{1000}(0,66)(37,66)(75,66)%
                \bezier{1000}(-75,44)(-102,28)(-150,12)%

                %
                \put(155,0){\line(-1,1){5}}\put(165,0){\line(-1,1){15}}
                \put(175,0){\line(-1,1){17}}\put(185,0){\line(-1,1){9}}
                \put(195,0){\line(-1,1){4}}\put(205,0){\line(-1,1){1}}

                \put(5,0){\line(-1,1){5}}\put(15,0){\line(-1,1){15}}
                \put(25,0){\line(-1,1){25}}\put(35,0){\line(-1,1){35}}
                \put(45,0){\line(-1,1){45}}\put(55,0){\line(-1,1){55}}
                \put(65,0){\line(-1,1){65}}\put(75,0){\line(-1,1){66}}
                \put(75,10){\line(-1,1){56}}\put(75,20){\line(-1,1){46}}
                \put(75,30){\line(-1,1){36}}\put(75,40){\line(-1,1){26}}
                \put(75,50){\line(-1,1){16}}\put(75,60){\line(-1,1){6}}

                \put(-75,0){\line(-1,1){29}}\put(-75,10){\line(-1,1){22.5}}
                \put(-75,20){\line(-1,1){16}}\put(-75,30){\line(-1,1){9}}
                \put(-75,40){\line(-1,1){3}}
                \put(-85,0){\line(-1,1){26.7}}\put(-95,0){\line(-1,1){24}}
                \put(-105,0){\line(-1,1){21}}\put(-115,0){\line(-1,1){18}}
                \put(-125,0){\line(-1,1){15}}\put(-135,0){\line(-1,1){13}}
                \put(-145,0){\line(-1,1){5}}
            }
            \put(310,65){ %
                \bezier{1000}(-75,66)(-90,66)(-150,22)%
                \bezier{1000}(0,66)(37,66)(75,44)%
                \bezier{1000}(220,0)(185,0)(150,12)%

                \linethickness{0.2pt}
                \put(-145,0){\line(0,1){26}}\put(-140,0){\line(0,1){29}}\put(-135,0){\line(0,1){33}}\put(-130,0){\line(0,1){36}}
                \put(-125,0){\line(0,1){40}}\put(-120,0){\line(0,1){43}}\put(-115,0){\line(0,1){46}}\put(-110,0){\line(0,1){50}}
                \put(-105,0){\line(0,1){53}}\put(-100,0){\line(0,1){56}}\put(-95,0){\line(0,1){59}}\put(-90,0){\line(0,1){61}}
                \put(-85,0){\line(0,1){63}}\put(-80,0){\line(0,1){65}}

                \put(5,0){\line(0,1){66}}\put(10,0){\line(0,1){65}}\put(15,0){\line(0,1){65}}\put(20,0){\line(0,1){64}}
                \put(25,0){\line(0,1){63}}\put(30,0){\line(0,1){62}}\put(35,0){\line(0,1){61}}\put(40,0){\line(0,1){60}}
                \put(45,0){\line(0,1){58}}\put(50,0){\line(0,1){56}}\put(55,0){\line(0,1){54}}\put(60,0){\line(0,1){52}}
                \put(65,0){\line(0,1){49}}\put(70,0){\line(0,1){47}}

                \put(155,0){\line(0,1){10}}\put(160,0){\line(0,1){9}}\put(165,0){\line(0,1){8}}\put(170,0){\line(0,1){6}}
                \put(175,0){\line(0,1){5}}\put(180,0){\line(0,1){4}}\put(185,0){\line(0,1){3}}\put(190,0){\line(0,1){2}}
                \put(195,0){\line(0,1){1}}\put(200,0){\line(0,1){1}}

            }
        }
\end{picture}
\end{center}}
\caption{Example that shows how the DT FRESH properizer with reference frequency $1/M$ works in the frequency domain, when the input is a deterministic signal $s[n]$.}
\label{Fig: FRESH_p_FD}
\end{figure}

Fig.~\ref{Fig: FRESH_p_FD} shows how the DT FRESH properizer works in the frequency domain, especially when the input is a deterministic signal $s[n]$ with the DTFT $S(f)\triangleq \mathscr{F}\{s[n]\}$, the outputs of the upper and the lower branches in Fig.~\ref{Fig: FRESH properizer} are denoted by $s_1[n]$ with the DTFT $S_1(f)\triangleq \mathscr{F}\{s_1[n]\}$ and $s_2[n]$ with the DTFT $S_2(f)\triangleq \mathscr{F}\{s_2[n]\}$, respectively, and the output signal is denoted by $t[n]$ with the DTFT $T(f)\triangleq \mathscr{F}\{t[n]\}$.
Note that $S(f)$ is processed by the FD-RSW filter to generate the first term $S_1(f)$ of the DTFT of the DT FRESH properizer output, while $S(-f)^*$ is processed by the FD-RSW filter and shifted in the frequency domain to generate the second term $S_2(f)$.
Thus, $S_1(f)$ contains all the frequency components of $s[n]$ on the support $\mathscr{G}$ of the FD-RSW filter, while $S_2(f)$ contains all the remaining frequency components.
Since the supports of $S_1(f)$ and $S_2(f)$ do not overlap, the DTFT of the output $T(f)$ of the DT FRESH properizer contains all the frequency components of the input signal $S(f)$ without any distortion.
Note also that, due to the periodicity of the DTFT with period $1$, the frequency shift of $S_2(f)$ by any $k/M$ for $k\in \mathbb{Z}$ generates the output signal that contains the same information as the input does.
The following lemma makes this invertibility argument more precise.

\begin{lemma}
From the output $Y[n]$ of the DT FRESH properizer with reference frequency $1/M$, the input $X[n]$ of the DT FRESH properizer can be recovered as
\begin{equation}\label{eq: def_output_inverse}
X[n] = Y[n]*g_M[n] + \left\{ \left(Y[n]e^{{\rm j}2\pi n/(2M)}\right)*g_M[n]\right\}^*.
\end{equation}
\end{lemma}

\begin{IEEEproof}
Straightforward by substituting (\ref{eq: def_output}) into the right side of (\ref{eq: def_output_inverse}).
\end{IEEEproof}

Note that, at this point, the DT FRESH properizer may not be more than one of many possible invertible operators.
The reason why this operator is named the properizer will become clear once the second-order property of the output is analyzed as follows when its input is a zero-mean SOCS random process.

\begin{theorem}\label{theorem: 1}
If the input $X[n]$ to the DT FRESH properizer with reference frequency $1/M$ is a zero-mean SOCS random process with cycle period $M$, then the output $Y[n]$ becomes a zero-mean proper-complex SOCS random process with cycle period $2M$, i.e., the mean, the auto-correlation, and the complementary auto-correlation functions of $Y[n]$ satisfy
\begin{IEEEeqnarray}{rCl}
\mu_{Y}[n]&\triangleq &\mathbf{E}\{Y[n]\}=0, \IEEEeqnarraynumspace\IEEEyessubnumber\label{eq: zeromean}\\
r_{Y}[n,m]&\triangleq  &\mathbf{E}\{Y[n]Y[m]^*\}\!=\! r_{Y}[n\!+\!2M, m\!+\!2M],\;\rm{and}\;\; \IEEEeqnarraynumspace\IEEEyessubnumber \label{eq: correlation_function_Y}\\
\tilde{r}_{Y}[n,m]& \triangleq &\mathbf{E}\{Y[n]Y[m]\}=0, \IEEEeqnarraynumspace\IEEEyessubnumber\label{eq: pseudocorrelation_function_Y}
\end{IEEEeqnarray}
$\forall m, \forall n$, respectively.
\end{theorem}

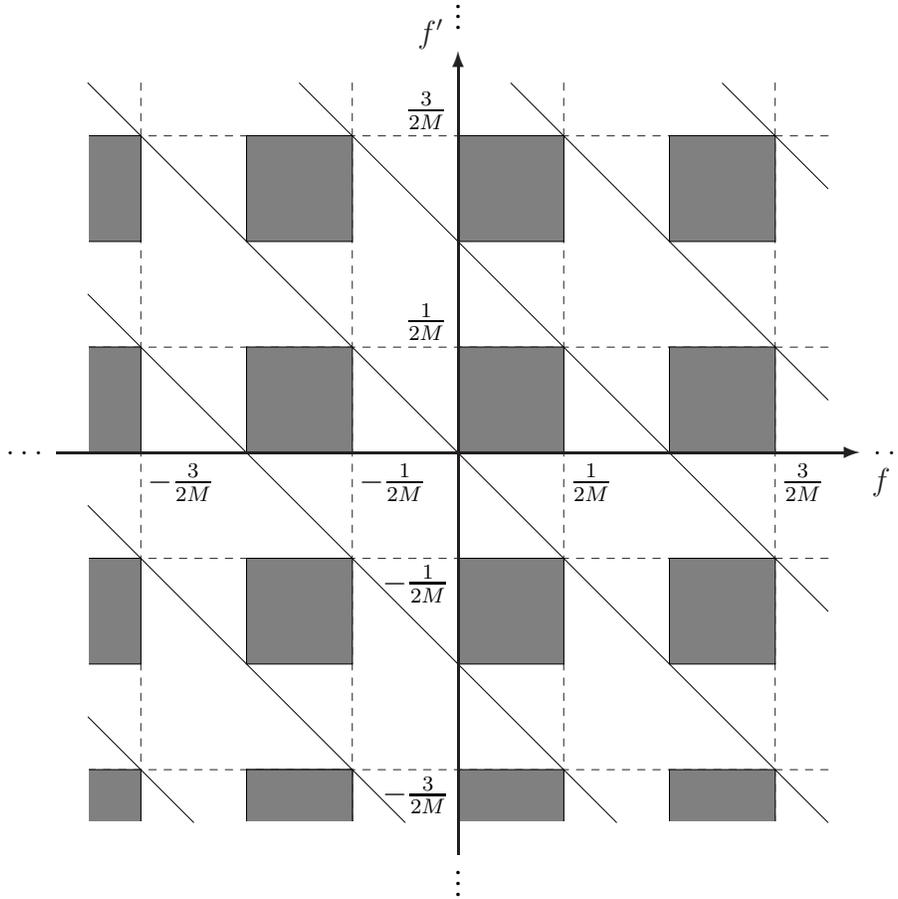
\begin{figure}[tbp]
\setlength{\unitlength}{0.8pt}
{
\begin{center}
    \begin{picture}(400,400)(-200,-190)

        \thinlines
        \put(200,-14){\makebox(0,0){$ f$}}%
        \put(-13,198){\makebox(0,0){$ f'$}}%

        \dashline[30]{4}(-150,-175)(-150, 175)
        \dashline[30]{4}(-50,-175)(-50, 175)
        \dashline[30]{4}(50,-175)(50, 175)
        \dashline[30]{4}(150,-175)(150, 175)

        \dashline[30]{4}(-175,150)(175, 150)
        \dashline[30]{4}(-175,50)(175, 50)
        \dashline[30]{4}(-175,-50)(175, -50)
        \dashline[30]{4}(-175,-150)(175, -150)

        \put(0,0){\shade\path(0,0)(50,0)(50,50)(0,50)(0,0)}
        \put(100,0){\shade\path(0,0)(50,0)(50,50)(0,50)(0,0)}
        \put(100,100){\shade\path(0,0)(50,0)(50,50)(0,50)(0,0)}
        \put(0,100){\shade\path(0,0)(50,0)(50,50)(0,50)(0,0)}

        \put(-100,100){\shade\path(0,0)(50,0)(50,50)(0,50)(0,0)}
        \put(-100,0){\shade\path(0,0)(50,0)(50,50)(0,50)(0,0)}

        \put(-100,-100){\shade\path(0,0)(50,0)(50,50)(0,50)(0,0)}

        \put(0,-100){\shade\path(0,0)(50,0)(50,50)(0,50)(0,0)}
        \put(100,-100){\shade\path(0,0)(50,0)(50,50)(0,50)(0,0)}

        \put(-175,100){\shade\path(0,0)(25,0)(25,50)(0,50)(0,0)}
        \put(-175,0){\shade\path(0,0)(25,0)(25,50)(0,50)(0,0)}
        \put(-175,-100){\shade\path(0,0)(25,0)(25,50)(0,50)(0,0)}
        \put(-175,-175){\shade\path(0,0)(25,0)(25,25)(0,25)(0,0)}
        \put(-100,-175){\shade\path(0,0)(50,0)(50,25)(0,25)(0,0)}
        \put(-100,-175){\shade\path(0,0)(50,0)(50,25)(0,25)(0,0)}
        \put(0,-175){\shade\path(0,0)(50,0)(50,25)(0,25)(0,0)}
        \put(100,-175){\shade\path(0,0)(50,0)(50,25)(0,25)(0,0)}

        \put(175,-175){\line(-1,1){350}}%
        \put(175,-75){\line(-1,1){250}}%
        \put(175,25){\line(-1,1){150}}%
        \put(175,125){\line(-1,1){50}}%
        \put(75,-175){\line(-1,1){250}}%
        \put(-25,-175){\line(-1,1){150}}%
        \put(-125,-175){\line(-1,1){50}}%

        \put(63,-14){\makebox(0,0){$\frac{1}{2M}$}}
        \put(163,-14){\makebox(0,0){$\frac{3}{2M}$}}
        \put(-31,-14){\makebox(0,0){$-\frac{1}{2M}$}}
        \put(-131,-14){\makebox(0,0){$-\frac{3}{2M}$}}

        \put(-15,62){\makebox(0,0){$\frac{1}{2M}$}}
        \put(-15,162){\makebox(0,0){$\frac{3}{2M}$}}
        \put(-20,-62){\makebox(0,0){$-\frac{1}{2M}$}}
        \put(-20,-162){\makebox(0,0){$-\frac{3}{2M}$}} 

        \put(-205,0){\makebox(0,0){$\hdots$}}
        \put(205,0){\makebox(0,0){$\hdots$}}
        \put(0,210){\makebox(0,0){$\vdots$}}
        \put(0,-200){\makebox(0,0){$\vdots$}}

        {\color{white}
        \put(-175,99){\line(0,1){52}}%
        \put(-175,100){\line(0,1){49}}%
        \put(-175,0.38){\line(0,1){50}}%
        \put(-175,-101){\line(0,1){52}}%
        \put(-175,-176){\line(0,1){27}}%
        \put(-176,-175){\line(1,0){27}}%
        \put(-101,-175){\line(1,0){52}}%
        \put(0.38,-175){\line(1,0){51}}%
        \put(99,-175){\line(1,0){52}}%
        }

        \thicklines
        \put(-190,0){\line(1,0){373}} %
        \put(190,0){\vector(1,0){0}}  %
        \put(0,-190){\line(0,1){373}} %
        \put(0,190){\vector(0,1){0}}  %
    \end{picture}
\end{center}}
\caption{Solid lines represent the impulse fences of $R_X( f, -f')$, $R_X( -f, f')$, $\tilde{R}_X( f, -f')$, or $\tilde{R}_X( -f, f')$. Shaded area represents the integration area $\mathscr{G}\times \mathscr{G}$.}
\label{Fig: pseudo-correlation function}
\end{figure}

\begin{IEEEproof}
It is straightforward to show \eqref{eq: zeromean} by using $\mu_{X}[n]=0,\forall n$.
Let $X_1[n]$ and $X_2[n]$ be defined again as shown in Fig.~\ref{Fig: FRESH properizer}.
Then, the auto-correlation function $r_{Y}[n,m]$ of $Y[n]$ can be written as
\begin{IEEEeqnarray}{rCl}\label{eq: output_correlation}
r_{Y}[n,m]&=&\mathbf{E}\{X_1[n]X_1[m]^*\} + \mathbf{E}\{X_1[n]X_2[m]^*\} +  \mathbf{E}\{X_2[n]X_1[m]^*\} +  \mathbf{E}\{X_2[n]X_2[m]^*\}.\IEEEeqnarraynumspace
\end{IEEEeqnarray}
The first term on the right side of (\ref{eq: output_correlation}) can be rewritten as
\begin{IEEEeqnarray}{rCl}
\mathbf{E}\{ X_1[n] X_1[m]^* \}&=&\int_{0}^{1} \int_{0}^{1}
G_{M}(f) e^{{\rm j}2\pi f n} R_X(f,f') G_{M}(f')^* e^{-{\rm j}2\pi f'm}d f d f'\IEEEeqnarraynumspace\IEEEyessubnumber\label{eq: 11_R_X1}\\
&=& \sum_{k=0}^{M-1}\left( \int_{\mathscr{G}} R_X^{(k)}\left( f-\frac{k}{M}\right) e^{{\rm j}2\pi f(n-m)}d f\right) e^{{\rm j}2\pi \frac{k}{M}m},\IEEEeqnarraynumspace\IEEEyessubnumber\label{eq: 11_R_X2}
\end{IEEEeqnarray}
where (\ref{eq: 11_R_X1}) holds by Parseval's relation and (\ref{eq: 11_R_X2}) holds by substituting (\ref{eq: R_X_a}) into (\ref{eq: 11_R_X1}).
It turns out in (\ref{eq: 11_R_X2}) that $\mathbf{E}\{ X_1[n] X_1[m]^* \}$ is periodic in both $n$ and $m$ with period $M$.
Similarly, the second term of (\ref{eq: output_correlation}) can be rewritten as
\begin{IEEEeqnarray}{l}
\mathbf{E}\{ X_1[n] X_2[m]^* \}= \sum_{k=0}^{M-1}\left( \int_{\mathscr{G}} \tilde{R}_X^{(k)}\left( f-\frac{k}{M}\right) e^{{\rm j}2\pi f(n-m)}d f\right) e^{{\rm j}2\pi \frac{2k+1}{2M}m}.\IEEEeqnarraynumspace\label{eq: 12_R_X2}
\end{IEEEeqnarray}
It also turns out in (\ref{eq: 12_R_X2}) that $\mathbf{E}\{ X_1[n] X_2[m]^* \}$ is periodic in both $n$ and $m$ with period $2M$.
In the same way, the other two terms can be obtained, which turn out to be periodic in $n$ and $m$ with period $2M$ and $M$, respectively.
Thus, the auto-correlation function $r_{Y}[n,m]$ of $Y[n]$ satisfies (\ref{eq: correlation_function_Y}).

On the other hand, the complementary auto-correlation function $\tilde{r}_{Y}[n,m]$ of $Y[n]$ can be written as
\begin{IEEEeqnarray}{rCl}\label{eq: output_com_correlation}
\tilde{r}_{Y}[n,m]&=&\mathbf{E}\{X_1[n]X_1[m]\} + \mathbf{E}\{X_1[n]X_2[m]\}  +  \mathbf{E}\{X_2[n]X_1[m]\} +  \mathbf{E}\{X_2[n]X_2[m]\}\}.\IEEEeqnarraynumspace
\end{IEEEeqnarray}
The first and the second terms of the right side of (\ref{eq: output_com_correlation}) can be rewritten, respectively, as 
\begin{subequations}
\begin{equation}
\mathbf{E}\{ X_1[n] X_1[m] \}\!=\!\! \iint_{\mathscr{G}\times\mathscr{G}} \!\! \tilde{R}_X( f, -f') e^{{\rm j}2\pi f n} e^{{\rm j}2\pi f'm} d f d f' \qquad
\end{equation}
and
\begin{equation}
\mathbf{E}\{ X_1[n] X_2[m] \}\!= \!\! \iint_{\mathscr{G}\times\mathscr{G}} \!\! R_X( f, -f') e^{{\rm j}2\pi \left(f n+f'm-\frac{m}{2M}\right)} d f d f',
\end{equation}
\end{subequations}
$\forall n, \forall m$ by using Parseval's relation.
These two terms are all zeros because the impulse fences of $R_X( f, -f')$ and $\tilde{R}_X(f,-f')$ along the line $f=-f'-k/M$ for any $k$ do not cross the integration area $\mathscr{G}\times \mathscr{G}$.
Fig.~\ref{Fig: pseudo-correlation function} illustrates these lines and the integration area.
Similarly, the other two terms of $\tilde{r}_{Y}[n,m]$ can be obtained, which again turn out to be all zeros.
Thus, the complementary auto-correlation function $\tilde{r}_{Y}[n,m]$ of $Y[n]$ satisfies (\ref{eq: pseudocorrelation_function_Y}).
Therefore, the conclusion follows.
\end{IEEEproof}

This theorem shows that the DT FRESH properizer in general doubles the cycle period at the cost of the propriety of the output.
However, it does not double the cycle period if the input is already proper.

\begin{corollary}\label{corollary: 1}
If the input $X[n]$ to the DT FRESH properizer with reference frequency $1/M$ is a zero-mean proper-complex SOCS random process with cycle period $M$, then the output $Y[n]$ is a zero-mean proper-complex SOCS random process with cycle period $M$.
\end{corollary}

\begin{IEEEproof}
It suffices to show that $r_{Y}[n,m]$ is periodic in $n$ and $m$ with period $M$.
It is already shown in Theorem~\ref{theorem: 1} that the first and the forth terms of $r_{Y}[n,m]$ are periodic in $n$ and $m$ with period $M$.
As it can be easily seen in (\ref{eq: 12_R_X2}), the second term $\mathbf{E}\{ X_1[n] X_2[m]^* \}$ on the right side of (\ref{eq: output_correlation}) is zero, $\forall n, \forall m$, because the propriety of $X[n]$ implies $\tilde{R}_X(f,f')=0, \forall f, \forall f'$.
Similarly, the third term is zero.
Therefore, the conclusion follows.
\end{IEEEproof}

It is already shown that the amount of the frequency shift in the second term of (\ref{eq: def_output}) can be any $(2k+1)/(2M)$, for $k\in \mathbb{Z}$, to satisfy the invertibility of the DT FRESH properizer.
It can be also shown that Theorem~\ref{theorem: 1} holds for any frequency shift $(2k+1)/(2M)$, for $k\in \mathbb{Z}$.
Moreover, since any integer multiple of $M$ is also a cycle period of an SOCS random process with cycle period $M$, the random process can be FRESH properized by using any reference frequency $1/(kM) ,\forall k \in\mathbb{N}$.
Thus, it is not unique to FRESH properize an SOCS random process.

\section{Asymptotic FRESH Properizer}

In this section, the block processing of an SOCS random process is considered.
Motivated by how the DT FRESH properizer works in the frequency domain, an LCL block operator is proposed that converts a finite number of consecutive samples of an SOCS random process to an equivalent random vector.
Unlike the DT FRESH properizer proposed in the previous section, this invertible operator does not directly make the complementary covariance matrix of the output vector vanish.
Instead, it is shown that the LCL operator makes the complementary covariance matrix of the output vector approach all-zero matrix as the number of samples tends to infinity.
This is why it is named the asymptotic FRESH properizer.

\subsection{Asymptotic FRESH Properizer and Its Inverse Operator}

Let $\bm{x}$ be the length-$MN$ vector obtained by taking the $MN$ consecutive samples of a DT signal, where and in what follows it is assumed that $N$ is a positive even number.
Motivated by the DT FRESH properizer, we introduce in this subsection an LCL block operator and its inverse.

To proceed, some definitions are provided.

\begin{definition}
The centered DFT matrix $\bm{W}_{MN}$ is defined as an $(MN)$-by-$(MN)$ matrix whose $(m,n)$th entry, for $m,n \in \{1,\cdots,MN\}$, is given by
\begin{equation}\label{eq: def_CDFT}
[\bm{W}_{MN}]_{m,n}\triangleq \frac{1}{\sqrt{MN}} e^{-{\rm j}2\pi (m-c_{MN})(n-c_{MN})/(MN)},
\end{equation}
where $c_{MN}\triangleq (MN+1)/2$.
\end{definition}

It is well known that the matrix-vector multiplication with a centered DFT matrix can be implemented with low computational complexity \cite{Bi_98}, as the multiplication with an ordinary DFT matrix is efficiently implemented by using the fast Fourier transform algorithm.

\begin{definition}\label{definition: G}
Given $M$ and an even number $N$, the $(MN)$-by-$(MN)$ matrix $\bm{G}_{M,N}$ is defined as
\begin{equation}\label{eq: asymptotic properization}
\bm{G}_{M,N}\triangleq\bm{I}_{M}\otimes \begin{bmatrix}
\bm{O}_{N/2} & \bm{O}_{N/2}\\%
\bm{O}_{N/2} & \bm{I}_{N/2}\\%
\end{bmatrix}.
\end{equation}
\end{definition}

Similar to the FD-RSW pulse, the matrix $\bm{G}_{M,N}$ will be called the raised square wave (RSW) matrix because, when pre-multiplied to a column vector or a matrix, it turns the $((m-1)N + n)$th row, for $m=1,\cdots, M$ and $n=1, \cdots, N/2$, into all zeros, i.e., it alternately nulls every other band of $N/2$ consecutive rows.

\begin{definition}
Given $M$ and an even number $N$, the $(MN)$-by-$(MN)$ matrix $\bm{S}_{M,N}$ is defined as
\begin{equation}
\bm{S}_{M,N} \triangleq \left[ \begin{tabular}{cc}
$\bm{O}_{MN-N/2,N/2}$ & $\bm{I}_{MN-N/2}$ \\
$\bm{I}_{N/2}$ & $\bm{O}_{N/2,MN-N/2}$ \\%
\end{tabular}
\right].
\end{equation}
\end{definition}

Note that the matrix $\bm{S}_{M,N}$, when pre-multiplied to a column vector or a matrix, circularly shifts the rows by $N/2$, which corresponds to multiplying $e^{-{\rm j}2\pi n/(2M)}$ in the second term of the right side of (\ref{eq: def_output}).
Now, we are ready to introduce an LCL operator that is the block-processing counterpart to the DT FRESH propertizer.

\begin{definition}\label{definition: asymptotic properization}
Given $M$ and an even number $N$, the LCL operator $\bm{f}$ with input $\bm{x}$ and output $\bm{y}=\bm{f}(\bm{x})$, both of length $MN$, is called the asymptotic FRESH properizer if
\begin{equation}\label{eq: proposed vector}
\bm{f}(\bm{x}) \triangleq \bm{W}_{MN}^{\mathcal{H}}\big( \bm{G}_{M,N}\bm{W}_{MN}\bm{x} + \bm{S}_{M,N}\bm{G}_{M,N}\bm{W}_{MN}\bm{x}^* \big).
\end{equation}
\end{definition}%

Note that the input $\bm{x}$ is pre-multiplied by the centered DFT matrix $\bm{W}_{MN}$ and the RSW matrix $\bm{G}_{M,N}$, while the complex conjugate $\bm{x}^*$ is multiplied additionally by the circular shift matrix $\bm{S}_{M,N}$ to generate the frequency-domain output $\bm{W}_{MN}\bm{y}$.

Fig.~\ref{Fig: asymptotic_properization} shows $\bm{W}_{MN}\bm{x}$, $\bm{G}_{M,N}\bm{W}_{MN}\bm{x}$, and $\bm{S}_{M,N}$ $\bm{G}_{M,N}\bm{W}_{MN}\bm{x}^*$, when the $l$th entry of the $\bm{W}_{MN}\bm{x}$ is denoted by $x'_l$.
Note that, similar to the DT FRESH properization illustrated in Fig.~\ref{Fig: FRESH_p_FD}, the locations of all possible non-zero rows of $\bm{G}_{M,N}\bm{W}_{MN}\bm{x}$ and $\bm{S}_{M,N}\bm{G}_{M,N}\bm{W}_{MN}\bm{x}^*$ do not overlap, which makes $\bm{W}_{MN}\bm{y}$ contain all the entries of $\bm{W}_{MN}\bm{x}$ without any distortion.
Note also that the amount of the circular shift of $\bm{S}_{M,N}\bm{G}_{M,N}\bm{W}_{MN}\bm{x}^*$ by any $kN$ for $k\in \mathbb{Z}$ generates the signal that contains the same information as $\bm{S}_{M,N}\bm{G}_{M,N}\bm{W}_{MN}\bm{x}^*$ does.
The following lemma makes this invertibility argument more precise.

\begin{figure}[tbp]
\setlength{\unitlength}{0.64pt}
{\small
\begin{center}
\begin{picture}(620, 660)(0,0)

	\put(50,-10){\makebox(0,0){{\footnotesize(a)} $\bm{W}_{MN}\bm{x}$}}
	\put(50, 0){
    		\put(-75,25){\path(10,-5)(0,-5)(0,610)(10,610)}
		\put(75,25){\path(-10,-5)(0,-5)(0,610)(-10,610)}
		\put(-75,138){\dashline[30]{4}(0,0)(150, 0)}
		\put(-75,260){\dashline[30]{4}(0,0)(150, 0)}
		\put(-75,390){\dashline[30]{4}(0,0)(150, 0)}
		\put(-75,510){\dashline[30]{4}(0,0)(150, 0)}
	}
	\put(50,330){\makebox(0,0){$
		\begin{matrix}
		x'_1\\
		\vdots \\
		x'_{N/2}\\ 
	    	x'_{N/2+1}\\
		\vdots \\
		x'_{N}\\ 
		{ }\\
		\vdots\\
		{ }\\ 
		x'_{MN-N+1}\\
		\vdots\\
		x'_{MN-N/2}\\ 
		x'_{MN-N/2+1}\\
		\vdots\\
		x'_{MN}\\
		\end{matrix}	
	$}
	}
	
	\put(290,-10){\makebox(0,0){{\footnotesize(b)} $\bm{G}_{M,N}\bm{W}_{MN}\bm{x}$}}
	\put(300, 0){
    		\put(-75,25){\path(10,-5)(0,-5)(0,610)(10,610)}
		\put(75,25){\path(-10,-5)(0,-5)(0,610)(-10,610)}
		\put(-75,138){\dashline[30]{4}(0,0)(150, 0)}
		\put(-75,260){\dashline[30]{4}(0,0)(150, 0)}
		\put(-75,390){\dashline[30]{4}(0,0)(150, 0)}
		\put(-75,510){\dashline[30]{4}(0,0)(150, 0)}
	}
	\put(300,330){\makebox(0,0){$
		\begin{matrix}
		0\\
		\vdots \\
		0\\ 
		x'_{N/2+1}\\
		\vdots \\
		x'_{N}\\ 
		{ }\\
		\vdots\\
		{ }\\ 
		0\\
		\vdots\\
		0\\ 
		x'_{MN-N/2+1}\\
		\vdots\\
		x'_{MN}\\
		\end{matrix}	
	$}
	}

	\put(540,-10){\makebox(0,0){{\footnotesize(c)} $\bm{S}_{M,N}\bm{G}_{M,N}\bm{W}_{MN}\bm{x}^*$}}
	\put(550, 0){
    		\put(-75,25){\path(10,-5)(0,-5)(0,610)(10,610)}
		\put(75,25){\path(-10,-5)(0,-5)(0,610)(-10,610)}
		\put(-75,138){\dashline[30]{4}(0,0)(150, 0)}
		\put(-75,260){\dashline[30]{4}(0,0)(150, 0)}
		\put(-75,390){\dashline[30]{4}(0,0)(150, 0)}
		\put(-75,510){\dashline[30]{4}(0,0)(150, 0)}
	}
	\put(550,330){\makebox(0,0){$
		\begin{matrix}
		{x'}^*_{\!\!MN-N/2}\\
		\vdots \\
		{x'}_{\!\!MN-N+1}^*\\ 
		0\\
		\vdots \\
		0\\ 
		{ }\\
		\vdots\\
		{ }\\ 
		{x'}_{\!\!N/2}^*\\
		\vdots\\
		{x'}_{\!\!1}^*\\ 
		0\\
		\vdots\\
		0\\
		\end{matrix}	
	$}
	}

\end{picture}
\end{center}}
\caption{Illustration that shows how the asymptotic FRESH properizer works in the frequency domain.}
\label{Fig: asymptotic_properization}
\end{figure}

\begin{lemma}
From the output $\bm{y}$ of the asymptotic FRESH properizer with parameters $M$ and $N$, the input $\bm{x}$ of the asymptotic FRESH properizer can be recovered as
\begin{IEEEeqnarray}{rCl}\label{eq: invertible_all}
\bm{x} &=& \bm{W}_{MN}^\mathcal{H} \big(\bm{G}_{M,N}\bm{W}_{MN}\bm{y} + \bm{P}_{MN}\bm{G}_{M,N}\bm{S}_{M,N}^\mathcal{T}\bm{W}_{MN}^*\bm{y}^*\big)\IEEEeqnarraynumspace\IEEEyessubnumber\label{eq: invertible}\\
& \triangleq &\bm{f}^{-1}(\bm{y}) \IEEEeqnarraynumspace\IEEEyessubnumber
\end{IEEEeqnarray}
\end{lemma}

\begin{IEEEproof}
By substituting (\ref{eq: proposed vector}) into the right side of (\ref{eq: invertible}), we have $\bm{W}_{MN}^\mathcal{H}\bm{G}_{M,N}^2\bm{W}_{MN}\bm{x} + \bm{W}_{MN}^\mathcal{H}\bm{G}_{M,N}$ $\bm{S}_{M,N}\bm{G}_{M,N}\bm{W}_{MN}\bm{x}^* + \bm{W}_{MN}^\mathcal{H}\bm{P}_{MN}\bm{G}_{M,N}\bm{S}_{M,N}^\mathcal{T}\bm{G}_{M,N}$ $\bm{W}_{MN}^*\bm{x}^* + \bm{W}_{MN}^\mathcal{H}\bm{P}_{MN}\bm{G}_{M,N}\cdot $ $ \bm{S}_{M,N}^\mathcal{T}\bm{S}_{M,N}\bm{G}_{M,N}\bm{W}_{MN}^*\bm{x}$.
The second and the third terms vanish since $\bm{G}_{M,N}\bm{S}_{M,N}\bm{G}_{M,N} = \bm{O}_{MN}$ and $\bm{G}_{M,N}\bm{S}_{M,N}^\mathcal{T}\bm{G}_{M,N} \! = \! \bm{O}_{MN}$, respectively.
Thus, the right side of (\ref{eq: invertible}) becomes $\bm{W}_{MN}^\mathcal{H}(\bm{G}_{M,N}^2 + \bm{P}_{MN}\bm{G}_{M,N}^2\bm{P}_{MN})\bm{W}_{MN}\bm{x} $.
Moreover, it can be easily shown that $\bm{G}_{M,N}^2= \bm{G}_{M,N}$ and $\bm{G}_{M,N} $ $+ \bm{P}_{MN}\bm{G}_{M,N}\bm{P}_{MN} = \bm{I}_{MN}$, because $\bm{P}_{MN}= \bm{P}_{M}\otimes\bm{P}_{N}$ by the properties of the Kronecker product.
Therefore, the conclusion follows.
\end{IEEEproof}


Note that, at this point, the asymptotic FRESH properizer may not be more than one of many possible invertible operators.
The reason why this operator is named the asymptotic properizer will become clear once the second-order property of the output is analyzed in the next subsection when its input is a finite number of consecutive samples of a zero-mean SOCS random process.

\subsection{Second-Order Properties of Output of Asymptotic FRESH Properizer}

Let $\bar{\bm{x}}$ be the length-$2MN$ augmented vector defined as
\begin{equation}\label{eq: def_augmented}
\bar{\bm{x}} \triangleq \begin{bmatrix}
\bm{x}\\
\bm{x}^*\\
\end{bmatrix},
\end{equation}
where and in what follows it is assumed that $\bm{x}$ consists of a finite number of consecutive samples of a \emph{zero-mean} SOCS random process $X[n]$ with cycle period $M\in \mathbb{N}$.
Then, the output $\bm{y}=\bm{f}(\bm{x})$ in (\ref{eq: proposed vector}) of the asymptotic FRESH properizer can be rewritten as
\begin{IEEEeqnarray}{rCl}
\bm{y}&=&\bm{W}_{MN}^\mathcal{H}\bar{\bm{G}}_{M,N}\bar{\bm{W}}_{MN} \bar{\bm{x}}\IEEEeqnarraynumspace\IEEEyessubnumber\\
&\triangleq& \bm{W}_{MN}^\mathcal{H}\hat{\bm{y}},\IEEEeqnarraynumspace\IEEEyessubnumber
\end{IEEEeqnarray}
where the $(MN)$-by-$(2MN)$ matrix $\bar{\bm{G}}_{M,N}$ and the $(2MN)$-by-$(2MN)$ matrix $\bar{\bm{W}}_{MN}$ are given by
\begin{IEEEeqnarray}{rCl}
\bar{\bm{G}}_{M,N}  &\triangleq& \begin{bmatrix}
\bm{G}_{M,N} & \bm{S}_{M,N}\bm{G}_{M,N}
\end{bmatrix}\;\rm{ and}\IEEEeqnarraynumspace\IEEEyessubnumber\\
\bar{\bm{W}}_{MN} &\triangleq& \begin{bmatrix}
\bm{W}_{MN} & \bm{O}_{MN}\\
\bm{O}_{MN} & \bm{W}_{MN}\\
\end{bmatrix},\IEEEeqnarraynumspace\IEEEyessubnumber
\end{IEEEeqnarray}
respectively, and $\hat{\bm{y}}$ is the centered DFT of $\bm{y}$.
Thus, the covariance matrix $\bm{R}_{\bm{y}}\triangleq \mathbf{E}\{\bm{y}\bm{y}^\mathcal{H}\}$ and the complementary covariance matrix $\tilde{\bm{R}}_{\bm{y}}\triangleq \mathbf{E}\{\bm{y}\bm{y}^\mathcal{T}\}$ of $\bm{y}$ are given by
\begin{IEEEeqnarray}{rCl}
\bm{R}_{\bm{y}}&=& \bm{W}_{MN}^\mathcal{H}\bar{\bm{G}}_{M,N}\bar{\bm{W}}_{MN} \bm{R}_{\bar{\bm{x}}} \bar{\bm{W}}_{MN}^\mathcal{H} \bar{\bm{G}}_{M,N}^\mathcal{H}\bm{W}_{MN}\IEEEeqnarraynumspace\IEEEyessubnumber\label{eq: proposed cov}\\
\noalign{\noindent{\text{and}}\vspace{\jot}}
\tilde{\bm{R}}_{\bm{y}}&=& \bm{W}_{MN}^\mathcal{H}\bar{\bm{G}}_{M,N}\bar{\bm{W}}_{MN}  \tilde{\bm{R}}_{\bar{\bm{x}}}\bar{\bm{W}}_{MN}^\mathcal{T} \bar{\bm{G}}_{M,N}^\mathcal{T}\bm{W}_{MN}^*,\IEEEeqnarraynumspace\IEEEyessubnumber\label{eq: proposed com cov}
\end{IEEEeqnarray}
respectively, where $\bm{R}_{\bar{\bm{x}}} \triangleq \mathbf{E}\{\bar{\bm{x}}\bar{\bm{x}}^\mathcal{H} \}$ and $\tilde{\bm{R}}_{\bar{\bm{x}}} \triangleq \mathbf{E}\{\bar{\bm{x}}\bar{\bm{x}}^\mathcal{T} \}$ denote the covariance and the complementary covariance matrices of the augmented vector $\bar{\bm{x}}$, respectively.

Throughout this paper, we call
\begin{IEEEeqnarray}{rCCCl}
\bm{R}_{\hat{\bm{y}}}&\triangleq&\mathbf{E}\{\hat{\bm{y}}\hat{\bm{y}}^\mathcal{H}\}&=& \bm{W}_{MN}\bm{R}_{\bm{y}}\bm{W}_{MN}^\mathcal{H} \IEEEeqnarraynumspace\IEEEyessubnumber\\
\noalign{\noindent{\text{and}}\vspace{\jot}}
\tilde{\bm{R}}_{\hat{\bm{y}}}&\triangleq&\mathbf{E}\{\hat{\bm{y}}\hat{\bm{y}}^\mathcal{T}\}&= & \bm{W}_{MN}\tilde{\bm{R}}_{\bm{y}}\bm{W}_{MN}^\mathcal{T}\IEEEeqnarraynumspace\IEEEyessubnumber
\end{IEEEeqnarray}
the frequency-domain covariance and the frequency-domain complementary covariance matrices, respectively.

To analyze the asymptotic second-order properties of $\bm{y}$, we briefly review definitions and related lemmas for asymptotic equivalence between two sequences of matrices.

\begin{definition}\cite{Gray_book}
Let $(\bm{A}_k)_k$ and $(\bm{B}_k)_k$ be two sequences of $N_k$-by-$N_k$ matrices with $N_k\rightarrow \infty$ as $k\rightarrow \infty$.
Then, $(\bm{A}_k)_k$ and $(\bm{B}_k)_k$ are asymptotically equivalent and denoted by $\bm{A}_k\sim \bm{B}_k$ if
1) the strong norms of $\bm{A}_k$ and $\bm{B}_k$ are uniformly bounded, i.e., there exists a constant $c$ such that $\|\bm{A}_k\|,\|\bm{B}_k\|\leq c < \infty, \forall k$, and
2) the weak norm of $\bm{A}_k-\bm{B}_{k}$ vanishes asymptotically, i.e., $\lim_{k\rightarrow\infty}|\bm{A}_k-\bm{B}_{k}|=0$, where the strong norm $\|\bm{A}\|$ and the weak norm $|\bm{A}|$ of an $N_k$-by-$N_k$ matrix $\bm{A}$ are defined as $\|\bm{A}\|\triangleq\max_{\bm{x}\ne \bm{0}} \sqrt{\bm{x}^{\mathcal{H}}\bm{A}^{\mathcal{H}}\bm{A}\bm{x}/\bm{x}^{\mathcal{H}}\bm{x}}$ and $|\bm{A}|\triangleq \sqrt{\tr ( \bm{A}\bm{A}^{\mathcal{H}} )/N_k}$, respectively.
\end{definition}

\begin{lemma}\label{lemma: multiply}
If two sequences $(\bm{A}_k)_k$ and $(\bm{B}_k)_k$ of $N_k$-by-$N_k$ matrices are asymptotically equivalent and if the strong norms of $N_k$-by-$N_k$ matrices $\bm{C}_k$ and $\bm{D}_k$ are uniformly bounded, then $\bm{C}_k\bm{A}_k\bm{D}_k\sim \bm{C}_k\bm{B}_k\bm{D}_k$.
\end{lemma}

\begin{IEEEproof}
Since $\bm{C}_k\sim \bm{C}_k$ and $\bm{D}_k\sim\bm{D}_k$, it is straightforward to show the conclusion by applying \cite[Theorem~1-(3)]{Gray_book} twice.
\end{IEEEproof}

\begin{lemma}\label{lemma: property_sum}
If $\bm{A}_k \sim \bm{B}_k$ and $\bm{C}_k \sim \bm{D}_k$, then ($\bm{A}_k +\bm{C}_k) \sim (\bm{B}_k + \bm{D}_k)$.
\end{lemma}

\begin{IEEEproof}
By applying the triangle inequality, it is straightforward to show that the strong norms of $\bm{A}_k +\bm{C}_k$ and $\bm{B}_k +\bm{D}_k$ are uniformly bounded and that the weak norm of $\bm{A}_k + \bm{C}_k - \bm{B}_k - \bm{D}_k$ vanishes asymptotically.
Therefore, the conclusion follows.
\end{IEEEproof}

For brevity, in what follows, two square matrices $\bm{A}_k$ and $\bm{B}_k$ of the same size will be said to be asymptotically equivalent if $\bm{A}_k\sim \bm{B}_k$.
Now, we introduce our definition of asymptotic propriety.

\begin{definition}\label{definition: asymp_proper}
A sequence $(\bm{y}_k)_k$ of length-$N_k$ complex-valued random vectors is asymptotically proper if the complementary covariance matrix of $\bm{y}_k$ is asymptotically equivalent to the $N_k$-by-$N_k$ all-zero matrix.
\end{definition}

For brevity, a random vector $\bm{y}_k$ will be said to be asymptotically proper if the sequence $(\bm{y}_k)_k$ is asymptotically proper.
Before showing that the output $\bm{y}$ of the asymptotic FRESH properizer is asymptotically proper, the second-order properties of the augmented vector $\bar{\bm{x}}$ are examined as follows.

\begin{proposition}\label{proposition: augmented asymptotic}
If the random vector $\bm{x}$ is obtained by taking the $MN$ consecutive samples from a zero-mean SOCS random process with cycle period $M$, then the covariance matrix $\bm{R}_{\bar{\bm{x}}}$ and the complementary covariance matrix $\tilde{\bm{R}}_{\bar{\bm{x}}}$ of the augmented vector $\bar{\bm{x}}$ defined in (\ref{eq: def_augmented}) satisfy
\begin{IEEEeqnarray}{rCl}\label{eq: asymptotic equivalence}
\bar{\bm{W}}_{MN}\bm{R}_{\bar{\bm{x}}}\bar{\bm{W}}_{MN}^\mathcal{H} &\sim& \bm{\Omega} \IEEEeqnarraynumspace\IEEEyessubnumber\label{eq: asymptotic equivalence_a}\\
\noalign{\noindent{\text{and}}\vspace{\jot}}
\bar{\bm{W}}_{MN}\tilde{\bm{R}}_{\bar{\bm{x}}}\bar{\bm{W}}_{MN}^\mathcal{T} &\sim& \tilde{\bm{\Omega}}, \IEEEeqnarraynumspace\IEEEyessubnumber\label{eq: asymptotic equivalence_b}
\end{IEEEeqnarray}
respectively, where the $(2MN)$-by-$(2MN)$ matrices $\bm{\Omega}$ and $\tilde{\bm{\Omega}}$ are given by
\begin{IEEEeqnarray}{rCl}\label{eq: block matrix}
\bm{\Omega} &\triangleq& (\bar{\bm{W}}_{MN}\bm{R}_{\bar{\bm{x}}}\bar{\bm{W}}_{MN}^\mathcal{H}) \odot ( \bm{1}_{2M} \otimes \bm{I}_{N} ) \IEEEeqnarraynumspace\IEEEyessubnumber\label{eq: block matrix_a}\\
\noalign{\noindent{\text{and}}\vspace{\jot}}
\tilde{\bm{\Omega}} &\triangleq& (\bar{\bm{W}}_{MN}\tilde{\bm{R}}_{\bar{\bm{x}}}\bar{\bm{W}}_{MN}^\mathcal{T})  \odot ( \bm{1}_{2M} \otimes \bm{P}_{N} ), \IEEEeqnarraynumspace\IEEEyessubnumber\label{eq: block matrix_b}
\end{IEEEeqnarray}
respectively.
\end{proposition}

\begin{IEEEproof}
By definition, the left sides of (\ref{eq: asymptotic equivalence_a}) and (\ref{eq: asymptotic equivalence_b}) can be rewritten, respectively, as
\begin{IEEEeqnarray}{rCL}
\bar{\bm{W}}_{MN}\bm{R}_{\bar{\bm{x}}}\bar{\bm{W}}_{MN}^\mathcal{H}&=& \begin{bmatrix}
\bm{W}_{MN}\bm{R}_x \bm{W}_{MN}^\mathcal{H} & \bm{W}_{MN}\tilde{\bm{R}}_x \bm{W}_{MN}^\mathcal{H}\\
\bm{W}_{MN}\tilde{\bm{R}}_x^* \bm{W}_{MN}^\mathcal{H} & \bm{W}_{MN}\bm{R}_x^* \bm{W}_{MN}^\mathcal{H}\\
\end{bmatrix}\nonumber\\
&& \IEEEeqnarraynumspace\IEEEyessubnumber\label{eq: aug_cov}\\
\noalign{\noindent{\text{and}}\vspace{\jot}}
\bar{\bm{W}}_{MN}\tilde{\bm{R}}_{\bar{\bm{x}}}\bar{\bm{W}}_{MN}^\mathcal{T} &= &\begin{bmatrix}
\bm{W}_{MN}\tilde{\bm{R}}_x \bm{W}_{MN}^\mathcal{T} & \bm{W}_{MN}\bm{R}_x \bm{W}_{MN}^\mathcal{T}\\
\bm{W}_{MN}\bm{R}_x^* \bm{W}_{MN}^\mathcal{T} & \bm{W}_{MN}\tilde{\bm{R}}_x^* \bm{W}_{MN}^\mathcal{T}\\
\end{bmatrix}.\nonumber\\
&& \IEEEeqnarraynumspace\IEEEyessubnumber\label{eq: aug_com_cov}
\end{IEEEeqnarray}
Since both $\bm{R}_x$ and $\tilde{\bm{R}}_x$ are $(MN)$-by-$(MN)$ block Toeplitz matrices with block size $M$-by-$M$, each submatrix on the right side of (\ref{eq: aug_cov}) is asymptotically equivalent to the Hadamard product of $(\bm{1}_{M} \otimes \bm{I}_{N} )$ and the submatrix itself as shown in \cite[Proposition~1]{Yoo_10}.
Thus, by using the fact that $\bm{A} \sim \bm{B}$ if each submatrix of $\bm{A}$ is asymptotically equivalent to the corresponding submatrix of $\bm{B}$ \cite[Lemma~5]{Yoo_10}, we obtain (\ref{eq: asymptotic equivalence_a}).
Similar to (\ref{eq: asymptotic equivalence_a}), it can be shown by using the property $\bm{W}_{MN}^\mathcal{T} = \bm{W}_{MN}^\mathcal{H}\bm{P}_{MN}$ of the centered DFT matrix that each submatrix on the right side of (\ref{eq: aug_com_cov}) is asymptotically equivalent to Hadamard product of $(\bm{1}_{M} \otimes \bm{P}_{N} )$ and the submatrix itself.
Thus, we obtain (\ref{eq: asymptotic equivalence_b}).
Therefore, the conclusion follows.
\end{IEEEproof}

\begin{figure}[!t]\centering
{
\subfigure[]{
\includegraphics[width=2.5in]{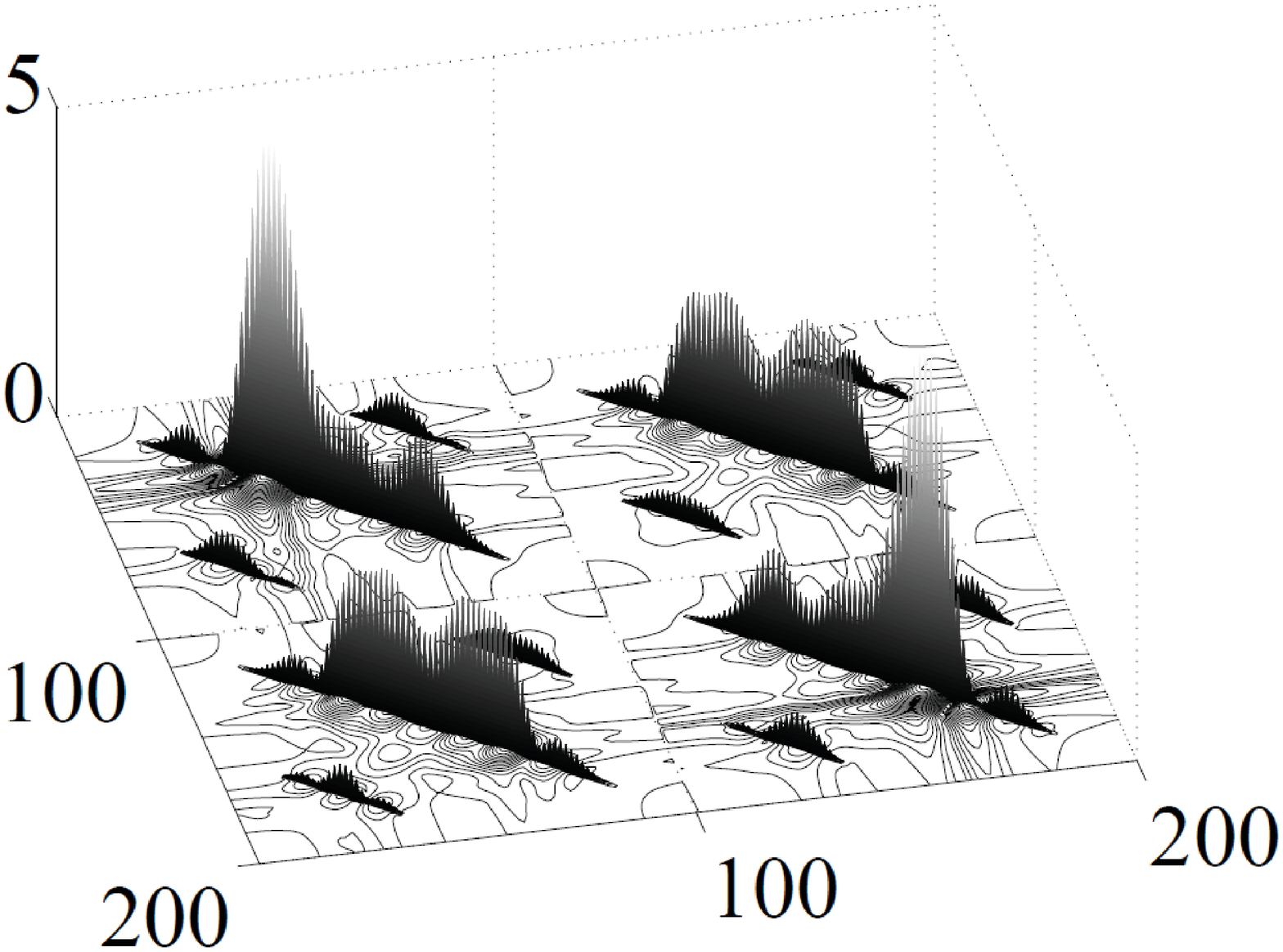}  
\label{Fig: approx_11}
}$\qquad$
\subfigure[]{
\includegraphics[width=2.5in]{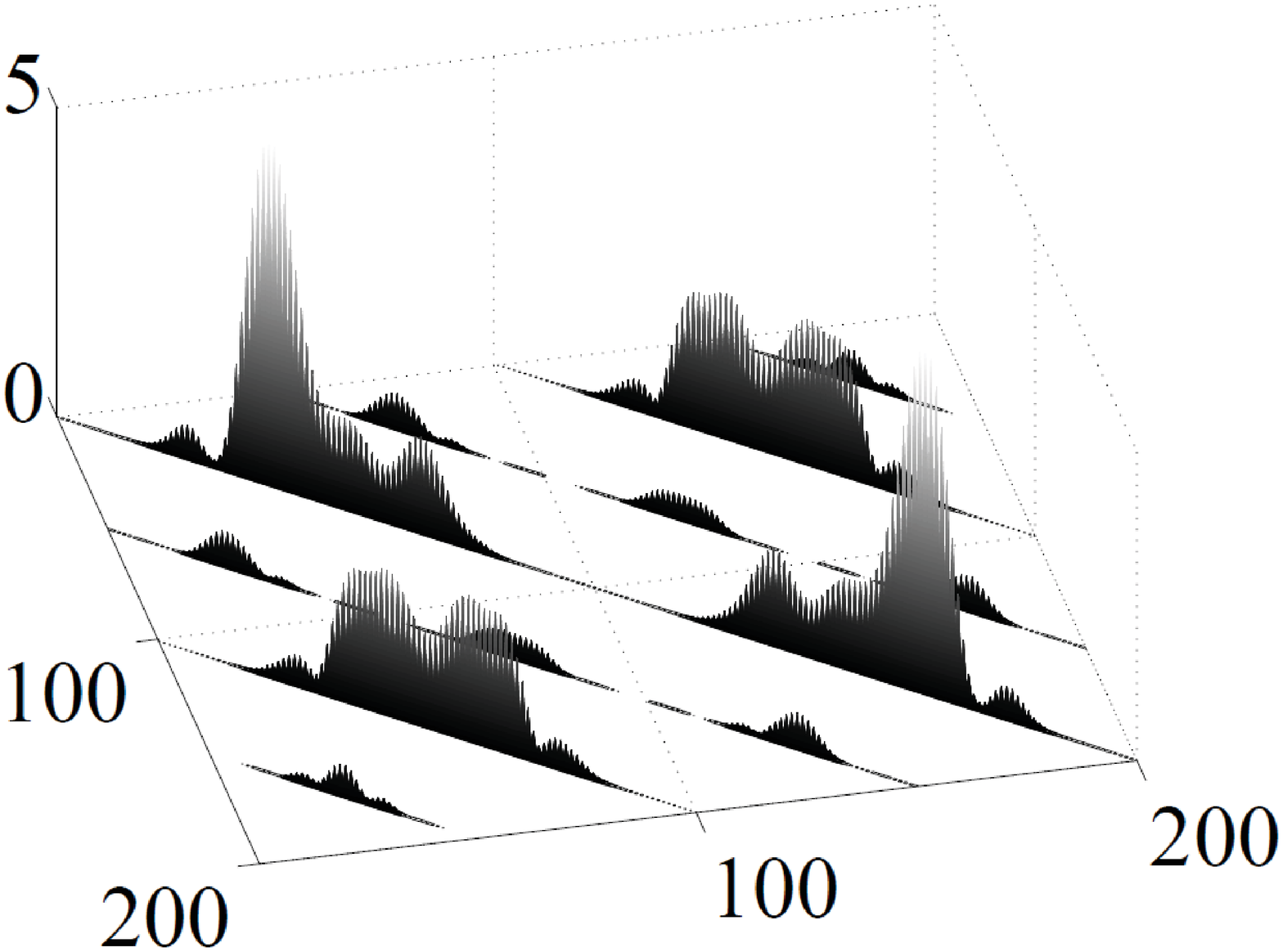}  
\label{Fig: approx_12}
}\\
\subfigure[]{
\includegraphics[width=2.5in]{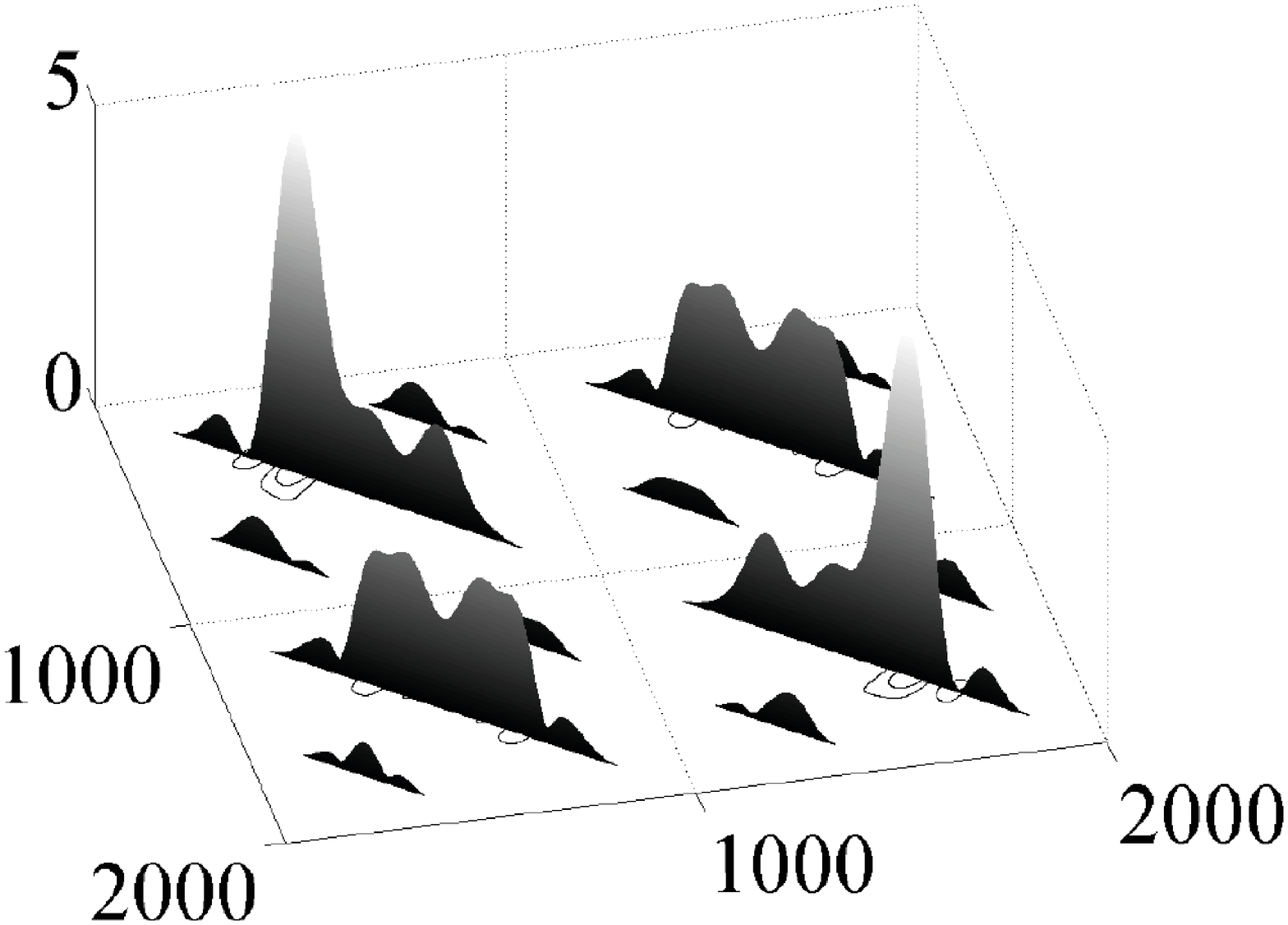}  
\label{Fig: approx_13}
}$\qquad$
 \subfigure[]{
\includegraphics[width=2.5in]{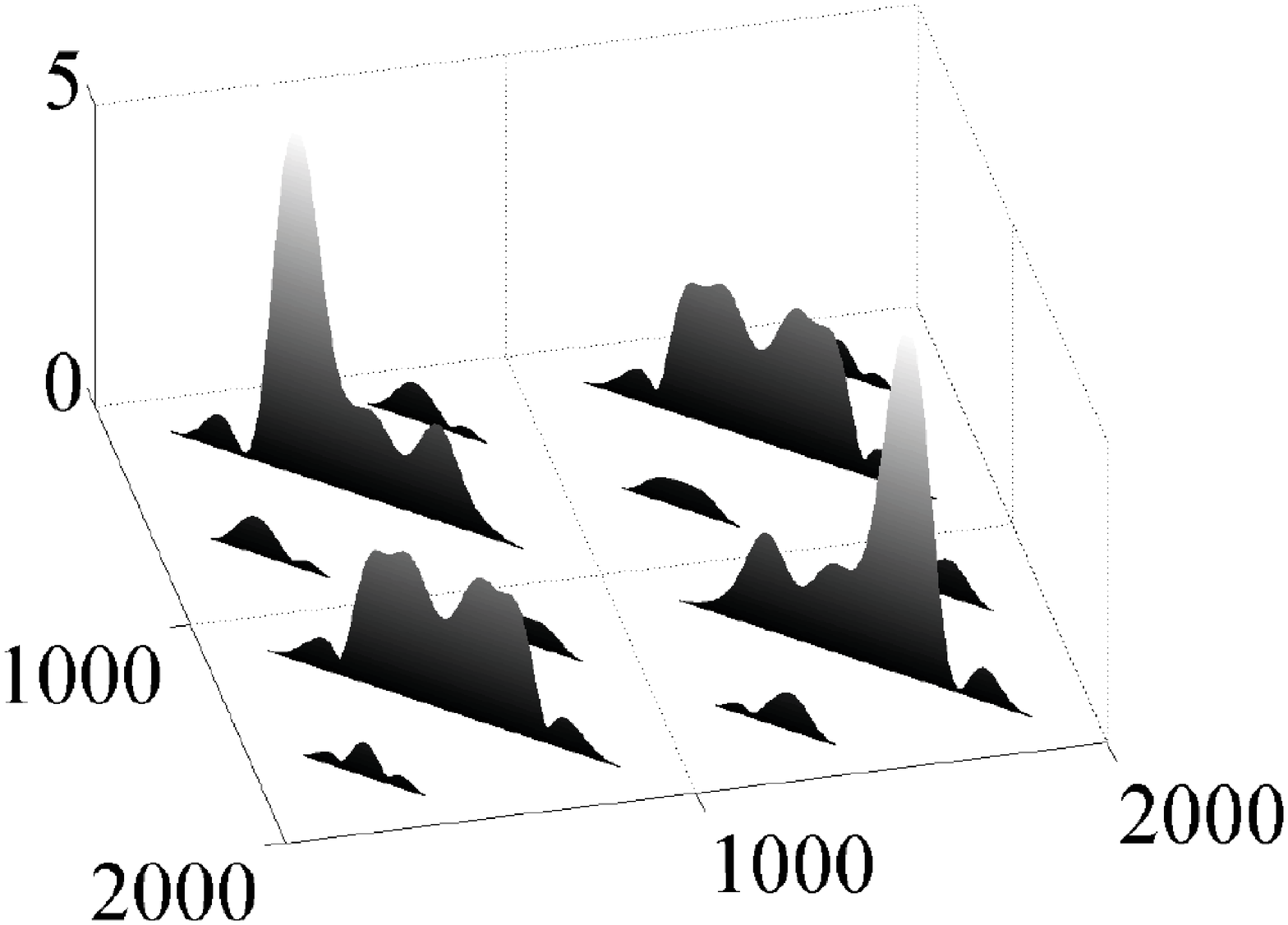}  
\label{Fig: approx_14}
}
\caption{Magnitudes of exemplary (a) $\bar{\bm{W}}_{MN}\bm{R}_{\bar{\bm{x}}}\bar{\bm{W}}_{MN}^\mathcal{H}$ for $N=50$, (b) $\bm{\Omega}$ for $N=50$, (c) $\bar{\bm{W}}_{MN}\bm{R}_{\bar{\bm{x}}}\bar{\bm{W}}_{MN}^\mathcal{H}$ for $N=500$, and (d) $\bm{\Omega}$ for $N=500$, when $M=2$.} \label{Fig: approx_covariance}
}
\end{figure}

\begin{figure}[!t]\centering
{
\subfigure[]{
\includegraphics[width=2.5in]{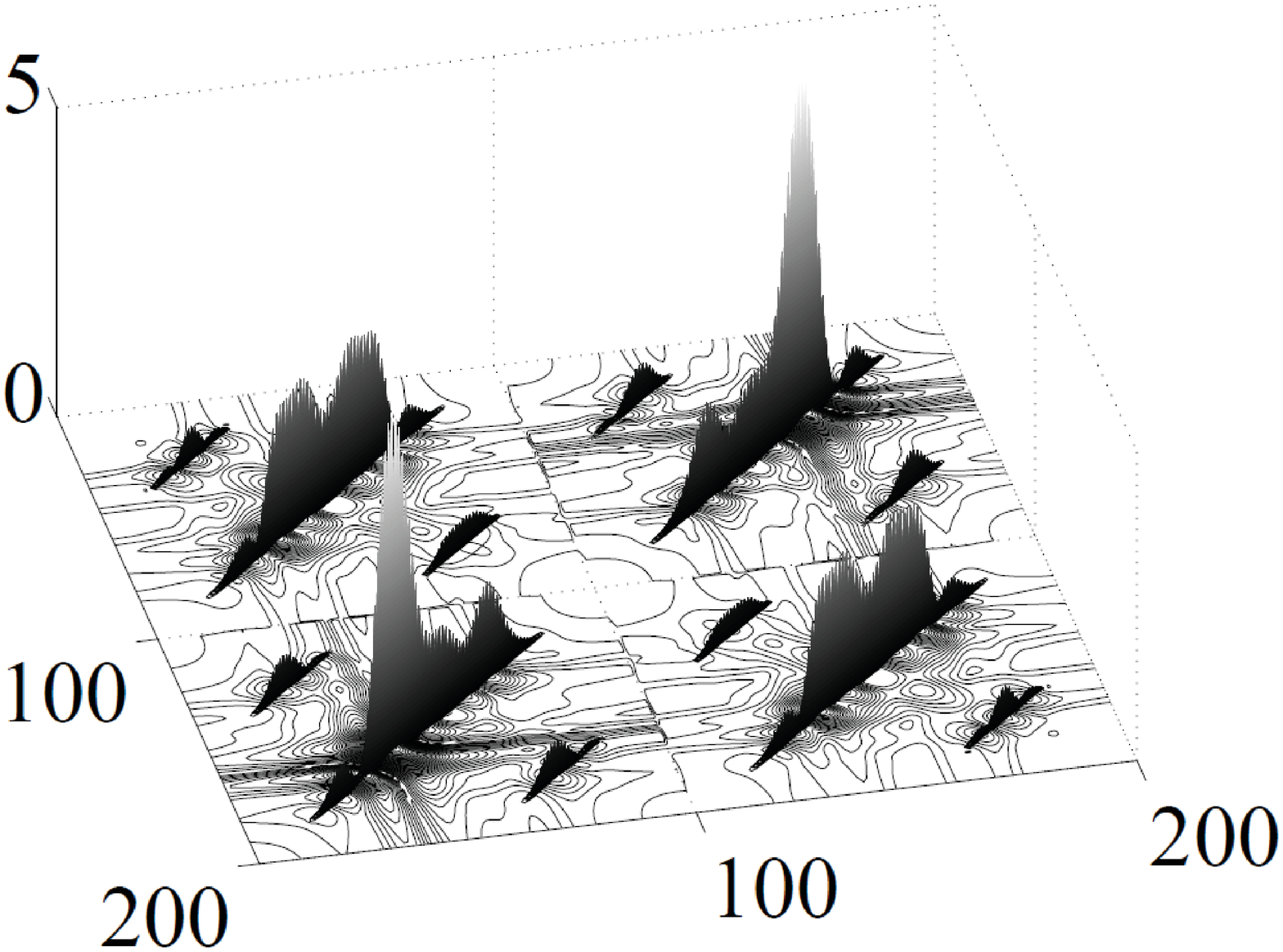}  
\label{Fig: approx_21}
}$\qquad$
\subfigure[]{
\includegraphics[width=2.5in]{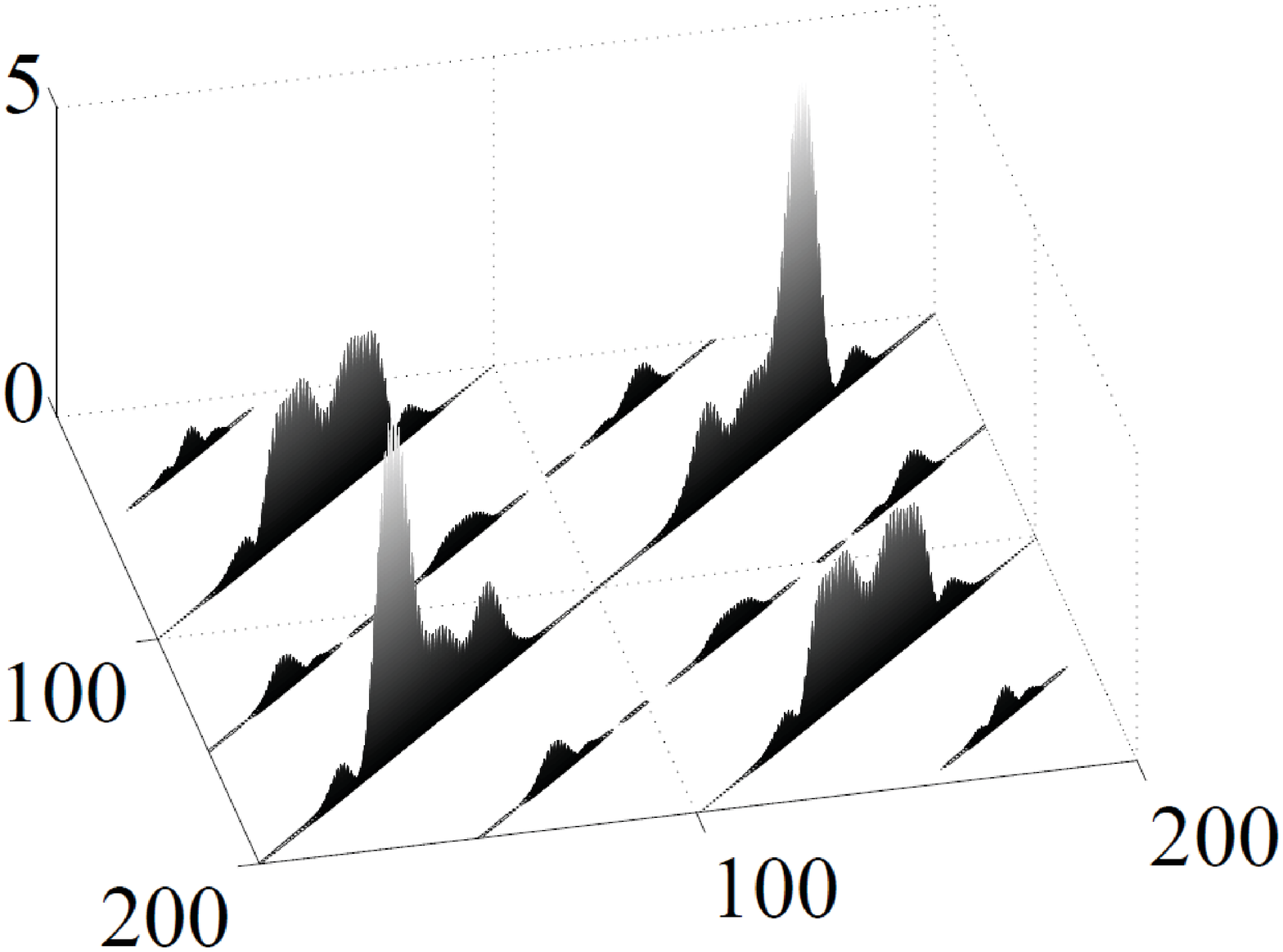}  
\label{Fig: approx_22}
}\\
\subfigure[]{
\includegraphics[width=2.5in]{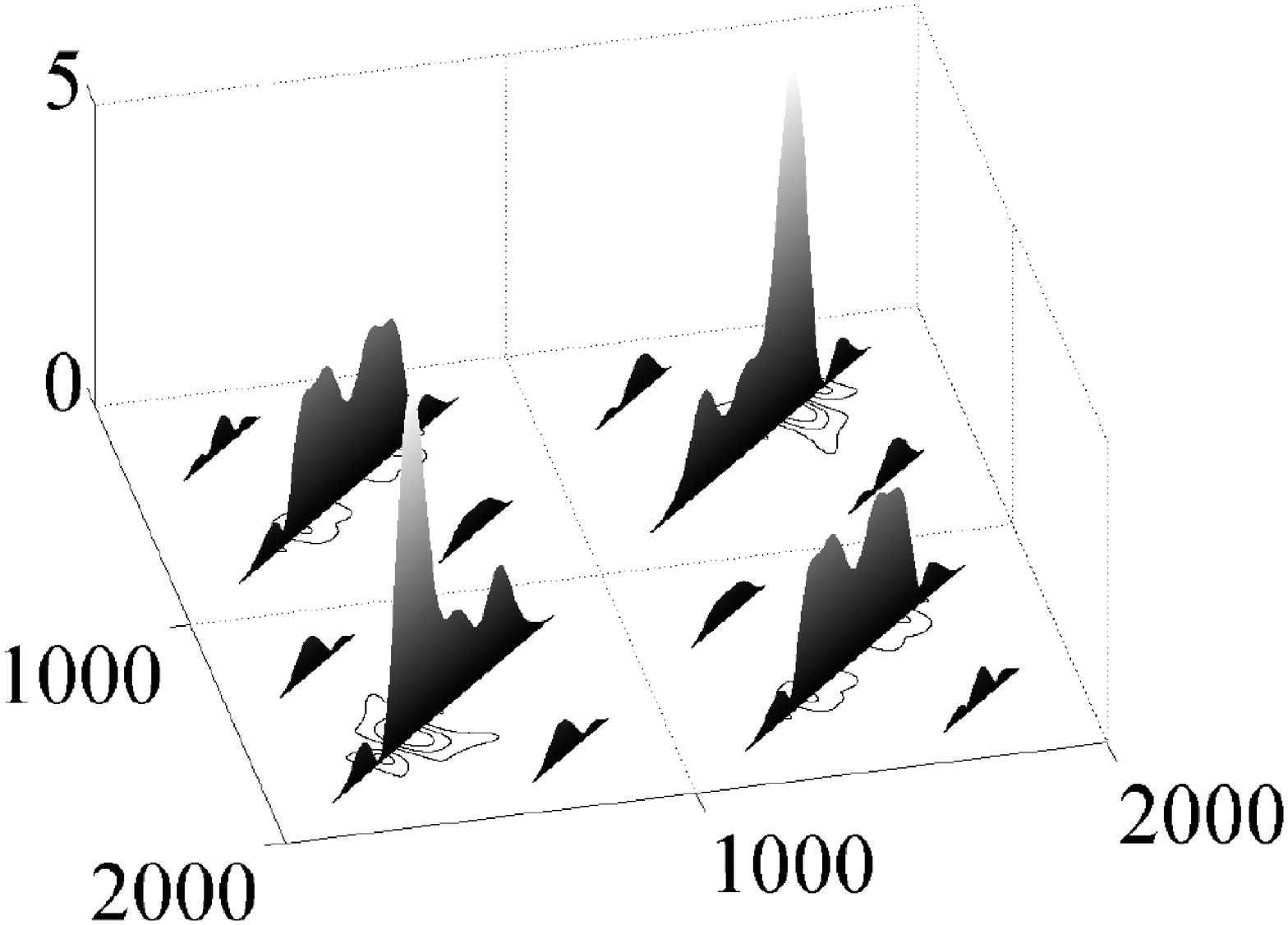}  
\label{Fig: approx_23}
}$\qquad$
\subfigure[]{
\includegraphics[width=2.5in]{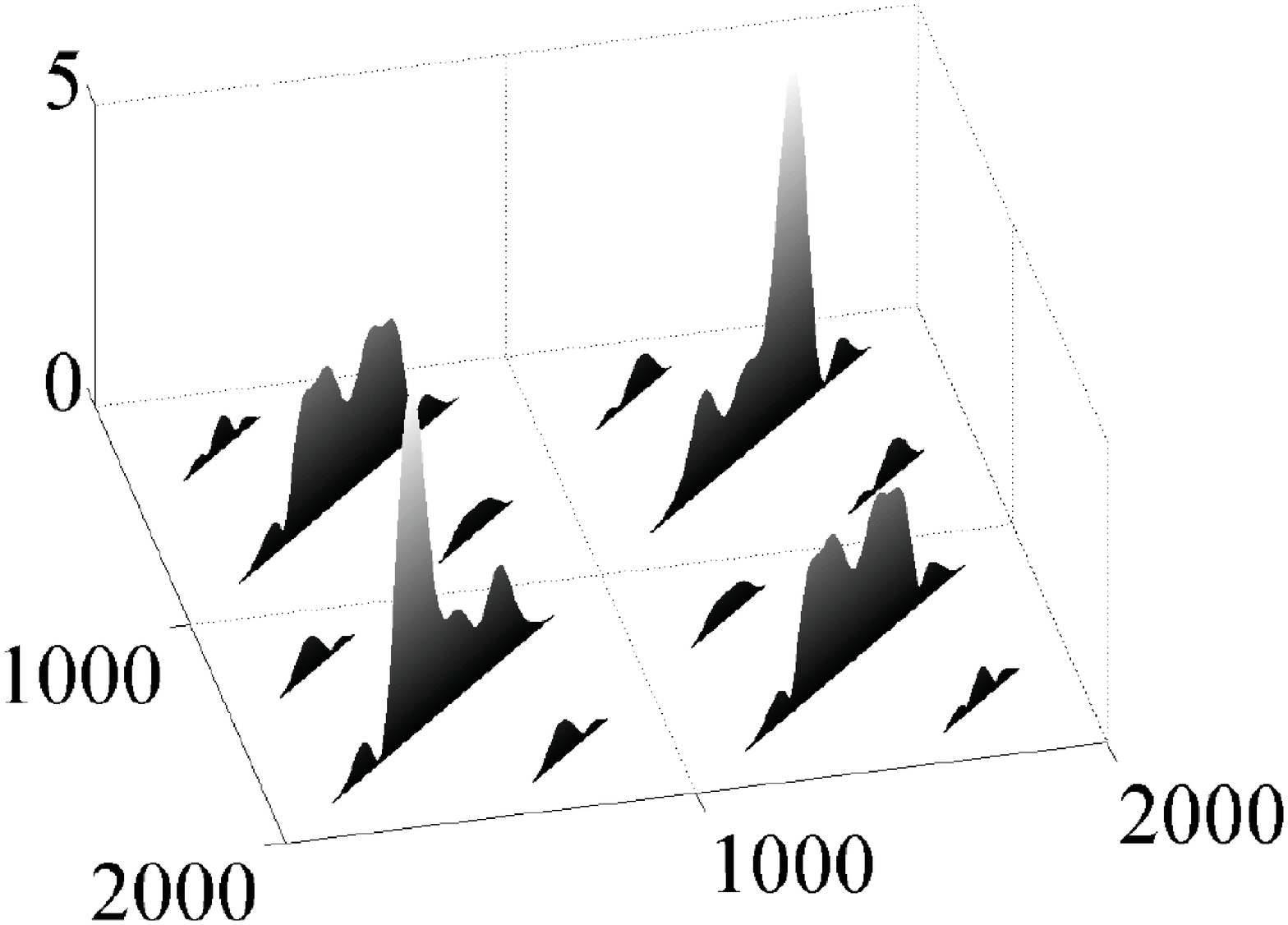}  
\label{Fig: approx_24}
}
\caption{Magnitudes of exemplary (a) $\bar{\bm{W}}_{MN}\tilde{\bm{R}}_{\bar{\bm{x}}}\bar{\bm{W}}_{MN}^\mathcal{T}$ for $N=50$, (b) $\tilde{\bm{\Omega}}$ for $N=50$, (c) $\bar{\bm{W}}_{MN}\tilde{\bm{R}}_{\bar{\bm{x}}}\bar{\bm{W}}_{MN}^\mathcal{T}$ for $N=500$, and (d) $\tilde{\bm{\Omega}}$ for $N=500$, when $M=2$.} \label{Fig: approx_comp_covariance}
}
\end{figure}

Figs.~\ref{Fig: approx_covariance} and \ref{Fig: approx_comp_covariance} show the entry-by-entry magnitudes of the exemplary frequency-domain covariance and complementary covariance matrices.
To generate a zero-mean improper-complex SOCS random process, equally-likely BPSK symbols are linearly modulated with the square-root raised cosine (SRRC) pulse having roll-off factor $0.5$ and transmitted at the symbol rate of $1/T_s$ [symbols/sec].
After the BPSK signal passes through the frequency selective channel with impulse response $h(t) =  0.22 e^{{\rm j}\pi /3}\delta(t-5T_s) + 0.44 e^{{\rm j}\pi /2}\delta(t-3T_s) + 0.87 e^{{\rm j}\pi /6}\delta(t-T_s)$, the SOCS random process $X[n]$ is obtained by $2$-times over-sampling the received signal, i.e., the sampling rate is $2/T_s$ [samples/sec].
Thus, the zero-mean SOCS random process $X[n]$ has the cycle period $2$.
As shown in (\ref{eq: asymptotic equivalence_a}), it can be seen in Fig.~\ref{Fig: approx_covariance} that $\bar{\bm{W}}_{MN}\bm{R}_{\bar{\bm{x}}}\bar{\bm{W}}_{MN}^\mathcal{H}$ approaches $\bm{\Omega}$ as $N$ increases.
Similarly, as shown in (\ref{eq: asymptotic equivalence_b}), it can be seen in Fig.~\ref{Fig: approx_comp_covariance} that $\bar{\bm{W}}_{MN}\tilde{\bm{R}}_{\bar{\bm{x}}}\bar{\bm{W}}_{MN}^\mathcal{T}$ approaches $\tilde{\bm{\Omega}}$ as $N$ increases.

Now, the asymptotic second-order properties of the output $\bm{y}$ of the asymptotic FRESH properizer are examined as follows.

\begin{theorem}\label{theorem: properizer}
If the input $\bm{x}$ to the asymptotic FRESH properizer with parameters $M$ and $N$ is obtained by taking the $MN$ consecutive samples from an SOCS random process with cycle period $M$, then the covariance and the complementary covariance matrices of the output $\bm{y}$ of the asymptotic FRESH properizer satisfy
\begin{IEEEeqnarray}{rCl}
\bm{R}_{y} &\sim& \bm{W}_{MN}^{\mathcal{H}}\bm{\Sigma}\bm{W}_{MN}\IEEEeqnarraynumspace\IEEEyessubnumber\label{eq: proposed_cov}\\
\noalign{\noindent{\text{and}}\vspace{\jot}}
\tilde{\bm{R}}_{\bm{y}} &\sim& \bm{O}_{MN},\IEEEeqnarraynumspace\IEEEyessubnumber\label{eq: proposed_com_cov}
\end{IEEEeqnarray}
where an $(MN)$-by-$(MN)$ matrix $\bm{\Sigma}$ is given by
\begin{IEEEeqnarray}{rCl}\label{eq: approx_proposed}
\bm{\Sigma} &\triangleq &\bar{\bm{G}}_{M,N} \bm{\Omega} \bar{\bm{G}}_{M,N}^\mathcal{H}\IEEEeqnarraynumspace\IEEEyessubnumber\label{eq: approx_proposed_a}\\
&=& \bm{R}_{\hat{\bm{y}}} \odot ( \bm{1}_{2M} \otimes \bm{I}_{N/2}).\IEEEeqnarraynumspace\IEEEyessubnumber\label{eq: approx_proposed_b}
\end{IEEEeqnarray}
\end{theorem}

\begin{IEEEproof}
By applying Lemmas~\ref{lemma: multiply} and \ref{lemma: property_sum} to (\ref{eq: proposed cov}) and (\ref{eq: asymptotic equivalence_a}), we obtain $\bm{R}_{\hat{\bm{y}}} \sim \bar{\bm{G}}_{M,N} \bm{\Omega} \bar{\bm{G}}_{M,N}^\mathcal{H}$.
Since $\bar{\bm{G}}_{M,N} \bm{\Omega} \bar{\bm{G}}_{M,N}^\mathcal{H}$ can be rewritten as  $(\bar{\bm{G}}_{M,N}\bar{\bm{W}}_{MN}\bm{R}_{\bar{\bm{x}}}\bar{\bm{W}}_{MN}^\mathcal{H}\bar{\bm{G}}_{M,N}^\mathcal{H}) \odot (\bar{\bm{G}}_{M,N}( \bm{1}_{2M} \otimes \bm{I}_{N} )\bar{\bm{G}}_{M,N}^\mathcal{H})$ by the definition (\ref{eq: block matrix_a}), we obtain $\bm{R}_{\hat{\bm{y}}} \sim \bm{\Sigma}$ from $\bar{\bm{G}}_{M,N}( \bm{1}_{2M}\! \otimes \! \bm{I}_{N} )\bar{\bm{G}}_{M,N}^\mathcal{H} \!\! =  \!\! \bm{1}_{2M} \! \otimes  \!\bm{I}_{N/2}$.
Similarly, $\tilde{\bm{R}}_{\hat{\bm{y}}} \sim (\bar{\bm{G}}_{M,N}\bar{\bm{W}}_{MN}\tilde{\bm{R}}_{\bar{\bm{x}}}\bar{\bm{W}}_{MN}^\mathcal{T}\bar{\bm{G}}_{M,N}^\mathcal{T}) \odot (\bar{\bm{G}}_{M,N}( \bm{1}_{2M} \otimes \bm{P}_{N} )\bar{\bm{G}}_{M,N}^\mathcal{T})$ by (\ref{eq: proposed com cov}), (\ref{eq: asymptotic equivalence_b}), and (\ref{eq: block matrix_b}).
Thus, we obtain $\tilde{\bm{R}}_{\hat{\bm{y}}} \sim \bm{O}_{MN}$ from $\bar{\bm{G}}_{M,N}( \bm{1}_{2M} \otimes \bm{P}_{N} )\bar{\bm{G}}_{M,N}^\mathcal{T} = \bm{O}_{MN}$.
Therefore, the conclusion follows by Lemma~\ref{lemma: multiply}.
\end{IEEEproof}

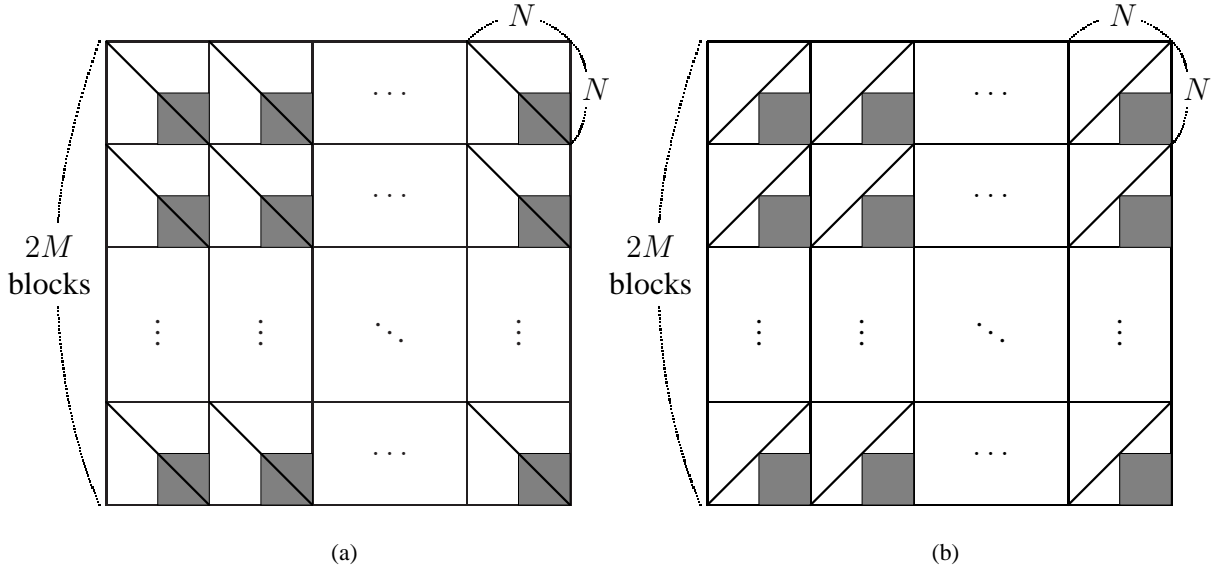
\begin{figure}[tbp]
\setlength{\unitlength}{0.65pt}
{
\begin{center}$\quad$
\subfigure[][]{\label{Fig: Cov}
    \begin{picture}(300,300)(10,15)

        \put(15,285){\line(1,0){270}}
        \put(15,225){\line(1,0){270}}
        \put(15,165){\line(1,0){270}}
        \put(15,75){\line(1,0){270}}
        \put(15,15){\line(1,0){270}}

        \put(15,15){\line(0,1){270}}
        \put(75,15){\line(0,1){270}}
        \put(135,15){\line(0,1){270}}
        \put(225,15){\line(0,1){270}}
        \put(285,15){\line(0,1){270}}

        \put(45,120){\makebox(0,0){$\vdots$}}
        \put(105,120){\makebox(0,0){$\vdots$}}
        \put(255,120){\makebox(0,0){$\vdots$}}
        \put(180,255){\makebox(0,0){$\hdots$}}
        \put(180,195){\makebox(0,0){$\hdots$}}
        \put(180,120){\makebox(0,0){$\ddots$}}
        \put(180,45){\makebox(0,0){$\hdots$}}

        \put(45,15){\shade\path(0,0)(30,0)(30,30)(0,30)(0,0)}
        \put(105,15){\shade\path(0,0)(30,0)(30,30)(0,30)(0,0)}
        \put(255,15){\shade\path(0,0)(30,0)(30,30)(0,30)(0,0)}
        \put(45,165){\shade\path(0,0)(30,0)(30,30)(0,30)(0,0)}
        \put(105,165){\shade\path(0,0)(30,0)(30,30)(0,30)(0,0)}
        \put(255,165){\shade\path(0,0)(30,0)(30,30)(0,30)(0,0)}
        \put(45,225){\shade\path(0,0)(30,0)(30,30)(0,30)(0,0)}
        \put(105,225){\shade\path(0,0)(30,0)(30,30)(0,30)(0,0)}
        \put(255,225){\shade\path(0,0)(30,0)(30,30)(0,30)(0,0)}

        \bezier{12}(225,285)(231,292)(242,297)
        \put(257,300){\makebox(0,0){$N$}}
        \bezier{12}(285,285)(279,292)(268,297)

        \bezier{15}(285,225)(292,231)(294,246)
        \put(300,257){\makebox(0,0){$N$}}
        \bezier{12}(285,285)(292,279)(294,268)

        \bezier{50}(11,285)(-10,225)(-13,180)
        \put(-18,166){\makebox(0,0){$2M$}}
        \put(-18,144){\makebox(0,0){blocks}}
        \bezier{60}(11,15)(-10,65)(-13,130)

        {\linethickness{1pt}
        \thicklines
        \put(15,285){\line(1,-1){120}}
        \put(75,285){\line(1,-1){60}}
        \put(225,285){\line(1,-1){60}}
        \put(15,225){\line(1,-1){60}}
        \put(225,225){\line(1,-1){60}}
        \put(15,75){\line(1,-1){60}}
        \put(75,75){\line(1,-1){60}}
        \put(225,75){\line(1,-1){60}}
        }
    \end{picture}}$\qquad$
\subfigure[][]{\label{Fig: Comple_Cov}
    \begin{picture}(300,290)(10,15)
        \put(15,285){\line(1,0){270}}
        \put(15,225){\line(1,0){270}}
        \put(15,165){\line(1,0){270}}
        \put(15,75){\line(1,0){270}}
        \put(15,15){\line(1,0){270}}

        \put(15,15){\line(0,1){270}}
        \put(75,15){\line(0,1){270}}
        \put(135,15){\line(0,1){270}}
        \put(225,15){\line(0,1){270}}
        \put(285,15){\line(0,1){270}}

        \put(45,120){\makebox(0,0){$\vdots$}}
        \put(105,120){\makebox(0,0){$\vdots$}}
        \put(255,120){\makebox(0,0){$\vdots$}}
        \put(180,255){\makebox(0,0){$\hdots$}}
        \put(180,195){\makebox(0,0){$\hdots$}}
        \put(180,120){\makebox(0,0){$\ddots$}}
        \put(180,45){\makebox(0,0){$\hdots$}}

        \put(45,15){\shade\path(0,0)(30,0)(30,30)(0,30)(0,0)}
        \put(105,15){\shade\path(0,0)(30,0)(30,30)(0,30)(0,0)}
        \put(255,15){\shade\path(0,0)(30,0)(30,30)(0,30)(0,0)}
        \put(45,165){\shade\path(0,0)(30,0)(30,30)(0,30)(0,0)}
        \put(105,165){\shade\path(0,0)(30,0)(30,30)(0,30)(0,0)}
        \put(255,165){\shade\path(0,0)(30,0)(30,30)(0,30)(0,0)}
        \put(45,225){\shade\path(0,0)(30,0)(30,30)(0,30)(0,0)}
        \put(105,225){\shade\path(0,0)(30,0)(30,30)(0,30)(0,0)}
        \put(255,225){\shade\path(0,0)(30,0)(30,30)(0,30)(0,0)}

        \bezier{12}(225,285)(231,292)(242,297)
        \put(257,300){\makebox(0,0){$N$}}
        \bezier{12}(285,285)(279,292)(268,297)

        \bezier{15}(285,225)(292,231)(294,246)
        \put(300,257){\makebox(0,0){$N$}}
        \bezier{12}(285,285)(292,279)(294,268)

        \bezier{50}(11,285)(-10,225)(-13,180)
        \put(-18,166){\makebox(0,0){$2M$}}
        \put(-18,144){\makebox(0,0){blocks}}
        \bezier{60}(11,15)(-10,65)(-13,130)

        {\linethickness{1pt}
        \thicklines
        \put(15,165){\line(1,1){120}}
        \put(15,225){\line(1,1){60}}
        \put(225,225){\line(1,1){60}}
        \put(75,165){\line(1,1){60}}
        \put(225,165){\line(1,1){60}}
        \put(15,15){\line(1,1){60}}
        \put(75,15){\line(1,1){60}}
        \put(225,15){\line(1,1){60}}
        }
    \end{picture}}
\end{center}}
\caption{Illustrations that show how to construct $\bm{R}_{\hat{\bm{y}}}$ and $\tilde{\bm{R}}_{\hat{\bm{y}}}$ from $\bar{\bm{W}}_{MN}\bm{R}_{\bar{\bm{x}}}\bar{\bm{W}}_{MN}^\mathcal{H}\sim \bm{\Omega}$ and $\bar{\bm{W}}_{MN}\tilde{\bm{R}}_{\bar{\bm{x}}}\bar{\bm{W}}_{MN}^\mathcal{T}\sim \tilde{\bm{\Omega}}$, respectively.
Thick diagonal and anti-diagonal lines represent all possible non-zero entries of $\bm{\Omega}$ and $\tilde{\bm{\Omega}}$, respectively.}
\label{Fig: asymptotic properizer}
\end{figure}

By Definition~\ref{definition: asymp_proper} and the above theorem, the output $\bm{y}$ of the asymptotic FRESH properizer is indeed asymptotically proper.
Fig.~\ref{Fig: asymptotic properizer}-(a) illustrates how the frequency-domain covariance matrix $\bm{R}_{\hat{\bm{y}}}$ of the output $\bm{y}$ of the asymptotic FRESH properizer is constructed from $\bar{\bm{W}}_{MN}\bm{R}_{\bar{\bm{x}}}\bar{\bm{W}}_{MN}^\mathcal{H}$.
The thick diagonal lines in each block of size $N$-by-$N$ represent all possible non-zero entries of $\bm{\Omega}$ that is asymptotically equivalent to $\bar{\bm{W}}_{MN}\bm{R}_{\bar{\bm{x}}}\bar{\bm{W}}_{MN}^\mathcal{H}$.
Then, $\bm{R}_{\hat{\bm{y}}}$ that is asymptotically equivalent to $\bm{\Sigma}$ is obtained by collecting the shaded sub-blocks of size $(N/2)$-by-$(N/2)$.
Fig.~\ref{Fig: asymptotic properizer}-(b) illustrates how the frequency-domain complementary covariance matrix $\tilde{\bm{R}}_{\hat{\bm{y}}}$ of the output $\bm{y}$ of the asymptotic FRESH properizer is constructed from $\bar{\bm{W}}_{MN}\tilde{\bm{R}}_{\bar{\bm{x}}}\bar{\bm{W}}_{MN}^\mathcal{T}$.
The thick anti-diagonal lines in each block of size $N$-by-$N$ represent all possible non-zero entries of $\tilde{\bm{\Omega}}$ that is asymptotically equivalent to $\bar{\bm{W}}_{MN}\tilde{\bm{R}}_{\bar{\bm{x}}}\bar{\bm{W}}_{MN}^\mathcal{T}$.
Then, $\tilde{\bm{R}}_{\hat{\bm{y}}}$ that is asymptotically equivalent to $\bm{O}_{MN}$ is obtained by collecting the shaded sub-blocks of size $(N/2)$-by-$(N/2)$.

It is already shown that the amount of the circular shift of $\bm{G}_{M,N}\bm{W}_{MN}\bm{x}^*$ in (\ref{eq: asymptotic properization}) can be any $N(2k+1)/2$, for $k\in \mathbb{Z}$, to satisfy the invertibility of the asymptotic FRESH properizer.
It can be also shown that Theorem~\ref{theorem: properizer} holds for every circular shift by $N(2k+1)/2$, for $k\in \mathbb{Z}$.
Moreover, since $\bm{R}_{\bm{x}}$ and $\tilde{\bm{R}}_{\bm{x}}$ that are $(MN)$-by-$(MN)$ block Toeplitz matrices with block size $M$-by-$M$ can be viewed as block Toeplitz matrices with block size $(lM)$-by-$(lM)$, $\forall l \in \mathbb{N}$, the sampled vector $\bm{x}$ can be asymptotically properized by using any parameters $lM$ and $N/l$, for all $l\in\mathbb{N}$ such that $N/l$ is even.
Thus, similar to the DT FRESH properizer, it is not unique to asymptotically properize a finite number of consecutive samples of a zero-mean SOCS random process with cycle period $M$.

In the following two sections, two representative estimation and detection problems are presented to demonstrate that low-complexity asymptotically optimal post-processors can be easily designed by exploiting the asymptotic second-order properties of the output of the asymptotic FRESH properizer.

\section{Application of Asymptotic FRESH Properizer to Signal Estimation Problem}
In this section, given a finite number of consecutive samples of a zero-mean improper-complex SOCS random process in additive proper-complex white noise, the minimum mean squared error estimation of the random process is considered when only the second-order statistics of the observation vector are provided.
It is well known \cite{Schreier_10} that an optimal estimator is the WLMMSE estimator that linearly processes the observation vector augmented by its complex conjugate.
This optimality of the use of the augmented vector still holds even if the output of the asymptotic FRESH properizer is processed, because the output as an equivalent observation vector is still improper.
Instead, motivated by Theorem~\ref{theorem: properizer}, we propose to use the LMMSE estimator that processes the output of the asymptotic FRESH properizer as if the signal component is a proper-complex random vector with its frequency-domain covariance matrix being equal to the masked version of the exact frequency-domain covariance matrix.
It turns out that this suboptimal linear estimator is asymptotically optimal in the sense that the difference of its average MSE performance from that of the WLMMSE estimator having higher computational complexity converges to zero as the number of samples tends to infinity.

\subsection{Asymptotically Optimal Low-Complexity Estimator}

Let $\bm{r}$ be the length-$MN$ observation vector modeled by
\begin{equation}\label{eq: signal_vector}
\bm{r} = \bm{x} + \bm{v},
\end{equation}
where $\bm{x}$ is the desired signal to be estimated that consists of the $MN$ consecutive samples of a zero-mean improper-complex SOCS random process with cycle period $M\in\mathbb{N}$, and $\bm{v}$ is the additive proper-complex white noise vector.

Given the observation model (\ref{eq: signal_vector}) and the second-order moments $\mathbf{E}\{\bm{x}\}=\bm{0}_{MN}$, $\mathbf{E}\{\bm{x}\bm{x}^\mathcal{H}\}=\bm{R}_{\bm{x}}$, $\mathbf{E}\{\bm{x}\bm{x}^\mathcal{T}\}=\tilde{\bm{R}}_{\bm{x}}$, $\mathbf{E}\{\bm{v}\}=\bm{0}_{MN}$, $\mathbf{E}\{\bm{v}\bm{v}^\mathcal{H}\}=\sigma^2 \bm{I}_{MN}$, and $\mathbf{E}\{\bm{v}\bm{v}^\mathcal{T}\}=\bm{O}_{MN}$ of $\bm{x}$ and $\bm{v}$, the WLMMSE estimation problem can be formulated as
\begin{equation}\label{eq: optimization_estimation}
\begin{array}{rl}
\underset{\bm{x}_{\rm{wl}}}{{\rm minimize}}& \;\; \varepsilon^{(N)}\\
{\text{subject to}} & \;\; \bm{x}_{\rm{wl}}= \bm{F}_{1}\bm{r} + \bm{F}_{2}\bm{r}^*,
\end{array}
\end{equation}
where the average MSE $\varepsilon^{(N)}$ is defined as
\begin{equation}
\varepsilon^{(N)} \triangleq \frac{1}{MN}\mathbf{E} \{ \| \bm{x} -  \bm{x}_{\rm{wl}} \|^2 \}.
\end{equation}

In the following lemma, the optimal solution to the above estimation problem is provided along with its average MSE performance.
It is well known \cite{Schreier_10} that the optimal estimator that minimizes the average MSE without the wide linearity constraint becomes this WLMMSE estimator if the desired signal vector and the additive noise vector are both Gaussian.

\begin{lemma}\label{lemma: WLMMSE}
The WLMMSE estimator $\bm{x}_{\rm{opt}}$ of $\bm{x}$ as the solution to the optimization problem (\ref{eq: optimization_estimation}) is given by
\begin{equation}\label{eq: WLMMSE}
\bm{x}_{\rm{opt}} = \bm{F}_{1,{\rm{opt}}}\bm{r} + \bm{F}_{2,{\rm{opt}}}\bm{r}^*
\end{equation}
where  
\begin{IEEEeqnarray}{rCl}
\bm{F}_{1,{\rm{opt}}} &\triangleq& (\bm{R}_{\bm{x}} - \tilde{\bm{R}}_{\bm{x}} {\bm{R}_{\bm{r}}^{-1}}^* \tilde{\bm{R}}_{\bm{x}}^* ) (\bm{R}_{\bm{r}} - \tilde{\bm{R}}_{\bm{x}} {\bm{R}_{\bm{r}}^{-1}}^* \tilde{\bm{R}}^*_{\bm{x}}  )^{-1}  \IEEEeqnarraynumspace\IEEEyessubnumber\label{eq: WLMMSE_F1}\\
\noalign{\noindent{\text{and}}\vspace{\jot}}
\bm{F}_{2,{\rm{opt}}} &\triangleq& (\tilde{\bm{R}}_{\bm{x}} - \tilde{\bm{R}}_{\bm{x}}^* \bm{R}_{\bm{r}}^{-1} \bm{R}_{\bm{x}} ) (\bm{R}_{\bm{r}}^* - \tilde{\bm{R}}_{\bm{x}}^* \bm{R}_{\bm{r}}^{-1} \tilde{\bm{R}}_{\bm{x}}  )^{-1} . \IEEEeqnarraynumspace\IEEEyessubnumber\label{eq: WLMMSE_F2}
\end{IEEEeqnarray}
with $\bm{R}_{\bm{r}}\triangleq \mathbf{E}\{\bm{r}\bm{r}^\mathcal{H}\}=\sigma^2\bm{I}_{MN}+\bm{R}_{\bm{x}}$.
Moreover, the average MSE $\varepsilon_{\rm{opt}}^{(N)}\triangleq \mathbf{E}\{\|\bm{x} -  \bm{x}_{\rm{opt}} \|^2 \}/(MN)$ is given by
\begin{equation}\label{eq: MSE_WL}
\varepsilon_{\rm{opt}}^{(N)} = \frac{1}{MN}\tr ( \bm{R}_{\bm{x}} - \bm{F}_{1,{\rm{opt}}}\bm{R}_x - \bm{F}_{2,{\rm{opt}}}\tilde{\bm{R}}_{\bm{x}}^* ).
\end{equation}
\end{lemma}

\begin{IEEEproof}
See \cite[Section 5.4]{Schreier_10} and the references therein.
\end{IEEEproof}

The WLMMSE estimator in (\ref{eq: WLMMSE}) requires the computation of two matrices $\bm{F}_{1,{\rm{opt}}}$ and $\bm{F}_{2,{\rm{opt}}}$ to be pre-multiplied to $\bm{r}$ and $\bm{r}^*$, respectively.
As it can be seen from (\ref{eq: WLMMSE_F1}) and (\ref{eq: WLMMSE_F2}), the major burden in computing these matrices comes from the multiplications and the inversions of $(MN)$-by-$(MN)$ matrices.
It can be shown that the matrix multiplications in computing $\bm{F}_{1,{\rm{opt}}}$ and $\bm{F}_{2,{\rm{opt}}}$ require $\mathcal{O}(M^3 N^3)$ complex-valued scalar multiplications because the matrix $\bm{R}_{\bm{r}}^{-1}$ does not have any structure to be exploited in matrix-matrix multiplications even though $\bm{R}_{\bm{r}}$ is block Toeplitz.
By the same reason, the matrix inversion $(\bm{R}_{\bm{r}} - \tilde{\bm{R}}_{\bm{x}} {\bm{R}_{\bm{r}}^{-1}}^* \tilde{\bm{R}}^*_{\bm{x}})$ in (\ref{eq: WLMMSE_F1}) and (\ref{eq: WLMMSE_F2}) requires $\mathcal{O}(M^3 N^3)$ complex-valued scalar multiplications.
Thus, the overall computational complexity of the WLMMSE estimator is $\mathcal{O}(M^3 N^3)$.

An alternative observation model that is equivalent to the original one in (\ref{eq: signal_vector}) can be obtained by applying the asymptotic FRESH properization to $\bm{r}$ as
\begin{equation}\label{eq: signal_vector_MMSE}
\bm{s} = \bm{y} + \bm{w},
\end{equation}
where $\bm{s}=\bm{f}(\bm{r})$, $\bm{y}=\bm{f}(\bm{x})$, and $\bm{w}=\bm{f}(\bm{v})$.
Since the equivalent observation vector $\bm{s}$ is not proper in general, the linear processing of $\bm{s}$ and $\bm{s}^*$ is still needed for the second-order optimal estimation of $\bm{x}$.
Instead, we propose to use a suboptimal linear estimator as follows, which is a function only of $\bm{s}$.

\begin{definition}
The proposed estimate $\bm{x}_{p}$ of $\bm{x}$ is defined by
\begin{equation}\label{eq: proposed_x}
\bm{x}_{p} = \bm{f}^{-1}(\bm{F}_{p}\bm{s}),
\end{equation}
where $\bm{F}_{p}$ is given by
\begin{equation}\label{eq: AP_estimator_b}
\bm{F}_{p}\triangleq \bm{W}_{MN}^{\mathcal{H}} \bm{\Sigma} (\sigma^2 \bm{I}_{MN} + \bm{\Sigma})^{-1} \bm{W}_{MN}.
\end{equation}
\end{definition}

Note from (\ref{eq: proposed_x}) and (\ref{eq: AP_estimator_b}) that
$\bm{\Sigma} (\sigma^2 \bm{I}_{MN} + \bm{\Sigma})^{-1}  \bm{W}_{MN}\bm{s}$ is the estimate of $\bm{W}_{MN}\bm{y}=\bm{W}_{MN}\bm{f}(\bm{x})$ and, consequently, that $\bm{F}_{p}\bm{s}$ is the estimate of $\bm{y}=\bm{f}(\bm{x})$.
Let $\bm{y}_{p} \triangleq  \bm{F}_{p}\bm{s}$.
Then, the estimate $\bm{x}_{p}$ of $\bm{x}=\bm{f}^{-1}(\bm{y})$ is obtained as (\ref{eq: proposed_x}) by applying the inverse operation $\bm{f}^{-1}$ to $\bm{y}_{p}$.

It can be immediately seen that $\bm{F}_{p}$ is chosen to make $\bm{y}_{p}$ the LMMSE estimate of $\bm{y}$ when the covariance and the complementary covariance matrices of $\bm{y}$ are set equal to $\bm{W}_{MN}^{\mathcal{H}}\bm{\Sigma}\bm{W}_{MN}$ and $\bm{O}_{MN}$ appearing in (\ref{eq: proposed_cov}) and (\ref{eq: proposed_com_cov}), respectively.
This idea of using the asymptotically equivalent matrices is motivated by Theorem~\ref{theorem: properizer}, which naturally leads to the asymptotic optimality of the proposed estimator as shown in what follows.

First, the invariance of the Euclidean norm under the asymptotic FRESH properization is shown.

\begin{lemma}\label{lemma: invariance}
The input $\bm{x}$ and the output $\bm{y}=\bm{f}(\bm{x})$ of the asymptotic FRESH properizer have the same Euclidean norm, i.e., $\|\bm{x} \| = \|\bm{y}\|$.
\end{lemma}

\begin{IEEEproof}
Since the centered DFT matrix $\bm{W}_{MN}$ is unitary, we have $\|\bm{x} \| = \|\bm{W}_{MN}\bm{x}\|$.
Let $y_l$ and $y'_l$ denote the $l$th entries of $\bm{y}$ and $\bm{W}_{MN}\bm{x}$, respectively.
Then, we have $\|\bm{W}_{MN}\bm{x}\|^2 = \sum_{l=1}^{MN}|y'_l |^2=\sum_{l=1}^{MN}|y_l |^2=\|\bm{y}\|^2$ by Definition~\ref{definition: asymptotic properization} of $\bm{f}(\bm{x})$ in (\ref{eq: proposed vector}), which is illustrated in Fig.~\ref{Fig: asymptotic_properization}.
Therefore, the conclusion follows.
\end{IEEEproof}

Second, the average MSE of the proposed estimator is derived.

\begin{lemma}\label{lemma: MSE_AP}
The average MSE $\varepsilon_{p}^{(N)}\triangleq \mathbf{E}\{\|\bm{x} -  \bm{x}_{p} \|^2 \}/(MN)$ of the proposed estimator is given by
\begin{equation}\label{eq: MSE_AP}
\varepsilon_{p}^{(N)} = \frac{1}{MN} \tr \{ \bm{\Sigma} - \bm{\Sigma} (\sigma^2 \bm{I}_{MN} + \bm{\Sigma})^{-1} \bm{\Sigma} \}.
\end{equation}
\end{lemma}

\begin{IEEEproof}
The linearity of the asymptotic FRESH properizer leads to $\bm{y} -  \bm{y}_{p}=\bm{f}(\bm{x} -  \bm{x}_{p})$.
Then, by Lemma~\ref{lemma: invariance}, the average MSE performance can be written as $\varepsilon_{p}^{(N)} = \mathbf{E}\{\|\bm{y} - \bm{y}_{p} \|^2 \}/(MN)$.
Since $\bm{W}_{MN}$ is unitary and $\bm{y}_{p}=\bm{F}_{p}\bm{s} $, it can be shown that we have
\begin{IEEEeqnarray}{l}\label{eq: MSE_proposed_original}
\varepsilon_{p}^{(N)} = \frac{1}{MN} \tr \{ \bm{R}_{\hat{\bm{y}}} - 2\bm{\Sigma} (\sigma^2 \bm{I}_{MN}+ \bm{\Sigma})^{-1} \bm{R}_{\hat{\bm{y}}} \nonumber\\
\qquad\quad +  \bm{\Sigma}(\sigma^2 \bm{I}_{MN} + \bm{\Sigma})^{-1}(\sigma^2 \bm{I}_{MN} + \bm{R}_{\hat{\bm{y}}}) (\sigma^2 \bm{I}_{MN} + \bm{\Sigma})^{-1} \bm{\Sigma} \}.\IEEEeqnarraynumspace
\end{IEEEeqnarray}
It can be also shown that if $\tilde{\bm{A}} \triangleq \bm{A} \odot ( \bm{1}_{2M} \otimes \bm{I}_{N/2})$ and $\tilde{\bm{B}} \triangleq \bm{B} \odot ( \bm{1}_{2M} \otimes \bm{I}_{N/2})$ for $(MN)$-by-$(MN)$ matrices $\bm{A}$ and $\bm{B}$ then $\tr (\bm{A}) = \tr ( \tilde{\bm{A}} ) $ and $\tr ( \tilde{\bm{A}}\bm{B} ) =\tr ( \tilde{\bm{A}}\tilde{\bm{B}} )$.
Thus, the right side of (\ref{eq: MSE_proposed_original}) is invariant under replacing $\bm{R}_{\hat{\bm{y}}}$ with $\bm{\Sigma} = \bm{R}_{\hat{\bm{y}}} \odot ( \bm{1}_{2M} \otimes \bm{I}_{N/2})$ in (\ref{eq: approx_proposed_b}).
Therefore, (\ref{eq: MSE_proposed_original}) can be simplified to (\ref{eq: MSE_AP}).
\end{IEEEproof}

Now, the asymptotic optimality of the proposed estimator is provided.

\begin{theorem}\label{theorem: MSE_same}
The average MSE of the proposed estimator approaches that of the WLMMSE estimator as the number of samples tends to infinity in the sense that
\begin{equation}\label{eq: asymptotic_MSE}
\lim_{N \rightarrow \infty} \Big(\varepsilon_{p}^{(N)} - \varepsilon_{\rm{opt}}^{(N)} \Big) = 0.
\end{equation}
\end{theorem}

\begin{IEEEproof}
It is shown in \cite[Section 5.4]{Schreier_10} that the average MSE $\varepsilon_{\rm{opt}}^{(N)}$ in (\ref{eq: MSE_WL}) of the WLMMSE estimator can be simplified as
\begin{equation}\label{eq: MSE_modify}
\varepsilon_{\rm{opt}}^{(N)} = \frac{1}{2MN} \tr \{ \bm{R}_{\bar{\bm{x}}} - \bm{R}_{\bar{\bm{x}}} (\sigma^2 \bm{I}_{2MN} + \bm{R}_{\bar{\bm{x}}})^{-1}\bm{R}_{\bar{\bm{x}}} \}.
\end{equation}
Now, we define $\hat{\bm{\Omega}}\triangleq \bar{\bm{W}}_{MN}^\mathcal{H} \bm{\Omega} \bar{\bm{W}}_{MN}$ and introduce
\begin{IEEEeqnarray}{rCl}
\hat{\varepsilon}_{\rm{opt}}^{(N)} & \triangleq & \frac{1}{2MN} \tr \{ \hat{\bm{\Omega}} - \hat{\bm{\Omega}} (\sigma^2 \bm{I}_{2MN} + \hat{\bm{\Omega}})^{-1}\hat{\bm{\Omega}} \} \IEEEeqnarraynumspace\IEEEyessubnumber\label{eq: MSE_WL_approx}\\
&=& \frac{1}{2MN} \tr \{ \bm{\Omega} - \bm{\Omega} (\sigma^2 \bm{I}_{2MN} + \bm{\Omega})^{-1}\bm{\Omega} \},\IEEEeqnarraynumspace\IEEEyessubnumber
\end{IEEEeqnarray}
which is obtained by replacing $\bm{R}_{\bar{\bm{x}}}$ in (\ref{eq: MSE_modify}) with $\hat{\bm{\Omega}}$ and by using the fact that $\bar{\bm{W}}_{MN}$ is unitary.
Note that $\bar{\bm{W}}_{MN}\bm{R}_{\bar{\bm{x}}}\bar{\bm{W}}_{MN}^\mathcal{H} \sim \bm{\Omega}$ as shown in (\ref{eq: asymptotic equivalence_a}).
This introduction of $\hat{\varepsilon}_{\rm{opt}}^{(N)}$ is motivated by the result in \cite[Theorem~1]{Yoo_10}, where a suboptimal estimator is proposed for the estimation of a proper-complex cyclostationary random signal and its average MSE is shown to approach that of the LMMSE estimator as the number of samples tends to infinity.
Since $\bm{R}_{\bar{\bm{x}}}$ is a covariance matrix, it is positive semidefinite, which implies that all the eigenvalues of $(\sigma^2 \bm{I}_{2MN} + \bm{R}_{\bar{\bm{x}}})^{-1}$ in (\ref{eq: MSE_modify}) are upper bounded by $1/\sigma^2$.
Note that $\bm{\Omega}$ and $\hat{\bm{\Omega}}$ are also positive semidefinite because $( \bm{1}_{2M} \otimes \bm{I}_{N/2})$ in (\ref{eq: block matrix_a}) is positive semidefinite.
Similarly, all the eigenvalues of $(\sigma^2 \bm{I}_{2MN} + \hat{\bm{\Omega}})^{-1}$ in (\ref{eq: MSE_WL_approx}) are also upper bounded by $1/\sigma^2$.
Thus, the strong norms $\|(\sigma^2 \bm{I}_{2MN} + \bm{R}_{\bar{\bm{x}}})^{-1}\|$ and $\|(\sigma^2 \bm{I}_{2MN} + \hat{\bm{\Omega}})^{-1}\|$ are uniformly upper bounded by $1/\sigma^2$ for any matrix size.
Recall that if $\bm{A}_k \sim \bm{B}_k$ and $\|\bm{A}_k^{-1} \|, \|\bm{B}_k^{-1} \| \leq c < \infty, \;\forall k$, for a positive constant $c$, then $\bm{A}_k^{-1} \sim \bm{B}_k^{-1}$ \cite[Theorem~1-(4)]{Gray_book}.
Thus, we have $(\sigma^2 \bm{I}_{2MN} + \bm{R}_{\bar{\bm{x}}})^{-1} \sim (\sigma^2 \bm{I}_{2MN} + \hat{\bm{\Omega}})^{-1}$.
Recall also that if two sequences $(\bm{A}_k)_k$ and $(\bm{B}_k)_k$ of $N_k$-by-$N_k$ matrices are asymptotically equivalent then $\lim_{k \rightarrow \infty} \tr ( \bm{A}_k - \bm{B}_k ) /N_k = 0$ \cite[Corollary~1]{Gray_book}.
Thus, combined with Lemmas~\ref{lemma: multiply} and \ref{lemma: property_sum}, we have $\lim_{N \rightarrow \infty} ( \varepsilon_{\rm{opt}}^{(N)} - \hat{\varepsilon}_{\rm{opt}}^{(N)} ) = 0$.
Therefore, in order to show (\ref{eq: asymptotic_MSE}), it now suffices to show $\hat{\varepsilon}_{\rm{opt}}^{(N)}=\varepsilon_{p}^{(N)}$.

Let $\bm{E}_{2M,N}$ be the $(2MN)$-by-$(2MN)$ matrix that permutes the rows of the post-multiplied matrix in such a way that the $(N(m-1)+n)$th row of the post-multiplied matrix becomes the $(2M(n-1)+m)$th row for $m=1,2, \cdots, 2M$ and $n=1,2, \cdots, N$, i.e., the rows having indexes $(N(m-1)+n)$, for $m=1,2, \cdots, 2M,$ are grouped for each $n$.
Then, $\bm{E}_{2M,N}\bm{\Omega}\bm{E}_{2M,N}^\mathcal{T}$ is a $(2MN)$-by-$(2MN)$ block diagonal matrix with block size $(2M)$-by-$(2M)$, because $\bm{\Omega}$ is a $(2MN)$-by-$(2MN)$ block matrix with diagonal blocks of block size $N$-by-$N$.
Similarly, the $(MN)$-by-$(MN)$ matrix $\bm{E}_{2M,N/2}$ is defined.
Let $\bm{E}_1$ and $\bm{E}_2$ be the $(MN)$-by-$(MN)$ matrix that are defined as $\bm{E}_1\triangleq \bm{P}_{MN}(\bm{I}_{N/2}\otimes \hat{\bm{E}}_{2M})\bm{E}_{2M,N/2}$ and $\bm{E}_2\triangleq (\bm{I}_{N/2}\otimes \hat{\bm{E}}_{2M})\bm{E}_{2M,N/2}$, respectively, where the $(2M)$-by-$(2M)$ matrix $\hat{\bm{E}}_{2M}$ permutes the rows of the post-multiplied matrix in such a way that the $(2m)$th row becomes the $m$th row and that the $(2m-1)$th row becomes the $(M+m)$th row for $m=1,2, \cdots, M$.
Then, the definition $\bm{\Sigma} \triangleq \bar{\bm{G}}_{M,N} \bm{\Omega} \bar{\bm{G}}_{M,N}^\mathcal{H}$ in (\ref{eq: approx_proposed_a}) leads to
\begin{equation}\label{eq: permute}
\bm{E}_{2M,N}\bm{\Omega}\bm{E}_{2M,N}^\mathcal{T} = \begin{bmatrix}
\bm{E}_1 \bm{\Sigma}^* \bm{E}_1^\mathcal{T} & \bm{O}_{MN} \\
\bm{O}_{MN} & \bm{E}_2 \bm{\Sigma} \bm{E}_2^\mathcal{T}
\end{bmatrix}.
\end{equation}
Since $\bm{E}_{2M,N}$, $\bm{E}_1$, and $\bm{E}_2$ are all permutation matrices, $\hat{\varepsilon}_{\rm{opt}}^{(N)}$ in (\ref{eq: MSE_WL_approx}) can be rewritten as $\hat{\varepsilon}_{\rm{opt}}^{(N)} = \tr \{ \bm{\Sigma}^*  -  \bm{\Sigma}^*  (\sigma^2 \bm{I}_{2MN} + \bm{\Sigma}^*)^{-1} \bm{\Sigma}^* + \bm{\Sigma}  -  \bm{\Sigma}  (\sigma^2 \bm{I}_{2MN} +  \bm{\Sigma} )^{-1} \bm{\Sigma} \} /(2MN)$, which leads to $\hat{\varepsilon}_{\rm{opt}}^{(N)} = \varepsilon_{p}^{(N)}$ because $\bm{\Sigma}$ is Hermitian symmetric.
Therefore, the conclusion follows.
\end{IEEEproof}

The proposed estimator requires the pre-processing of $\bm{r}$ to obtain $\bm{s}$, the computation and the multiplication of $\bm{F}_{p}$, and the inverse operation on $\bm{y}_{p}$ to obtain $\bm{x}_{p}$.
As it can be seen from (\ref{eq: proposed vector}) and (\ref{eq: invertible_all}), the major burden in the asymptotic FRESH properization and its inverse operation comes from the multiplications of the $(MN)$-by-$(MN)$ centered DFT matrix.
This can be efficiently implemented with only $\mathcal{O}(MN \log (MN) )$ complex-valued scalar multiplications by using, e.g., the fast algorithm in \cite{Bi_98}.
As it can be seen from (\ref{eq: AP_estimator_b}), the major burden in computing $\bm{F}_{{p}}$ comes from the inversion of $(\sigma^2 \bm{I}_{2MN}+\bm{\Sigma})$.
The inversion of an $(MN)$-by-$(MN)$ block matrix with diagonal blocks of size $N$-by-$N$ requires $\mathcal{O}(M^3 N)$ complex-valued scalar multiplications \cite{Yoo_10}.
Thus, this inversion of $(\sigma^2 \bm{I}_{2MN}+\bm{\Sigma})$ has the same order $\mathcal{O}(M^3 N)$ of computational complexity because $\bm{\Sigma}$ is an $(MN)$-by-$(MN)$ block matrix with diagonal blocks of size $(N/2)$-by-$(N/2)$.

Since we consider the block processing where $M$ is a fixed small number and $N$ is much larger than $M$, the overall complexities of the WLMMSE and the proposed estimators as functions only of $N$ can be rewritten now as $\mathcal{O}(N^3)$ and $\mathcal{O}(N \log N)$, respectively.
Thus, the computational complexity of the proposed estimator is much lower than that of the WLMMSE estimator.
Though not major, an additional complexity reduction comes from the fact that the proposed estimator is linear that requires the computation and multiplication of one matrix $\bm{F}_{p}$ instead of two matrices $\bm{F}_{1,{\rm{opt}}}$ and $\bm{F}_{2,{\rm{opt}}}$, all with $MN (\gg 1)$ columns.

\subsection{Numerical Results}\label{sec: estimator_numerical}

In this subsection, numerical results are provided that show the computational efficiency and the asymptotic optimality of the proposed estimator.

\begin{figure}[tbp]\centering
\includegraphics[width=6in]{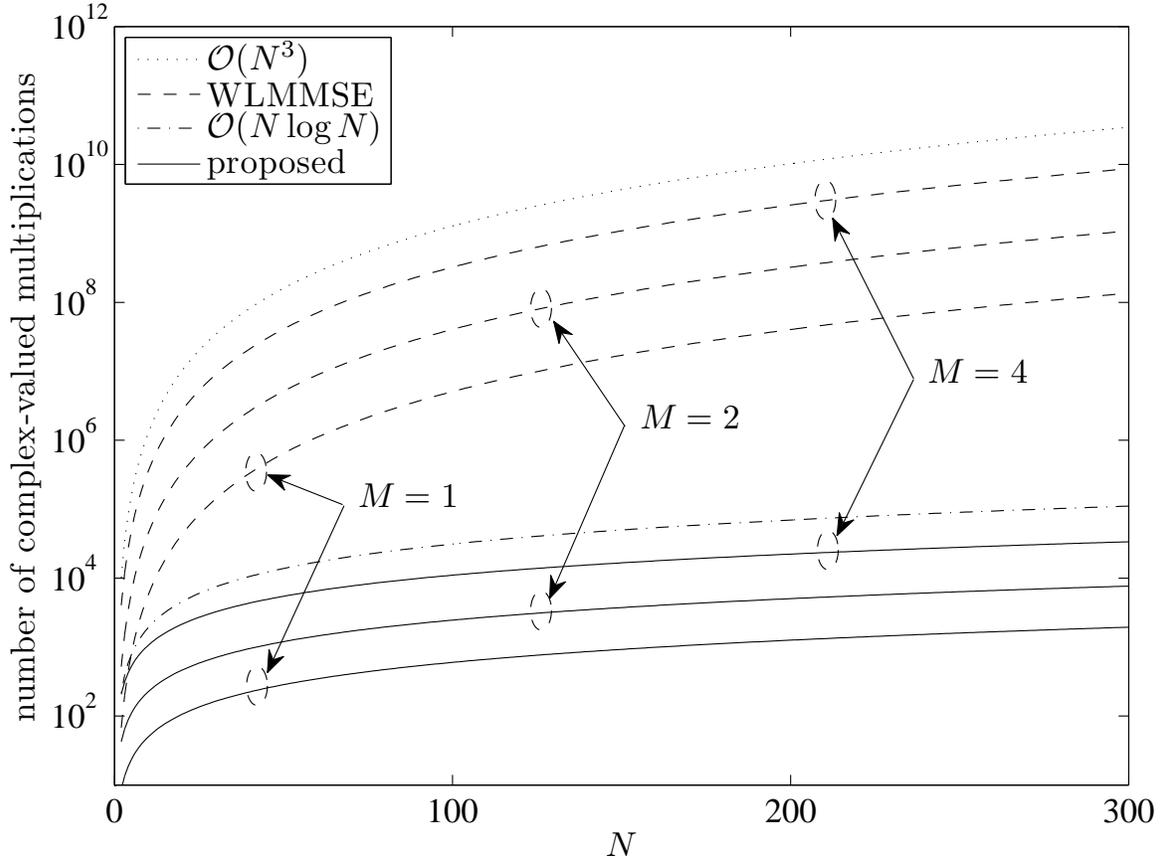} 
\caption{Computational complexities of the WLMMSE and the proposed estimators.} \label{Fig: numerical_complexity}
\end{figure}

The first result is to compare the complexity of the proposed estimator with that of the WLMMSE estimator.
Fig.~\ref{Fig: numerical_complexity} shows that the number of complex-valued multiplications needed in computing the WLMMSE and the proposed estimators for cycle periods $M=1$, $2$, and $4$.
Recall that the computational complexities of the WLMMSE and the proposed estimators are $\mathcal{O}(N^3)$ and $\mathcal{O}(N\log N)$ for a fixed integer $M$, respectively.
It can be seen from Fig.~\ref{Fig: numerical_complexity} that the computational complexity of the proposed estimator is much lower than that of the WLMMSE estimator.
As predicted, the approximation motivated by Theorem~\ref{theorem: properizer} leads to this significant complexity reduction.

\begin{figure}[tbp]\centering
\includegraphics[width=6in]{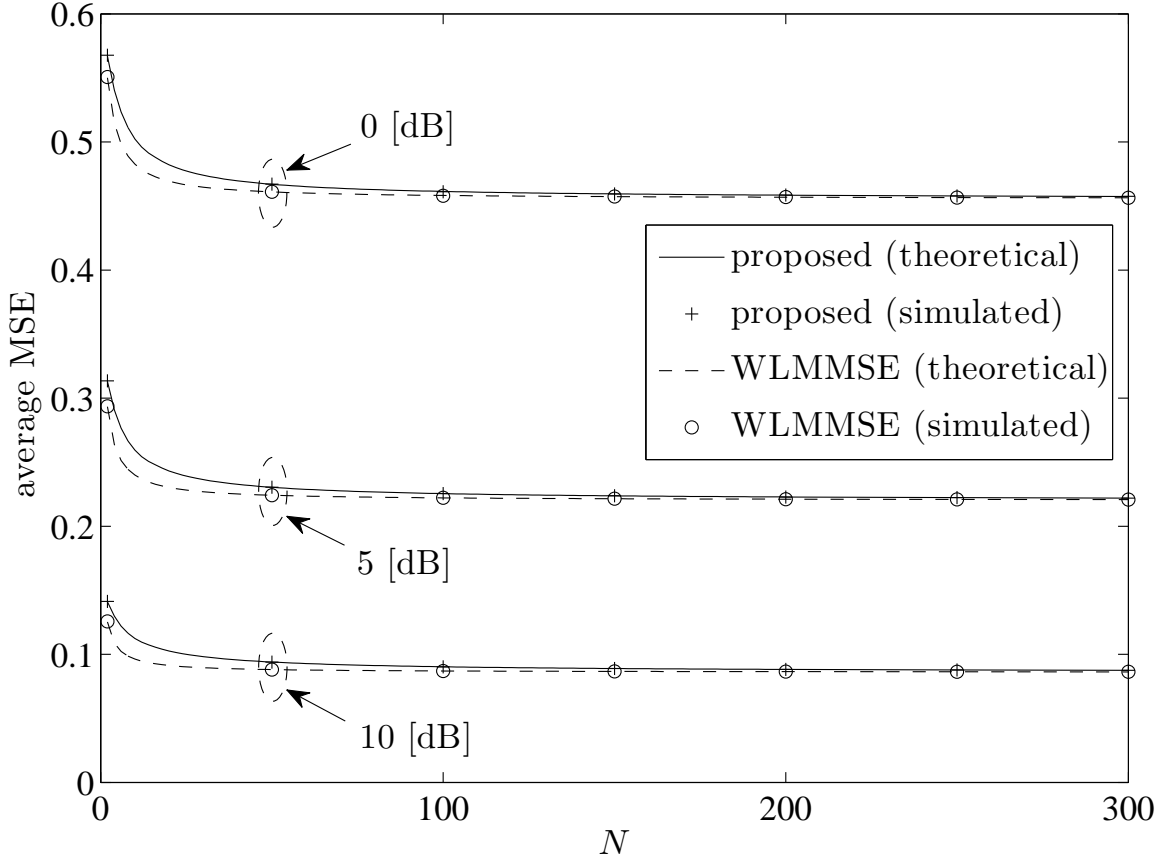} 
\caption{Average MSEs of the WLMMSE and the proposed estimators.} \label{Fig: numerical_MSE} 
\end{figure}

The second result is to show the asymptotic optimality of the proposed estimator.
We consider the case where an improper-complex SOCS random process is obtained by uniformly sampling a CT zero-mean improper-complex SOCS random process.
The CT random process is, e.g., a Gaussian jamming signal, generated by OQPSK modulating two independent real-valued independent and identically distributed zero-mean symbol sequences with the SRRC pulse having roll-off factor $0.22$.
This signal is sampled at $2$-times the symbol rate of the OQPSK symbols, which results in the DT zero-mean improper-complex SOCS random process with cycle period $M=2$.
Thus, the entries of $\bm{x}$ in (\ref{eq: signal_vector}) are the $MN$ consecutive samples of the random process.
Fig.~\ref{Fig: numerical_MSE} shows that the average MSEs of the WLMMSE and the proposed estimators versus $N$ for symbol energy per noise density $E_s/N_0=0$, $5$, and $10$ [dB], where the theoretical results are evaluated by (\ref{eq: MSE_WL}) and (\ref{eq: MSE_AP}) while the simulated results are obtained from $10^5$ Monte-Carlo runs.
It can be seen that, as shown in Theorem~\ref{theorem: MSE_same}, the average MSE of the proposed estimator approaches that of the WLMMSE estimator as $N$ increases.

\section{Application of Asymptotic FRESH Properizer to Signal Presence Detection Problem}

In this section, again given a finite number of consecutive samples, now the signal presence presence detection of a zero-mean improper-complex SOCS Gaussian random process is considered in additive proper-complex white Gaussian noise.
It is well known \cite{Poor_94} that the likelihood ratio is a sufficient statistic for all binary hypothesis tests under any optimality criterion.
In \cite{Schreier_10}, it is shown that the exact LRT statistic can be written as a quadratic function of the augmented observation vector when the improper-complex random process is \emph{Gaussian}.
Similar to the estimation problem in the previous section, this optimality of the use of the augmented vector still holds even if the output of the asymptotic FRESH properizer as an equivalent observation vector is processed.
Again motivated by Theorem~\ref{theorem: properizer}, we propose to use this equivalent observation vector of half the length of the augmented vector as if the signal component is a proper-complex random vector with its frequency-domain covariance matrix being equal to the masked version of the exact frequency-domain covariance matrix.
It turns out that this suboptimal test statistic is asymptotically optimal in the sense that its difference from the exact LRT statistic having higher computational complexity converges to zero w.p.~$1$ as the number of samples tends to infinity.

\subsection{Asymptotically Optimal Low-Complexity Detector}

Let $\bm{r}$ be the length-$MN$ observation vector modeled by
\begin{IEEEeqnarray}{lrCl}
\mathcal{H}_0: &\bm{r} &=& \bm{v}\nonumber\\
\rm{versus} & & &\label{eq: detection_observation}\\%
\mathcal{H}_1: &\bm{r} &=& \bm{x} + \bm{v},\nonumber
\end{IEEEeqnarray}
where $\mathcal{H}_0$ and $\mathcal{H}_1$ are the null and the alternative hypotheses, respectively.
The desired signal $\bm{x}$ to be detected consists of the $MN$ consecutive samples of a zero-mean improper-complex SOCS Gaussian random process with cycle period $M\in\mathbb{N}$, and $\bm{v}$ is the additive proper-complex white Gaussian noise vector.

Throughout this section, the definitions and the notations of the second-order statistics of $\bm{x}$, $\bm{v}$, and $\bm{r}$ follow those in the previous section except that under the null hypothesis the second-order statistic of $\bm{r}$ does not contain the desired signal component.
Similar to (\ref{eq: def_augmented}), the augmented observation vector is denoted by $\bar{\bm{r}}$.
Then, the LRT statistic of the observation vector $\bm{r}$ is provided as follows.

\begin{lemma}
Given the observation model (\ref{eq: detection_observation}), the LRT statistic of $\bm{r}$ is given by
\begin{equation}\label{eq: WL_LRT}
T_{\rm LRT}^{(N)} = \frac{1}{2MN}\bar{\bm{r}}^\mathcal{H} \bar{\bm{R}}_{\bar{\bm{x}}} \bar{\bm{r}},
\end{equation}
where $\bar{\bm{R}}_{\bar{\bm{x}}}\triangleq \sigma^{-2}\bm{I}_{2MN}-(\sigma^{2}\bm{I}_{2MN}+\bm{R}_{\bar{\bm{x}}})^{-1}$.
\end{lemma}

\begin{IEEEproof}
It is straightforward by using the probability density function (PDF) of the augmented improper-complex Gaussian random vector \cite{Bos_95}.
For details, see \cite[Section 7.4]{Schreier_10}.
\end{IEEEproof}

Note that, since the length-$MN$ observation vector $\bm{r}$ is improper and Gaussian, the LRT statistic $T_{\rm LRT}^{(N)}$ in (\ref{eq: WL_LRT}) is a quadratic form of the length-$2MN$ augmented observation vector $\bar{\bm{r}}$.
Under any optimality criterion such as the Neyman-Pearson, the Bayesian, and the minimax criteria, the optimal detector computes $T_{\rm LRT}^{(N)}$, compares it with an optimal threshold $\eta$, and then declares $\mathcal{H}_1$ if $T_{\rm LRT}^{(N)} > \eta$ and $\mathcal{H}_0$ otherwise.

The computation of $T_{\rm LRT}^{(N)}$ requires the matrix inversion of $(\sigma^{2}\bm{I}_{2MN}+\bm{R}_{\bar{\bm{x}}})$, which is the $(2MN)$-by-$(2MN)$ covariance matrix of the augmented observation vector $\bar{\bm{r}}$ under $\mathcal{H}_1$, and the matrix-vector multiplication.
The major burden in computing $T_{\rm LRT}^{(N)}$ comes from the matrix inversion and it requires $\mathcal{O}(M^3 N^3)$ complex-valued scalar multiplications.
The matrix-vector multiplication requires only $\mathcal{O}(M^2 N^2)$ complex-valued scalar multiplications.
Thus, the overall computational complexity of the LRT statistic is $\mathcal{O}(M^3 N^3)$.

An alternative observation model that is equivalent to the original one in (\ref{eq: detection_observation}) can be obtained by applying the asymptotic FRESH properization to $\bm{r}$ as
\begin{IEEEeqnarray}{lrCl}
\mathcal{H}_0: &\bm{s} &=& \bm{w}\nonumber\\
\rm{versus} & & &\label{eq: detection_problem_FRESH}\\%
\mathcal{H}_1: &\bm{s} &=& \bm{y} + \bm{w},\nonumber
\end{IEEEeqnarray}
where $\bm{s}=\bm{f}(\bm{r})$, $\bm{y}=\bm{f}(\bm{x})$, and $\bm{w}=\bm{f}(\bm{v})$.
Since the equivalent observation vector $\bm{s}$ is not proper in general, the augmentation of $\bm{s}$ and $\bm{s}^*$ is still needed to compute the exact LRT statistic of $\bm{s}$.
Instead, we propose to use a suboptimal LRT statistic as follows, which is a quadratic function only of $\bm{s}$.

\begin{definition}
Given the observation model (\ref{eq: detection_problem_FRESH}), the proposed test statistic of $\bm{s}$ is defined by
\begin{equation}\label{eq: proposed_LRT}
T_{p}^{(N)} \triangleq \frac{1}{MN}\bm{s}^\mathcal{H} \bm{W}_{MN}^{\mathcal{H}}\bar{\bm{\Sigma}} \bm{W}_{MN}\bm{s},
\end{equation}
where $\bar{\bm{\Sigma}} \triangleq\sigma^{-2}\bm{I}_{MN}-(\sigma^{2}\bm{I}_{MN}+\bm{\Sigma})^{-1}$.
\end{definition}

It can be immediately seen that $T_{p}^{(N)}$ can be viewed as the exact LRT statistic of $\bm{s}$ when the covariance and the complementary covariance matrices of $\bm{y}$ are set equal to $\bm{W}_{MN}^{\mathcal{H}}\bm{\Sigma}\bm{W}_{MN}$ and $\bm{O}_{MN}$ appearing in (\ref{eq: proposed_cov}) and (\ref{eq: proposed_com_cov}), respectively.
Similar to the estimation problem in the previous section, this idea of using the asymptotically equivalent matrices is again motivated by Theorem~\ref{theorem: properizer}, which naturally leads to the asymptotic equivalence of the LRT statistic $T_{\rm LRT}^{(N)}$ in (\ref{eq: WL_LRT}) and the proposed test statistic $T_{p}^{(N)}$ in (\ref{eq: proposed_LRT}).

Now, the asymptotic optimality of the proposed test statistic is provided.

\begin{theorem}\label{theorem: LRT_same}
The proposed test statistic approaches the exact LRT statistic as the number of samples tends to infinity in the sense that
\begin{equation}\label{eq: asymptotic_same_detector}
\lim_{N \rightarrow \infty} (T_{p}^{(N)} - T_{\rm LRT}^{(N)}) = 0, \text{w.p.~$1$},
\end{equation}
under $\mathcal{H}_0$ and $\mathcal{H}_1$.
\end{theorem}

\begin{IEEEproof}
We define $\bar{\bm{\Omega}} \triangleq\sigma^{-2}\bm{I}_{2MN}-(\sigma^{2}\bm{I}_{2MN}+\bm{\Omega})^{-1}$ and introduce a suboptimal test statistic
\begin{equation}\label{eq: WL_LRT_approx}
\hat{T}_{\rm LRT}^{(N)} \triangleq \frac{1}{2MN}\bar{\bm{r}}^\mathcal{H} \bar{\bm{W}}_{MN}^\mathcal{H}\bar{\bm{\Omega}} \bar{\bm{W}}_{MN} \bar{\bm{r}},
\end{equation}
which is obtained by replacing $\bar{\bm{R}}_{\bar{\bm{x}}}$ in (\ref{eq: WL_LRT}) with $\bar{\bm{W}}_{MN}^\mathcal{H} \bar{\bm{\Omega}} \bar{\bm{W}}_{MN} $ and by using the fact that $\bar{\bm{W}}_{MN}$ is unitary.
This introduction of $\hat{T}_{\rm LRT}^{(N)}$ is motivated by $\bar{\bm{W}}_{MN}\bm{R}_{\bar{\bm{x}}}\bar{\bm{W}}_{MN}^\mathcal{H} \sim \bm{\Omega}$ in (\ref{eq: asymptotic equivalence_a}).
Similar to the proof of Theorem~\ref{theorem: MSE_same}, it can be shown that $\bar{\bm{R}}_{\bar{\bm{x}}} \sim \bar{\bm{W}}_{MN}^\mathcal{H} \bar{\bm{\Omega}} \bar{\bm{W}}_{MN}$.
In \cite{Zhang_10}, the LRT problems are considered where the covariance matrix of a proper-complex Gaussian signal vector is either Toeplitz or block Toeplitz.
It is shown under both $\mathcal{H}_0$ and $\mathcal{H}_1$ that a suboptimal test statistic, which is obtained by replacing the covariance matrix in the quadratic form of the LRT statistic with its asymptotically equivalent one, converges to the exact LRT statistic w.p.~$1$ as the number of samples tends to infinity \cite[Propositions~1 and 3]{Zhang_10}.
By applying this result, we have $\lim_{N\rightarrow \infty} (T_{\rm LRT}^{(N)}- \hat{T}_{\rm LRT}^{(N)}) = 0$ w.p.~$1$ because $\bm{R}_{\bar{\bm{x}}} \sim \bar{\bm{W}}_{MN}^\mathcal{H} \bar{\bm{\Omega}} \bar{\bm{W}}_{MN}$.
Therefore, in order to show (\ref{eq: asymptotic_same_detector}), it now suffices to show $\hat{T}_{\rm LRT}^{(N)}= T_{p}^{(N)}$.

Let $\hat{\bm{s}}$ be the frequency-domain equivalent observation vector that is defined as $\hat{\bm{s}} \triangleq \bm{W}_{MN}\bm{s}$.
Then, by using the permutation matrices $\bm{E}_{2M,N}$, $\bm{E}_1$, and $\bm{E}_1$ that are defined in the proof of Theorem~\ref{theorem: MSE_same} in Section~IV, we have
\begin{equation}\label{eq: permute_ob}
\bm{E}_{2M,N}\bar{\bm{W}}_{MN}\bar{\bm{r}} = \begin{bmatrix}
\bm{E}_1\hat{\bm{s}}^*\\
\bm{E}_2\hat{\bm{s}}
\end{bmatrix}.
\end{equation}
By using (\ref{eq: permute}) and (\ref{eq: permute_ob}), we can rewrite $\hat{T}_{\rm LRT}^{(N)}$ defined in (\ref{eq: WL_LRT_approx}) as $\hat{T}_{\rm LRT}^{(N)}=(\hat{\bm{s}}^\mathcal{T}\bar{\bm{\Sigma}}^* \hat{\bm{s}}^* +\hat{\bm{s}}^\mathcal{H}\bar{\bm{\Sigma}} \hat{\bm{s}})/(2MN) = T_{p}^{(N)}$ because $\hat{\bm{s}}^\mathcal{H}\bar{\bm{\Sigma}} \hat{\bm{s}}$ is real-valued and (\ref{eq: permute}) leads
\begin{equation}
\bm{E}_{2M,N}\bar{\bm{\Omega}}\bm{E}_{2M,N}^\mathcal{T} = \begin{bmatrix}
\bm{E}_1 \bar{\bm{\Sigma}}^* \bm{E}_1^\mathcal{T} & \bm{O}_{MN} \\
\bm{O}_{MN} & \bm{E}_2 \bar{\bm{\Sigma}} \bm{E}_2^\mathcal{T}
\end{bmatrix}.
\end{equation}
Therefore, the conclusion follows.
\end{IEEEproof}

The computation of the proposed test statistic $T_{p}^{(N)}$ requires the pre-processing of $\bm{r}$ to obtain $\bm{s}$, the matrix inversion of $(\sigma^{2}\bm{I}_{MN}+\bm{\Sigma})$, and the matrix-vector multiplication to compute the quadratic form.
As shown in Section~IV, the asymptotic FRESH properization of the length-$MN$ vector $\bm{r}$ and the matrix inversion of $(\sigma^{2}\bm{I}_{MN}+\bm{\Sigma})$ require $\mathcal{O}(MN \log (MN) )$ and $\mathcal{O}(M^3 N)$ complex-valued scalar multiplications, respectively.
In addition, the matrix-vector multiplication requires $\mathcal{O}(M^2 N)$ complex-valued scalar multiplications, because $(\sigma^{2}\bm{I}_{MN}+\bm{\Sigma})$ is an $(MN)$-by-$(MN)$ block matrix with diagonal blocks of size $(N/2)$-by-$(N/2)$.
Since we consider the block processing where $M$ is a fixed small number and $N$ is much larger than $M$, the overall complexity in computing the exact LRT statistic and the proposed test statistic as functions only of $N$ can be rewritten now as $\mathcal{O}(N^3)$ and $\mathcal{O}(N \log N )$, respectively.
Thus, the computational complexity of the proposed test statistic is much lower than that of the exact LRT statistic.

\subsection{Numerical Results}

In this subsection, numerical results are provided that show only the asymptotic optimality of the proposed test statistic because its computational efficiency can be similarly shown as Fig.~\ref{Fig: numerical_complexity}.
Throughout this subsection, the zero-mean improper-complex SOCS Gaussian random process to be detected is obtained by uniformly sampling the OQPSK-like signal generated in a similar way to that described in Section~\ref{sec: estimator_numerical}.

\begin{figure}[tbp]\centering
\includegraphics[width=6in]{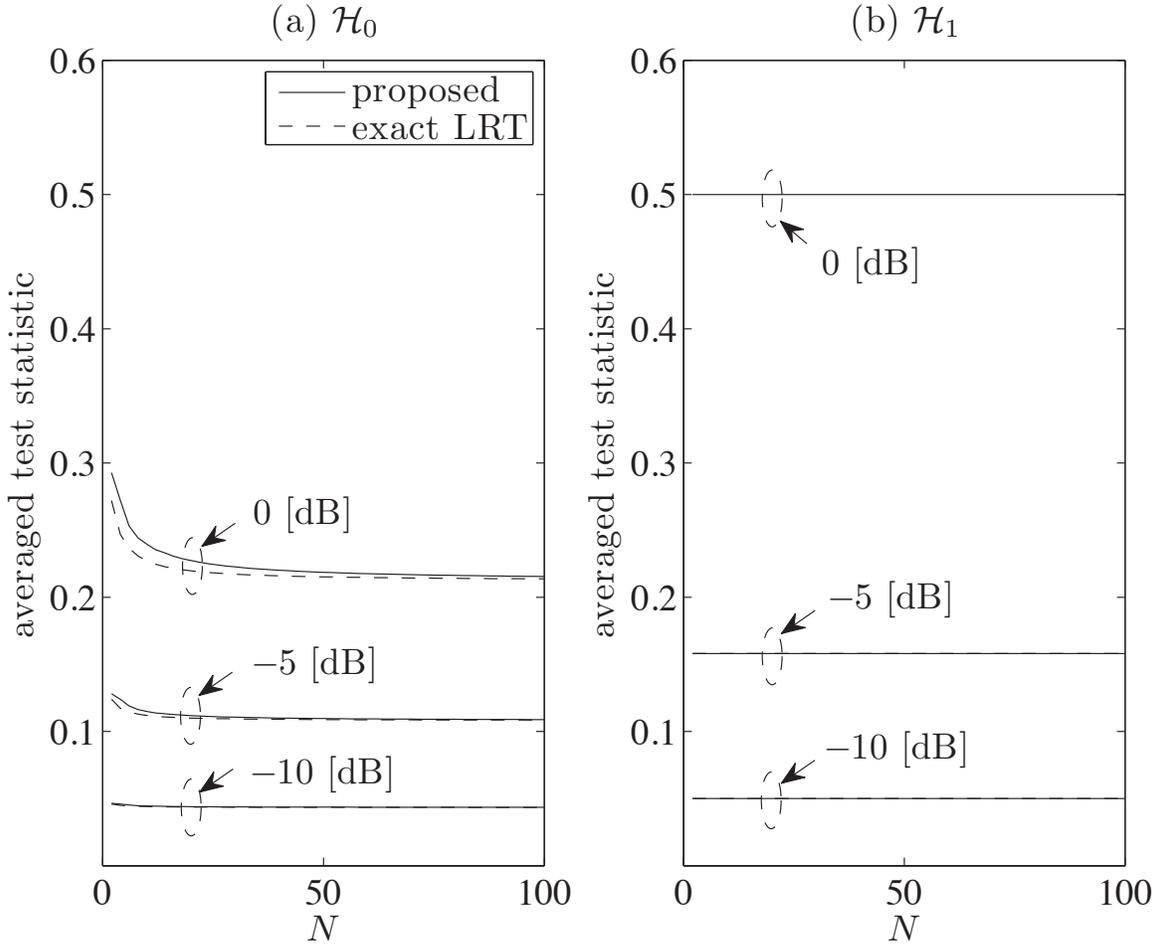} %
\caption{Statistical averages of the exact LRT and the proposed test statistics versus $N$ for $E_s/N_0 = 0, -5,$ and $-10$ [dB], under (a) $\mathcal{H}_0$ and (b) $\mathcal{H}_1$.} \label{Fig: numerical_LRT_avg}
\end{figure}

\begin{figure}[tbp]\centering
\includegraphics[width=6in]{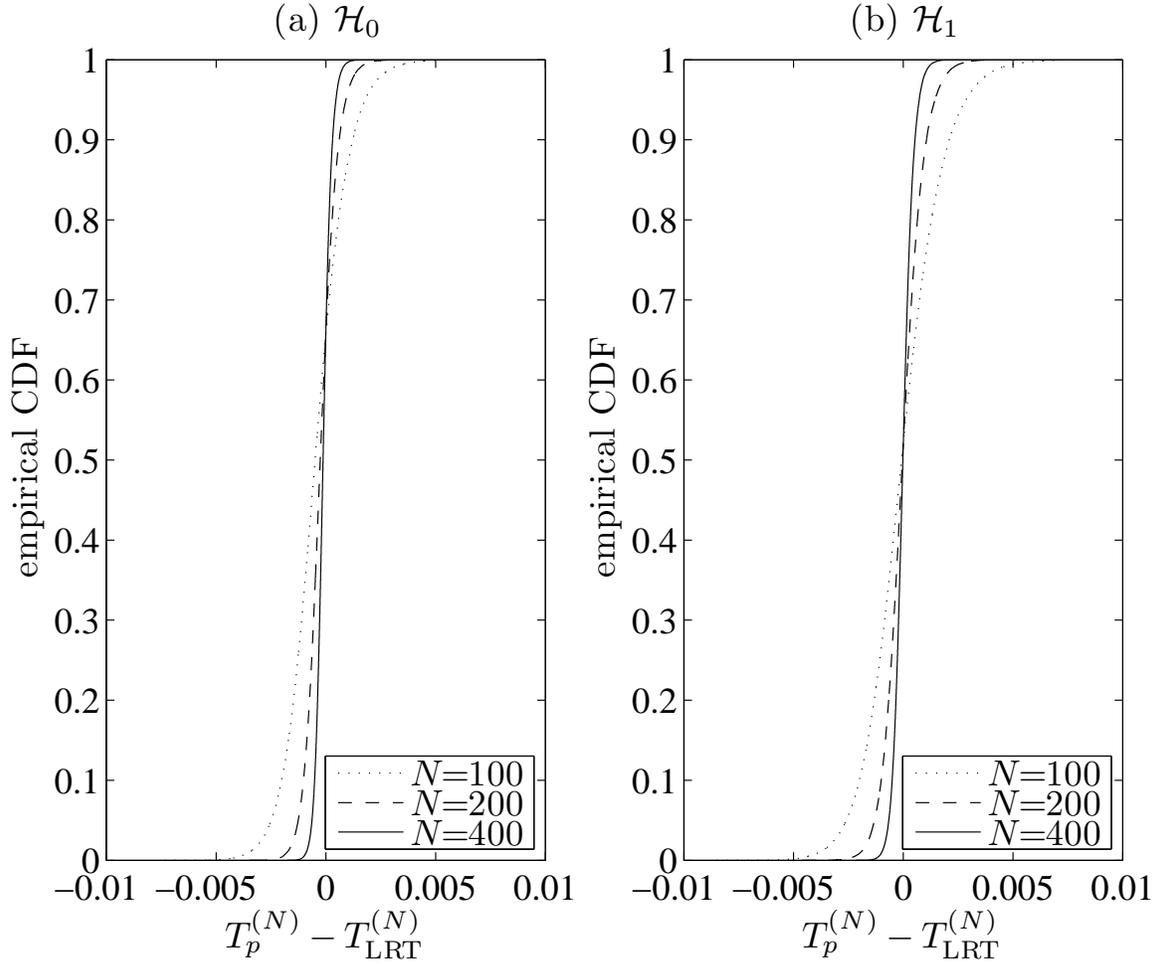}
\caption{Empirical CDFs of the difference of the proposed test statistic from the exact LRT statistic for $N=100,200,$ and $400$, under (a) $\mathcal{H}_0$ and (b) $\mathcal{H}_1$.} \label{Fig: numerical_CDF}
\end{figure}

The first result is to show the convergence of the proposed test statistic to the exact LRT statistic.
Fig.~\ref{Fig: numerical_LRT_avg} shows the statistical average of the exact LRT and that of the proposed test statistics for symbol energy per noise density $E_s/N_0=0, -5,$ and $-10$ [dB].
It can be seen that the statistical averages of the exact LRT and the proposed test statistics approach the same non-zero value as the number of samples tends to infinity under both hypotheses.
In particular, they coincide under $\mathcal{H}_1$ because the trace of $\bar{\bm{\Sigma}}$ are the same as that of $\bar{\bm{R}}_{\bar{\bm{x}}}$ .
Since the convergence w.p.~$1$ is hard to show by using Monte-Carlo simulations, we instead show the convergence in probability that is implied by the convergence w.p.~$1$ \cite{Papoulis_02}.
Fig.~\ref{Fig: numerical_CDF} shows the empirical cumulative distribution function (CDF) of the difference of the proposed test statistic from the exact LRT statistic for symbol energy per noise density $E_s/N_0=-5$ [dB] and $N=100,200,$ and $400$, where the empirical CDF is obtained by $10^5$ Monte-Carlo runs.
It can be seen that, as implied by Theorem~\ref{theorem: LRT_same}, the empirical CDF of the difference quickly converges to the unit step function as the number of samples increases under both hypotheses.

\begin{figure}[tbp]\centering
\includegraphics[width=6in]{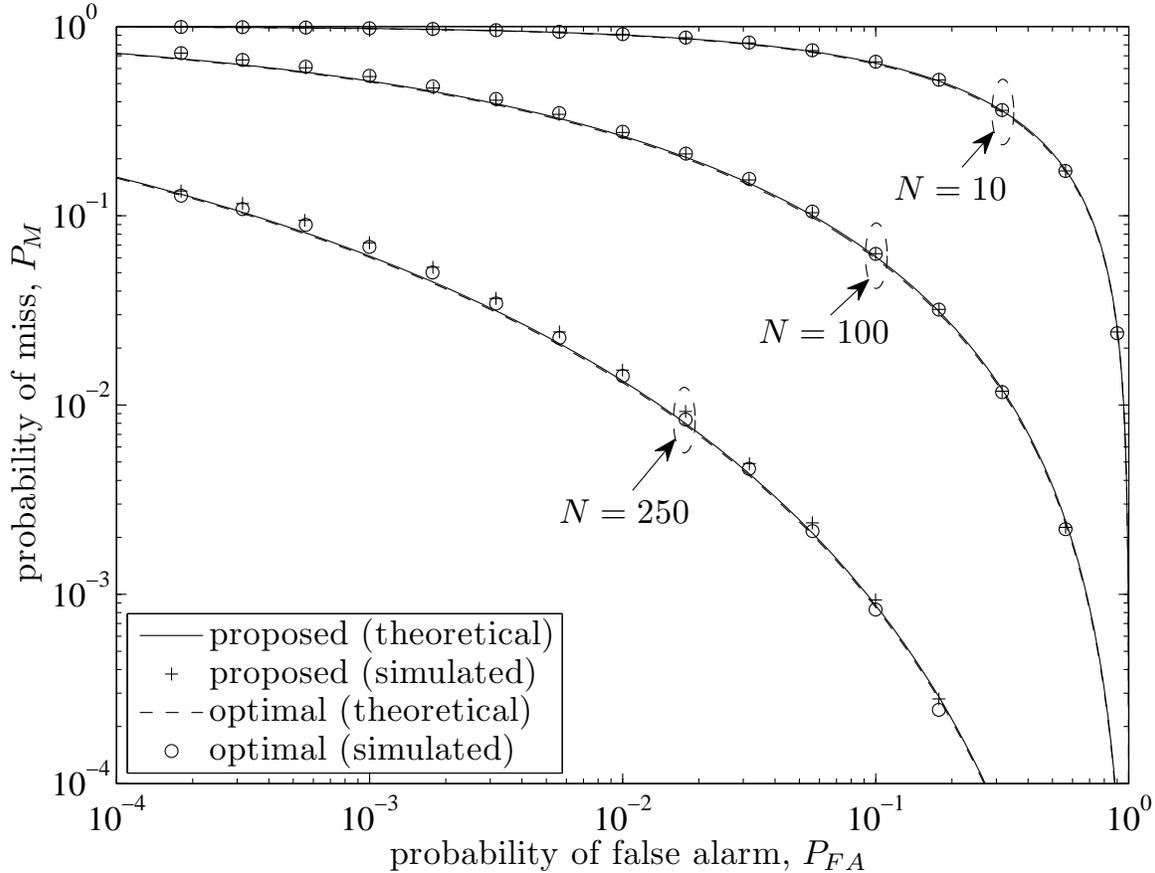}
\caption{ROC curves of the optimal and the proposed detectors.} \label{Fig: numerical_ROC}
\end{figure}

The second result is to compare the receiver operating characteristic (ROC) curves of the optimal detector that uses the exact LRT statistic (\ref{eq: WL_LRT}) and the proposed detector that uses the proposed test statistic (\ref{eq: proposed_LRT}).
As a common practice in computing the PDF of a quadratic function of Gaussian random vectors \cite{Grenander_59}, we approximate the statistics by gamma random variables to obtain the probability of miss $P_{M}\triangleq\Pr (\text{declare } \mathcal{H}_0|\mathcal{H}_1)$ and the probability of false alarm $P_{FA}\triangleq \Pr (\text{declare }\mathcal{H}_1|\mathcal{H}_0)$.
The CDF $F_X(x; a, b)$ of the gamma random variable $X$ with two parameters $a$ and $b$ is given by
\begin{equation}\label{eq: CDF_gamma}
F_X(x; a,b) = I\left(\frac{ax}{\sqrt{b}}, b-1\right), \;\; \forall x>0,
\end{equation}
where the Pearson's form of incomplete gamma function $I(u,p)$ is defined by
\begin{equation}
I(u,p)\triangleq\frac{1}{\Gamma(p+1)}\int_{0}^{u\sqrt{p+1}}t^p e^{-t}dt,
\end{equation}
and where the gamma function $\Gamma(x)$ is defined by $\Gamma(x)\triangleq\int_{0}^\infty t^{x-1}e^{-t}dt$ \cite{Papoulis_02}.
Note that the mean and the variance of the random variable whose CDF is $F_X(x; a, b)$ are given by $ab$ and $ab^2$, respectively.
Thus, $P_{M}$ and $P_{FA}$ of the optimal and the proposed detectors can be computed by using the conditional means and variances of the statistics under $\mathcal{H}_0$ and $\mathcal{H}_1$.
Fig.~\ref{Fig: numerical_ROC} shows the ROC curves of the optimal and the proposed detectors for symbol energy per noise density $E_s/N_0=-5$ [dB] and $N=10$, $100$, and $250$, where the theoretical results are evaluated by using (\ref{eq: CDF_gamma}) while the simulated results are obtained from $10^5$ Monte-Carlo runs.
It can be seen that the performance of the detectors is accurately approximated by using the gamma distributions.
It can be also seen that the proposed detector performs almost the same as the optimal detector does even when $N=10$.

\section{Conclusions}
In this paper, the asymptotic FRESH properizer is proposed as a pre-processor for the block processing of a finite number of consecutive samples of a DT improper-complex SOCS random process.
It turns out that the output of this pre-processor can be well approximated by a proper-complex random vector that has a highly structured frequency-domain covariance matrix for sufficiently large block size.
The asymptotic propriety of the output allows the direct application  with negligible performance degradation of the conventional signal processing techniques and algorithms dedicated to the block processing of proper-complex random vectors.
Moreover, the highly-structured frequency-domain covariance matrix of the output facilitates the development of low-complexity post-processors.
By solving the signal estimation and signal presence detection problems, it is demonstrated that the asymptotic FRESH properizer leads to the simultaneous achievement of computational efficiency and asymptotic optimality.
Further research is warranted to apply this pre-processor to various communications and signal processing problems involving the block processing of a DT improper-complex SOCS random process and to derive such almost optimal low-complexity post-processors.


\end{document}